\documentclass[11pt,a4paper]{article}

\usepackage{jhepmacro}
\usepackage{braket}
\usepackage{subfig}
\newcommand{\<}{\langle}
\renewcommand{\>}{\rangle}

\newcommand{\toTeal}[1]{{\color{teal}#1}}
\newcommand{\toOrange}[1]{{\color{orange}#1}}
\newcommand{\norm}[1]{\left\lVert#1\right\rVert}
\usepackage{pdflscape}


\title{\textbf{Improving 3d Ising OPE Coefficients with \\  Fuzzy Sphere Conformal Generators}}

\vspace{1.6cm}

\author{Giulia Fardelli, A. Liam Fitzpatrick, Emanuel Katz}

\affiliation[]{
\bigskip
Boston University, Boston, Massachusetts 02215, USA
}

\makeatletter
\renewcommand{\@email}[1]{#1}
\makeatother

\abstract{We use the $K$ special conformal generator in the Fuzzy sphere setup of the Ising CFT to determine primary states.  For $\Delta \lesssim 8$, we recover the known primaries and find several new ones, including in the parity-odd sector.
We then use these primaries to compute OPE coefficients.  We find that using primaries constructed from special-$K$ allows for better extrapolation of OPE coefficients to the CFT limit, because of the existence of an $O(1)$ gap between primaries and descendants in the spectrum of eigenvalues of $|K|^2$ which protects the primaries from strongly mixing with descendants.  We compare the CFT data we obtain with the Eigenstate Thermalization Hypothesis.}

\begin{document}

\maketitle
\section{Introduction}
How does one characterize the intrinsic information that a QFT contains, and what are the properties of this information?  
It is probably fair to say that the vast majority of work tends to prioritize the information contained in the lowest energy states for a fixed choice of conserved charges, and especially in the features of vacuum correlation functions and S-matrix elements.  On the other hand, as one moves up in the spectrum to states with large local energy densities and high particle multiplicities, interesting non-equilibrium phenomena of the system become relevant, such as whether it is ergodic, or whether it contains long-lived complex excitations which can partly preserve memory.  For better or for worse, in chaotic systems it is, essentially by definition, extremely difficult to probe the detailed behavior of individual states in such regimes, creating a sort of barrier for numeric methods.

These considerations are particularly relevant in the case of CFTs, which possess a discrete set of dynamical data, namely the dimensions of operators and their OPE coefficients.  For any individual CFT, obtaining the full set of this ``CFT data'' is one of the main goals of the conformal bootstrap program, since knowledge of this data is equivalent to a complete solution of the theory and is the starting point for further applications.\footnote{For instance, OPE data are essential for Hamiltonian Truncation methods, where non-conformal QFTs are formulated as relevant deformations of CFTs.  For this kind of application, one can extract some low-energy aspects of such QFTs, even from OPE data for moderate dimension operators. Convergence of low-energy states is expected to be faster, the more relevant the deforming operator. }  In chaotic CFTs,  this goal can never succeed if taken literally, because the CFT data of sufficiently high dimension operators is equivalent to a detailed knowledge of chaotic energy eigenstates of the system.  Indeed, as the spectrum becomes increasingly dense, these eigenstates become incredibly sensitive to the detailed aspects of the computational scheme used to extract the data.  Thus, in practice, one will eventually reach a barrier beyond which one must settle for probabilistic rather than deterministic predictions.    In systems with a discrete set of degrees of freedom, there have been various proposals, most notably the Eigenstate Thermalization Hypothesis (ETH), suggesting concrete properties of matrix elements which lead to ergodic behavior, or alternatively to memory retention.    Consequently, properties of the spectrum and OPE coefficients of large-dimension operators could potentially offer a path to understanding thermalization in Lorentz invariant systems. By contrast, encountering regimes where ETH fails suggests that those regimes may optimistically be amenable to controlled analytic or semi-analytic (say, based on a suitable Effective Field Theory with numerically determined Wilson coefficients) expansions, or more conservatively that at least fully numeric methods may still achieve high accuracy.
  Until recently, however, it was only possible to compute large-dimension OPE coefficients for various integrable CFTs.  This has now changed with the introduction of the Fuzzy Sphere technology, which allows for systematic computation of large dimension CFT data~\cite{Zhu:2022gjc}.  It is therefore desirable to find the most reliable method for extracting OPE coefficients using the Fuzzy Sphere.

In the Fuzzy Sphere (FS) setup, one considers $N$ fermions on the sphere in the lowest Landau level (LLL) of a monopole background.  The LLL degeneracy is broken through four-Fermi interaction terms.  The coefficients in front of these terms are tuned such that the ground-state of the system is near a second-order phase transition.  By the state-operator correspondence, the low-energy states then approximate the states of the CFT characterizing the universality of the transition.  Thus, from an EFT perspective, the FS Hamiltonian is given by a target CFT dilatation operator plus an infinite set of irrelevant operators.  These irrelevant operators violate the conformal symmetry and cause mixing between CFT states.  The mixing is of course most severe between states with degenerate energies (or dimensions).  The challenge is then to ameliorate this mixing, especially at high energies, where degeneracies in the spectrum are expected to increase.  In this work, we suggest one strategy to address the mixing:  We attempt to disentangle a primary from a descendant of another primary of similar energy, by using the special conformal generators.  The idea is to compute the eigenstates of the square of the special conformal generators, $|K|^2$, in the low-energy sector, and thus to separate primaries with very small $K$-eigenvalues from descendants with moderate $K$-eigenvalues.  Indeed, we will argue that only a few special descendants have $|K|^2 <1$, with the majority obeying $|K|^2 \gtrsim 1$ in the CFT limit.  Thus, armed with special-$K$, we will show that it is both easier to identify primaries, and to extrapolate their OPE coefficients to an infinite number of fermions. 

The paper is organized as follows.  We begin in section~\ref{sec:review} with a review of the setup and conformal generator construction.  In section~\ref{sec:primaries} we explain the origin of the gap in $|K|^2$ and use to it to extract the primaries.  Then in section~\ref{sec:OPE} we use the primaries to determine the OPE coefficients.  We then compare the extracted CFT data  to expectations from ETH in section~\ref{sec:ETH}.  We conclude in section~\ref{Sec:summaryandfuture} with a summary of results and  a discussion of future directions.  Several appendices contain  technical details and tables  presenting the large-$N$ extrapolation of OPE coefficients  between primaries and the operator $\sigma$ and $\epsilon$.  An ancillary Mathematica file provides the same data for all $N=8, \cdots, 16$.

\section{Review of Conformal Generators and Definitions of Primaries}
\label{sec:review}
\subsection{Fuzzy Sphere Setup}
The realization of the 3d Ising model via a system of non relativistic fermions on a fuzzy sphere as been discussed in several works, see for example the seminal paper~\cite{Zhu:2022gjc}. Here we provide a  brief review,  mainly to establish conventions. 

We consider  a system of two-flavor interacting non-relativistic fermions $\psi_i$ on a sphere with radius $R$ and in the presence of a magnetic monopole with total flux $4\pi s$ ($2s \in \mathbb{Z}$).  In this setting,  energy eigenstates get quantized into spherical Landau levels.  Restricting to the lowest Landau Level (LLL), which has $2s+1$ degeneracy,  the fermions can be expanded into a complete orthonormal basis of monopole harmonics $\Phi_m$:
\twoseqn{
\psi_i(\Omega)&=\frac{1}{R}\sum_{m=-s}^s \Phi_m(\Omega)c_{m,i}\, , \qquad \qquad (i=\uparrow,\,\downarrow)\, ,
}[]
{
\Phi_m(\Omega)&=\sqrt{\frac{(2s+1)!}{4\pi (s+m)!(s-m)!}}e^{i m\phi} \cos^{s+m}\left(\frac{\theta}{2}\right)\sin^{s-m}\left(\frac{\theta}{2}\right)\,.
}[][]

We restrict to the set of states at half-filling,  i.e.~states where the number of fermions is $N=2s+1$. The Hamiltonian reads
\eqna{
H&=R^2\int d^2\Omega\,  \CH\, , \\
\CH&=\sum_{n=0,1}\lambda_n \left( n_0 \frac{{\nabla}^{2n}}{R^{2n}} n_0- n_z \frac{{\nabla}^{2n}}{R^{2n}} n_z \right)-h n_x\, ,
}[FSH]
where 
we have defined
\eqna{
n_i=\psi^\dagger \sigma_i \psi\, ,
}[niDef]
with  $\sigma_0=\mathds{1}$ and $\sigma_{x, y,z}$ the usual Pauli matrices.  There is a $\mathbb{Z}_2$ symmetry that flips the internal spin $\sigma^z \rightarrow - \sigma^z$ and a spacetime parity symmetry $\CP$, which acts within the space of $j_z=0$ states at half-filling by swapping filled and unfilled modes.

As shown in~\cite{Zhu:2022gjc},  this system flows in the infrared (IR) to the 3d Ising CFT at an appropriate fixed point, corresponding to suitable choices of $\lambda_n$ and $h$.   In~\cite{Fardelli:2024qla}, we related these microscopic parameters to the Wilson coefficients of relevant and slightly irrelevant CFT operators using conformal perturbation theory -- see also~\cite{Lauchli:2025fii}.
Schematically, we can view the UV Hamiltonian density $\CH$ of the fuzzy sphere as a linear combination of the CFT stress tensor $T^{0}_0$ and an infinite number of  CFT operators\footnote{All  operators $\CO$ in this expansion must be $\mathbb{Z}_2$ even  $SO(3)$ scalars.  We assume without loss of generality that none of them are time derivatives since those can be removed at linear order by a basis rotation \cite{Fardelli:2024qla,Lao:2023zis} of the Hilbert space.}
\twoseqn{
\CH&=\gamma\lsp T^{0}_0+\sum_{\CO}g_{\CO}\,  \CO\, ,
}[HamiltonianDensity]
{
H&=\gamma \lsp H_{\rm CFT}+\sum_{\CO \text{ primary}} g_{\CO}\int d^2\lsp \Omega \CO( \Omega)\, , 
}[]
where in the second line the sum is restricted to conformal primaries
because descendant operators are total derivatives and can be integrated by parts. One can, in principle, tune the parameters of the microscopic Hamiltonian to set certain Wilson coefficients $g_{\CO}$ to zero, thereby reproducing the spectrum of known CFT operators. At first order in conformal perturbation theory,
\eqna{
H \ket{\CO_i}&=E_i \ket{\CO_i}\, ,  \qquad {E_i-E_0}=\gamma\Delta_i+ g_{\CO}\sum_{\CO}\delta E_i^{(\CO)} \, ,  \\
\delta E_i^{(\CO)}&=\bra{\CO_i} \int d^2\Omega \lsp  \CO(\Omega) \ket{\CO_i}\, , 
}[CPT]
where  $\delta E_i^{(\CO)}$ depends on the details of the external operators, and its exact form can be found in~\cite{Lao:2023zis,Lauchli:2025fii}. So given some known CFT dimensions $\Delta_i$ we can tune the parameters in the microscopic Hamiltonian such that we can minimize $\delta E_i^{({\CO})}$ for some operators $\CO_i$'s, whose dimensions are known for instance from numerical boostrap computations.  In our previous work we showed that,  at a fixed value of $N$, setting to zero $g_{\CO}$ for $\CO=\epsilon$ and $\epsilon^\prime$ correspond to the following choice of parameters in~\eqref{FSH}
\eqna{
V_0 &\equiv \frac{(2s+1)^2}{2\pi(4s+1)} \left( \frac{\lambda_0}{R^2}-s\frac{\lambda_1}{R^4}\right)=4.825\, ,  \\
V_1&\equiv  \frac{s(2s+1)^2}{2\pi(4s-1)}\frac{\lambda_1}{R^4}=1\, ,  \\
h&=3.158\, .
}[FPparameters]
Before moving on to the construction of the remaining conformal generators,  it is instructive to return to~\eqref{CPT} and express the energies in units of $R$
\eqna{
R (E_i-E_0)=\gamma \Delta_{i}+\sum_{\CO} g_{\CO} R^{3-\Delta_{\CO}}\lsp \delta E_i^{(\CO)}\, , 
}[]
where $\gamma$ can be interpreted as the stress-tensor Wilson coefficient and we fix it as a function of $N$ such that $\Delta_{\sigma}=0.518149$.  However,  this expression is incomplete for a theory defined on $\mathbb{R}\times S^2$, since curvature couplings must also be included \cite{Lauchli:2025fii}.\footnote{We thank Slava Rychkov and Yin-Chen He for discussions about such terms in the context of  fuzzy sphere calculations.}  In particular, the presence of the Ricci  scalar $\CR=\frac{2}{R^2}$ allows additional contributions that can be organized as higher-curvature corrections to each Wilson coefficient. In Appendix~\ref{appendix:CurvatureTerms} we show how such terms arise when integrating out a massive scalar field coupled to a light one  through a cubic interaction.  The complete expansion therefore takes the form
\eqna{
R(E_i-E_0)&=\Delta_{i}(\gamma + \gamma_{\CR} \CR + \cdots) +\sum_{\CO} \delta E_i^{(\CO)} R^{3-\Delta_{\CO}}\left( g_{\CO}+g_{\CR \CO}\CR+g_{\CR^2\CO} \CR^2+\cdots\right)\\
&= \Delta_{i}\left(\gamma +\frac{2 \gamma_{\CR}}{R^2}  + \cdots\right) +\sum_{\CO} \delta E_i^{(\CO)} R^{3-\Delta_{\CO}}\left( g_{\CO}+\frac{2g_{\CR \CO}}{R^2}+\frac{2g_{\CR^2\CO}}{ R^4}+\cdots\right)\, .
}[WC]
The sphere radius $R$ is related to the fermion number $N$ by
\eqna{
4\pi R^2 |B|=4\pi s \Rightarrow R^2=\frac{N-1}{2|B|} .
}[]

In practice, since $B$ is a dimensionful quantity, its value is equivalent to a choice of units.
For convenience, define the UV scale $\ell_{\rm UV} \equiv (\frac{N-1}{2 NB})^{1/2}$, 
in order to obtain the relation
\begin{equation}
R^2 = N \ell_{\rm UV}^2.
\end{equation}
When we write down the Wilson coefficients $g_\CO$ of irrelevant deformations around the CFT,  it is natural and convenient to set $\ell_{\rm UV}=1$, so that the $g_\CO$s are typically $O(1)$ (in the absence of tuning).  
From this perspective, however, any $O(1)$ choice of $\ell_{\rm UV}$ is equally well-motivated, and in fact 
 changing the definition of $R(N)$ can be absorbed into a change in the Wilson coefficients. Moreover, if we hold fixed the leading order relation $R^2 = N + O(N^0) + \dots$, changing the subleading $O(N^0)$ or higher terms in this relation can be absorbed into a change solely in the Wilson coefficients for the curvature terms.  For this reason, from this moment on we will set $R=\sqrt{N}$ and absorb any ambiguity into redefinition of $g_{ \CR^n \CO}$. 

\subsection{Review of Conformal Generators on the Fuzzy Sphere}
In a generic CFT, all conformal generators can be constructed from the stress tensor, the simplest case being the dilatation operator  (equivalent to the Hamiltonian).  In particular, the conserved charges associated with rotations, translations, and special conformal transformations arise from integrating appropriate components of the stress tensor against the corresponding conformal Killing vectors. On $\mathbb{R}\times S^2$ these generators takes the schematic form
\eqna{
Q_{\xi} =\int d^2\Omega \lsp  T^{0\mu} \xi_\mu\, , 
}[]
with $\xi_\mu$  the relevant conformal Killing vector.  Using embedding space coordinates $A=1,2,3$ for $S^2\subset \mathbb{R}^3$, translation and special conformal transformation generators read
\eqna{
P^{A}=\int d^2\Omega \left( \hat{x}^A T^{0}_0+i T^{0A}\right)\, , \qquad K^{A}=\int d^2\Omega \left( \hat{x}^A T^{0}_0-i T^{0A}\right)\,,
}[]
with $\sum_A (\hat{x}^A)^2=1$. An efficient way to construct $P^A$ and $K^A$ is to first consider the  combination\footnote{See~\cite[Appendix C]{Fardelli:2024qla} for details on the numerical implementation on the fuzzy sphere. }
\eqna{
\Lambda^A&=P^A+K^A=2 \int d^2\Omega\lsp\hat{x}^A T^{0}_0\, , \\
\Lambda_z&= \Lambda^3\, , \quad\, \Lambda_{\pm}=\Lambda^1\pm i\Lambda^2\, .
}[] 
The advantage of this construction is that $\Lambda^A$ depends only on the component $T^{0}_0$,   allowing us to construct it directly on the fuzzy sphere using the Hamiltonian density, in analogy with the construction of the dilatation operator.  The orthogonal combination can, in principle, be recovered from the algebra:
\eqna{
P^A-K^A=[D,\Lambda^A]\, .
}[]
However,  this is not the most efficient  or accurate approach for numerical purposes, and we will ultimately use a different criterion to disentangle $P^A$ and $K^A$.  Since we do not have access to the exact CFT stress tensor,  we construct $\Lambda^A$ directly on the fuzzy sphere starting from the Hamiltonian density $\CH$ as in~\eqref{HamiltonianDensity}.  Unlike the dilation generator, descendant operators now contribute non-trivially because they are no longer total derivatives in the presence of the $\hat{x}^A$ factor. Concretely, we define 
\eqna{
\tilde{\Lambda}^A=2 \int d^2 \Omega \hat{x}^A \CH=\Lambda^A+2 \sum_{\CO \text{ primary}} \int d^2\Omega \lsp \hat{x}^A \left( g_{\CO}\lsp \CO+\sum_{n=0}^\infty g_{ \nabla^{2n}  \CO} \nabla^{2n} \CO
\right)\,,
}[]
where $\nabla^{2n}\CO$ stands for a generic descendant and we have suppressed the curvature terms for compactness. Naively, if we want to improve $\tilde{\Lambda}^A$ in order to subtract out some of the descendants on the RHS above, we have to subtract a new microscopic fuzzy sphere operator from $\CH$ for each descendant that we want to remove.  However, it turns out that such a procedure it not necessary.  To see why, integrate  by parts and use $\nabla^2 \hat{x}^A=-2\hat{x}^A$ to find
\eqna{
\tilde{\Lambda}^A=\Lambda^A+2\sum_{\CO \text{ primary}} \underbrace{\left(g_{\CO}-\sum_n (-2)^n g_{\nabla^{2n} \CO} \right)}_{g_{\CO}^\prime}  \int d^2\Omega \lsp \hat{x}^A  \CO(\Omega)\, .
}[]
The point of this expression is that the contribution of all the descendants of a primary is proportional to the contribution of the primary itself.  Therefore, every descendant of a single primary operator $\CO$ can be tuned away in $\tilde{\Lambda}^A$ simultaneously using a single microscopic fuzzy sphere operator that flows to $\CO$.  Conveniently, the terms in the fuzzy sphere Hamiltonian itself mostly flow to $\epsilon, \epsilon'$, and $T^0_0$, which makes it easy to remove their descendants from $\tilde{\Lambda}^A$.  In practice, removing these extra descendants from $\tilde{\Lambda}^A$ therefore is equivalent to replacing the parameters $V_0, h$, and $\gamma$ in $\CH$ with new parameters 
\begin{equation}
V_0 \rightarrow  V_0+\delta V_0\,,\qquad  h\rightarrow h+ \delta h\, \qquad \gamma \rightarrow Z\times  \gamma \,
\end{equation} 
when we compute $\tilde{\Lambda}^A$.\footnote{To be clear, we  still use the parameters $V_0, h, \gamma$ in $\CH$ when we compute the Hamiltonian $H$, because there are no descendant contributions to be removed in that case. }

\begin{figure}\centering
\begin{minipage}{0.32\textwidth}
        \centering
        \includegraphics[width=1\textwidth]{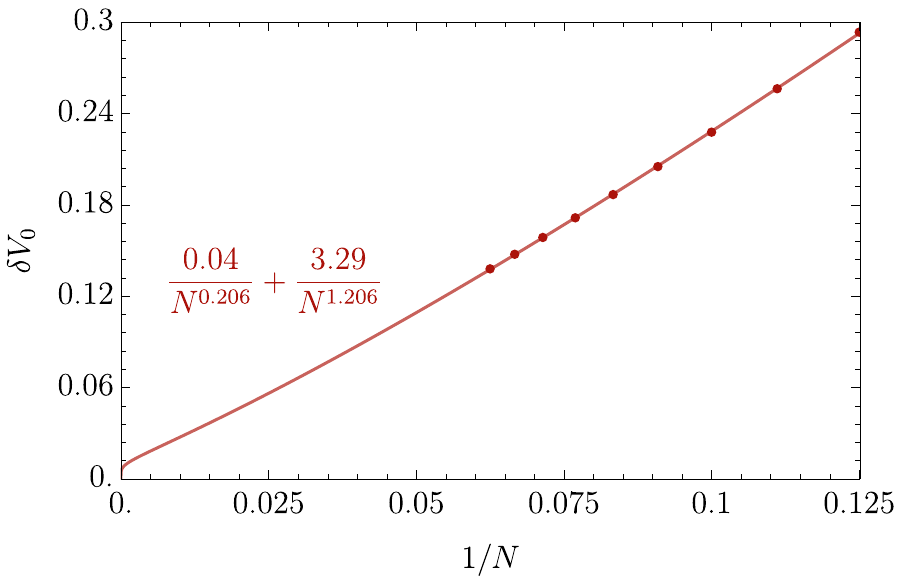}
    \end{minipage}
    \hfill
    % Second figure
    \begin{minipage}{0.32\textwidth}
        \centering
        \includegraphics[width=1\textwidth]{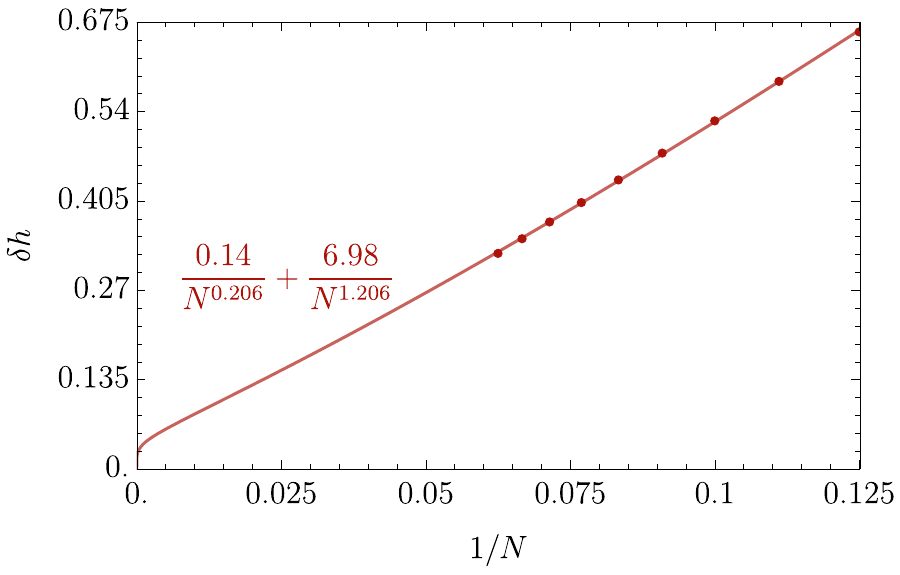}
    \end{minipage}
    \hfill
    % Third figure
    \begin{minipage}{0.32\textwidth}
        \centering
        \includegraphics[width=1\textwidth]{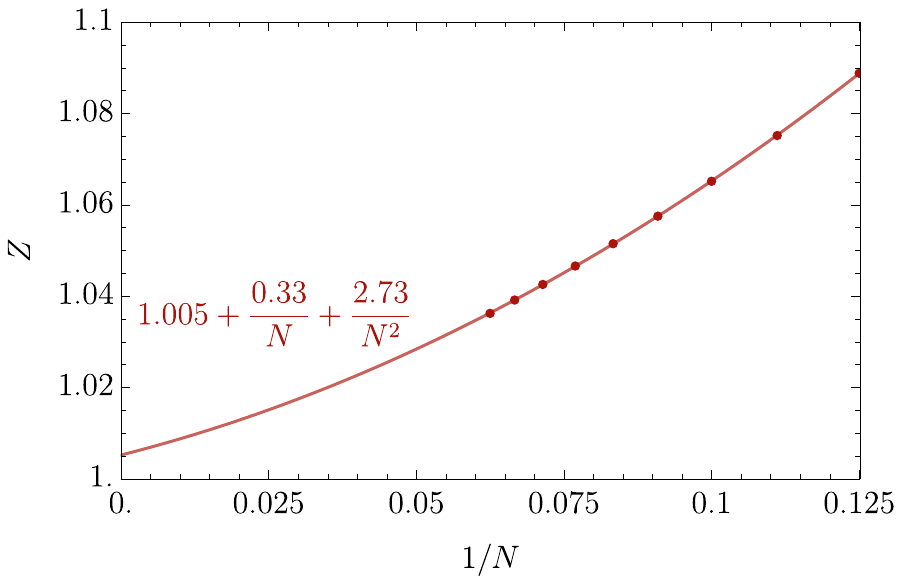}
    \end{minipage}
\caption{Plots of $\delta V_0$, $\delta h$ and $Z$ for different values of $N$. The leading power $N^{-0.206}$ in the first two plots is due to the descendant $\nabla^2 \epsilon$, and the leading power $1/N$ in the third plot is due to the descendant $\nabla^2 T^0_0$.  }\label{Fig:TuningGenerators}
\end{figure}

 In~\cite{Fardelli:2024qla}, we fixed these shifted parameters, for different values of $N$, by matching the following quantities to their CFT values:
\eqna{
\langle \text{vac}|\tilde{\Lambda}_z|\partial_A \epsilon \rangle=0\, , \qquad \langle \text{vac}|\tilde{\Lambda}_z|\partial_A \partial^2 \epsilon \rangle=0\, , \qquad \langle \epsilon|\tilde{\Lambda}_z|\partial_A \epsilon \rangle=\sqrt{2\Delta_\epsilon}\, .
}[]
In Fig.~\ref{Fig:TuningGenerators}, we show the result of this tuning procedure by plotting the shifts $\delta V_0$, $\delta h$ and $Z$ relative to their values in~\eqref{FPparameters} for various values of $N$.
Once these conditions on $\tilde{\Lambda}_z$ are imposed,  as anticipated before,  a more practical definition of $K_z$ and $P_z$ is to identify them with the matrix elements of $\tilde{\Lambda}_z$  that connect eigenstates  whose relative energies respectively decrease or increase
\eqna{
\langle E_i | K_z| E_j\rangle \equiv (K_z)_{ij}=\begin{cases}
(\tilde{\Lambda}_z)_{ij} &\quad \text{if}\quad \Delta_i-\Delta_j<0\,,  \\
0 &\quad \text{otherwise}\, ,
\end{cases}
\qquad \quad P_z=K_z^\dagger\,,
}[]
where $|E_i\rangle$ corresponds to an eigenstate with dimension $\Delta_i$, spin $\ell$ and in the $j_z=0$ sector.

\subsection{Building $|K|^2$}

To construct the other $K_A$ components,  we compute the matrix elements of $K_z$ in the $j_z=0$ as well as in the $j_z=1$ sectors, and use the Wigner-Eckart theorem to determine the $K_\pm$ matrix elements acting on the $j_z=0$ sector.  Specifically, 
 \begin{equation}
 \begin{aligned}
\frac{ \langle \ell_{\CO} , \pm 1  | K_\pm | \ell_{\CO'}, 0 \rangle}{\langle \ell_{\CO} , 0 | K_z | \ell_{\CO'}, 0 \rangle } &=  \begin{cases}
\frac{\ell_{\CO'}+2}{\sqrt{2(\ell_{\CO'}+2)(\ell_{\CO'}+1)} } &\quad \ell_{\CO} = \ell_{\CO'}+1  \, , \\
- \frac{\sqrt{\ell_{\CO'}-1}}{\sqrt{2 \ell_{\CO'}}} &\quad  \ell_{\CO} = \ell_{\CO'}-1 \, ,
\end{cases}\\
\frac{ \langle \ell_{\CO}, \pm 1 | K_\pm | \ell_{\CO'}, 0\rangle }{ \langle \ell_{\CO}, 1 | K_z | \ell_{\CO'}, 1\rangle} &= \eta_{\CO'} \eta_{\CO}^* \begin{cases}
 \frac{\sqrt{\ell_{\CO'}+1}}{\sqrt{2 \ell_{\CO'}}}  &\quad  \ell_{\CO} = \ell_{\CO'}+1 \, ,\\
- \frac{\sqrt{\ell_{\CO'}}}{\sqrt{2 (\ell_{\CO'}+1)}}  & \quad \ell_{\CO} = \ell_{\CO'}-1 \,  ,\\
\mp  \frac{\sqrt{\ell_{\CO'}(\ell_{\CO'}+1)}}{\sqrt{2}}  &\quad  \ell_{\CO} = \ell_{\CO'}\, ,
\end{cases}
\end{aligned}
\end{equation}
where the second line implicitly requires $\ell_\CO, \ell_{\CO'} \ge 1$.  The phase factor $\eta_\CO$ is defined by $|\ell_\CO, 1\rangle = \eta_\CO \frac{J_+ |\ell_\CO, 0\rangle}{ \norm{J_+ |\ell_\CO, 0\rangle}}$; the point is that numeric diagonalization of the fuzzy sphere Hamiltonian produces eigenvectors $|\ell_\CO, 0\rangle$ and $|\ell_\CO,1\rangle$ which are sometimes related to each other with $\eta_\CO=-1$ rather than $\eta_\CO=+1$.
  Then, $|K|^2$ is defined by
 \begin{equation}
| K|^2 = K^{\dagger A} K^A = K_z^\dagger K_z + K_+^\dagger K_+ + K_-^\dagger K_- .
\end{equation}
This matrix is a scalar and therefore diagonal in spin quantum numbers $\ell$ and $j_z$.   Consider the matrix element of $|K|^2$ between an in state with spin $\ell_{\CO}$ and an out state with spin $\ell_{\CO^\prime}$:
\begin{equation}
\begin{aligned}
\langle \ell_\CO, 0 | |K|^2 | \ell_{\CO'}, 0 \rangle &\equiv \sum_i \langle \ell_\CO , 0 | K_z^\dagger | \ell_i , 0  \rangle \langle \ell_i , 0 | K_z | \ell_{\CO'}, 0\rangle \\
& +\sum_i \langle \ell_\CO , 0 | K_+^\dagger | \ell_i, 1  \rangle \langle \ell_i, 1 | K_+ | \ell_{\CO'}, 0\rangle \\ 
& + \sum_i \langle \ell_\CO  , 0 | K_-^\dagger | \ell_i, -1  \rangle \langle \ell_i, -1 | K_- | \ell_{\CO'}, 0\rangle .
\end{aligned}
\end{equation}
When $\ell_{\CO}$ and $\ell_{\CO'}$ are both $\ge 2$, all the matrix elements that appear in the formula above can be inferred from the  matrix elements of $K_z$ in the $j_z=1$ sector:
\begin{equation}
\begin{aligned} 
& \langle \ell_\CO, 0 | |K|^2 | \ell_{\CO'}, 0 \rangle  \stackrel{\ell_{\CO}, \ell_{\CO'}\ge 2}{=} \delta_{\ell_{\CO}, \ell_{\CO'}}  \eta_{\CO}^* \eta_{\CO'} \\
 &\qquad  \times  \sum_i  {}\langle \ell_{\CO}, 1 | K_z^\dagger | \ell_i , 1\rangle  \langle \ell_i , 1 | K_z | \ell_{\CO'}, 1\rangle  
 \begin{cases}
 \frac{\ell_{\CO}(2\ell_{\CO}-1)}{(\ell_{\CO}-1)(\ell_{\CO}+1)} &\quad \ell_i = \ell_{\CO}-1  \, ,\\
  \frac{(\ell_{\CO}+1)(2\ell_{\CO}+3)}{\ell_{\CO}(\ell_{\CO}+2)} &\quad  \ell_i = \ell_{\CO}+1 \\
  \ell_{\CO}(\ell_{\CO}+1) &\quad  \ell_i = \ell_{\CO}
 \end{cases}
\end{aligned}
\end{equation}
 and therefore the phases factors $\eta_\CO$ can be rotated away by rephasing the basis states.
 A more general formula that applies to all spins is
 \begin{equation}
\begin{aligned} 
&\langle \ell_\CO, 0 | |K|^2 | \ell_{\CO'}, 0 \rangle = \\
 &\qquad \delta_{\ell_{\CO}, \ell_{\CO'}} \sum_i 
 \begin{cases}
  (2-\frac{1}{\ell_{\CO}} ) \langle \ell_{\CO} , 0 | K_z^\dagger | \ell_i , 0  \rangle \langle \ell_i , 0 | K_z | \ell_{\CO'}, 0\rangle &\quad  \ell_i = \ell_{\CO}-1\, , \\
  (2+\frac{1}{\ell_i} ) \langle \ell_{\CO} , 0 | K_z^\dagger | \ell_i , 0  \rangle \langle \ell_i , 0 | K_z | \ell_{\CO'}, 0\rangle &\quad  \ell_i = \ell_{\CO}+1\, ,  \\
  \ell_{\CO}(\ell_{\CO}+1) \langle \ell_{\CO}, 1 | K_z^\dagger | \ell_i , 1\rangle  \langle \ell_i , 1 | K_z | \ell_{\CO'}, 1\rangle \eta_\CO^* \eta_{\CO'} & \quad \ell_i = \ell_{\CO}\, ,
 \end{cases}
  \end{aligned}
\end{equation}
  but 
in this case the relative phases $\eta_\CO$  between $|\ell_\CO,\pm 1\>$ and $|\ell_\CO,0\>$
do not factor out   in the formula and it becomes necessary to  keep them for the purpose of determining the primary states.  In practice, we use both formulas and check that they agree.  Alternatively, one could simply compute the matrix elements of $K_\pm$ directly,  rather than inferring them from the matrix elements of $K_z$, however this increases the computational cost.

\section{Extracting Primaries}
\label{sec:primaries}
\begin{figure}
\begin{center}
\includegraphics[width=0.6\textwidth]{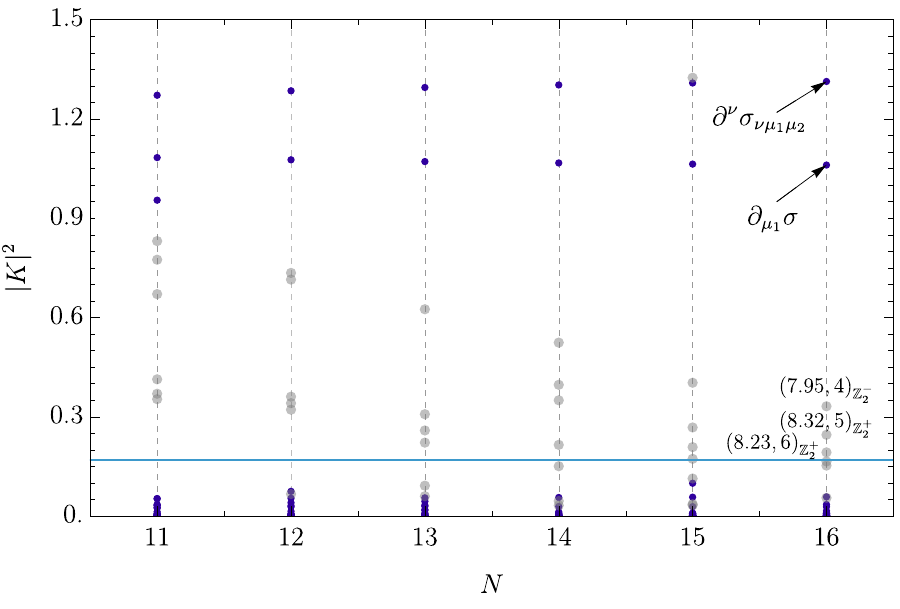}
\caption{$|K|^2$ values between 0 and 1.5 for increasing $N$; a gap visibly opens up in the range between small $|K|^2$ and $|K|^2=1$.  Gray points are for states with $\Delta>7.8$, which have the largest errors since they are closest to the cutoff.  The teal horizontal line is our $|K|^2$ threshold for keeping primaries.  A few descendants with $|K|^2$ close to 1 have been explicitly identified as descendants of $\sigma$ and $\sigma_{\mu_1\mu_2\mu_3}$, and the (dimension,spin) $(\Delta,\ell)$ is indicated for three states just above our $|K|^2$ threshold at $N=16$ (these discarded states are shown in more detail in Appendix \ref{app:discarded}).  Only the first 150 eigenvalues in the $j_z=0$ sector are used to make the plot.}
\label{fig:gapOpening}
\end{center}
\end{figure}
\subsection{Gap in $|K|^2$ Eigenvalues}

The key point that makes $|K|^2$ extremely effective for obtaining primaries is that in the CFT limit, there is an $O(1)$ gap in its spectrum of eigenvalues between primaries and descendants, as we will explain.  Actually, this statement requires a slight asterisk, because there are special descendant states that have small $|K|^2$ values.  Fortunately, these `almost-primary' descendant states are well-understood and completely classified. In the 3d Ising model, there is exactly one such almost-primary with $|K|^2 < 1$ at $\ell=0,3,5,7,9 \dots$ and none at $\ell=1,2,4,6,8, \dots$.

To begin,  the eigenvalues of $|K|^2$ in the CFT limit can easily be obtained using the fact that the conformal Casimir operator $\CC$ is
\begin{equation}
\CC = D(D-3) + J^2 - |K|^2
\end{equation}
and therefore for a descendent of dimension $\Delta_0+n$ (i.e.~a `level $n$' descendant) and spin $\ell$ of a primary with dimension $\Delta_0$ and spin $\ell_0$, we have (using $\CC = \Delta_0(\Delta_0-3)+ \ell_0(\ell_0+1)$, $D=\Delta_0+n$, and $J^2 =\ell (\ell+1)$)
\begin{equation}
\begin{aligned}
 |K|^2 | n, \ell\rangle &= (D(D-3)+J^2 - \CC) |n,\ell\rangle   = \lambda(\Delta_0, n, \ell, \ell_0) | n, \ell\rangle, \\
&   \lambda(\Delta_0,n,\ell, \ell_0)  \equiv (2 \Delta_0+n-3) n + (\ell+\ell_0+1)(\ell-\ell_0)  .
 \end{aligned} \label{Ksqrdvalues}
 \end{equation}
 Note that when $n\ge 2$ and $\ell_0 \ge 1$ (as well as when $n\ge 3$ and $\ell_0=0$),
 \begin{equation}
 \lambda(\Delta_0,n,\ell, \ell_0)   =  \left(2 \ell_0+1\right) \left(\ell-\ell_0+n\right)+\left(\ell-\ell_0\right){}^2+n \left(2(\tau _0-1) + (n-2) \right)\, ,
 \end{equation}
 (where $\tau_0 \equiv \Delta_0 - \ell_0$ is the twist of the primary, which by unitarity is $\tau_0 \ge 1$ for $\ell_0 \ge 1$ and $\tau_0 \ge 1/2$ for $\ell_0=0$)  is manifestly a sum over non-negative terms in unitary theories and therefore at least as large as each individual term.
One can show from (\ref{Ksqrdvalues}) that in any unitary 3d CFT, the eigenvalues of $|K|^2$ are all $\ge 1$ unless $|n,\ell\>$ is either a primary (in which case $|K|^2=0$), or a descendant  of the form $\partial_{\mu_1} C^{\mu_1 \mu_2 \dots \mu_{\ell+1}}$ of an almost-conserved current $C$ (in which case $|K|^2 = 2 (\tau_0 -1)$), or the descendant $\partial^2 \phi$ of an almost-free scalar field $\phi$ (in which case $ |K|^2 = 2 (2\Delta_0-1))$.\footnote{Here is the proof. The eigenvalue $ \lambda(\Delta_0,n,\ell, \ell_0)$ is a monotonically increasing function of $\Delta_0$ and $\ell$, so is minimized by $\Delta_0$ at the unitarity bound ($\Delta_0 \ge 1/2$ for $\ell_0=0$,  $\Delta_0 \ge \ell_0+1$ for $\ell_0 \ge 1$) and $\ell$ at the minimum allowed value by addition of angular momentum.  Consider $\ell_0 \ge 1$. At $n=1$, $|K|^2 \ge \lambda(\ell_0+1,1,\ell, \ell_0) = (\ell+\ell_0)(\ell-(\ell_0-1))$, which vanishes iff $\ell = \ell_0-1$ (the minimum $\ell$ allowed), the case where the state is the divergence of a conserved current, and otherwise is $\ge 2\ell+1 \ge 1$.  At $n\ge 2$, $|K|^2 \ge \lambda(\ell_0+1, n, \ell_0-n, \ell_0) = 2n(n-1) \ge 4$.  Next consider $\ell_0=0$.  At $n=1$, we have $\ell=1$, and $|K|^2 \ge \lambda(1/2,1,1,0) = 1$.  At $n \ge 2$, we have $|K|^2 \ge \lambda(1/2,n,\ell,0) = \ell(\ell+1)+ n(n-2)$, which vanishes iff $\ell=0, n=2$, the case where the state is $\partial^2 \phi$ of a free field $\phi$, and otherwise is $\ge 2$.  In this paper, we only consider integer spin states.}

Now consider the states in the 3d Ising model with $0 \le |K|^2 \le 1$, i.e.~the `almost-primary' descendants.  If such a state is a scalar, then it is either a $n=2$ descendant of a scalar primary, with $|K|^2 = 2(2\Delta_0-1)$, or a $n=1$ descendant of a vector $\ell_0=1$ primary, with $|K|^2 = 2(\Delta_0-2)$.  Vector primaries do not arise in the 3d Ising model until fairly high dimension, so there are no almost-primaries of the latter kind.  For the former, $|K|^2<1$ implies that the primary dimension must have $\Delta_0 < 0.75$, and therefore the only such case is $\partial^2 \sigma$ (with $|K|^2 = 0.0726$).  On the other hand, in order for a state with spin $\ell>1$ to be an almost-primary descendant, it must be a $n=1$ descendant $\partial \cdot C$ of a $\ell_0 = \ell+1$ primary $C$, with $|K|^2 = 2 (\tau_0-1)$.  Therefore, $|K|^2 <1$ requires the primary to have very low twist, $\tau_0<1.5$. In the 3d Ising model, there are two families of such states, one family with $\tau_0 \approx 1$ and another family with $\tau_0 \approx 1.5$.  The former family is the $\mathbb{Z}_2$-even leading  double-twist family $[\sigma, \sigma]_{\ell_0}$, $\ell_0 = 4,6, 8, \dots$, whose twists are known very accurately from the lightcone bootstrap \cite{Simmons-Duffin:2016wlq}. Its $(n=1, \ell=\ell_0-1)$ descendants have $|K|^2$ values that start at $0.045$ (for $\ell_0=4, \ell=3$) and approach $2(2\Delta_\sigma -1) \approx 0.0726$ at large $\ell$.  The second such family is the $\mathbb{Z}_2$-odd  leading triple-twist states $[\sigma, \sigma, \sigma]_{\ell_0}$, $\ell_0=6,7,8,\dots$.  Their twists are not known as accurately, but are expected to be close to $\tau_0 \approx 3 \Delta_\sigma \approx 1.5$ and therefore have $|K|^2 \approx 1$. Consequently, for $|K|^2 <1$, the only `almost-primary' states are $\partial^2 \sigma$ ($\ell=0$) and $\partial \cdot [\sigma , \sigma]_{\ell_0}$ ($\ell=3,5,7,9, \dots$). 

In Fig.~\ref{fig:gapOpening}, we show this gap in our numeric spectrum as a function of $N$. From the preceding discussion, in the CFT limit there should be no eigenvalues of $|K|^2$ between 0.0726 and 1.  Because of UV corrections from the fuzzy sphere, at finite $N$ we do see states in this range, but they are mostly from high energy states near the cutoff. In the plot, states with $\Delta >7.8$ are depicted in gray to make this more apparent.  Moreover, as $N$ is increased, even these states can be seen to follow a trend taking them towards $|K|^2 =0$.\footnote{For each $N$, states with $\Delta \gtrsim 8.5$ are completely discarded by the construction since we use only the first 150 eigenstates of $H$ when we construct $K^A$; the exact maximum value of $\Delta$ depends on $N$.}

Because $|K|^2$ is quadratic in the level $n$ and linear in twist $\tau_0$, most descendants actually have $|K|^2$ eigenvalues much larger than 1, and so decouple quickly from the $|K|^2 \ll 1$ subspace.  For instance, in the range $1< |K|^2 < 4$, the only  descendants  are again just more almost conserved currents, plus the state $\partial_\mu \sigma$ which has $|K|^2 = 2 \Delta_\sigma \approx 1.036$.  Most states with small $\tau_0$ lie on multi-twist trajectories  and so are mainly an issue at large spin.\footnote{Almost-conserved currents in the range $1<|K|^2<4$ include  the $(n=1, \ell=\ell_0-1)$ 
descendants of the $\mathbb{Z}_2$-odd double-twist family $[\sigma, \epsilon]_{\ell_0}$, with $|K|^2$ eigenvalues that oscillate for even vs odd spin within the range $2.3 > |K|^2 > 1.2$ and asymptote at large $\ell$ to $|K|^2 = 2(\Delta_\epsilon + \Delta_\sigma -1) \approx 1.86$,  and the $(n=1, \ell=\ell_0-1)$  descendants of the $\mathbb{Z}_2$-even double-twist family $[\epsilon, \epsilon]_{\ell_0}$, with $|K|^2$ values beginning at 2.84 ($\ell_0=4, \ell=3$) and asymptoting to $2(2\Delta_\epsilon-1) \approx 3.65$ at large $\ell$. There are also expected to be higher-multiplicity-twist trajectories, e.g.~$[\sigma, \sigma,\sigma,\sigma]_{\ell_0}$, in this range.}

The essential point here is that the spectrum of $|K|^2$ creates an $O(1)$ gap between primaries and descendants (modulo the well-understood `almost-primaries' mentioned above), and therefore the space of primary states picked out by selecting small $|K|^2$  is robust\footnote{There can still be mixing between states within the space of primaries, which we discuss further on.} to small UV effects from the fuzzy sphere regulator as long as we are not too close to the UV cutoff (where the UV effects cease to be small).

\subsection{Numeric Results for the Spectrum} 
\begin{figure}[t]
\center
    \begin{minipage}{0.48\textwidth}
        \centering
        \includegraphics[width=1\textwidth]{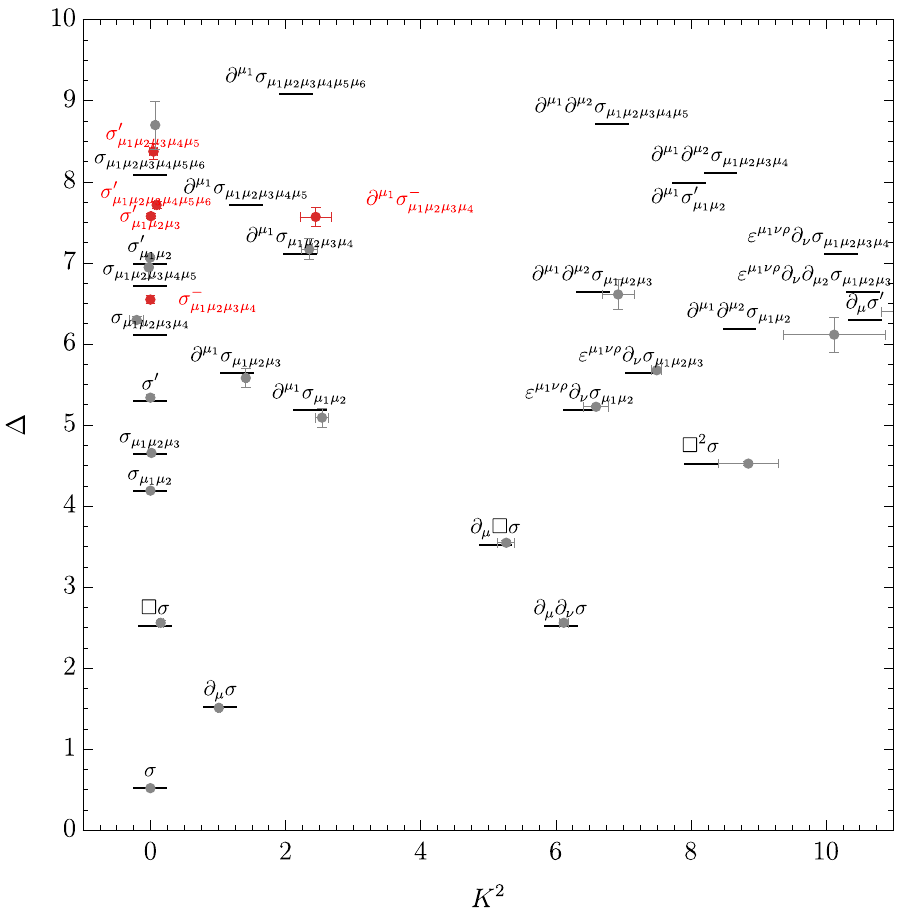}
    \end{minipage}
    \hfill
    % Third figure
    \begin{minipage}{0.48\textwidth}
        \centering
        \includegraphics[width=1\textwidth]{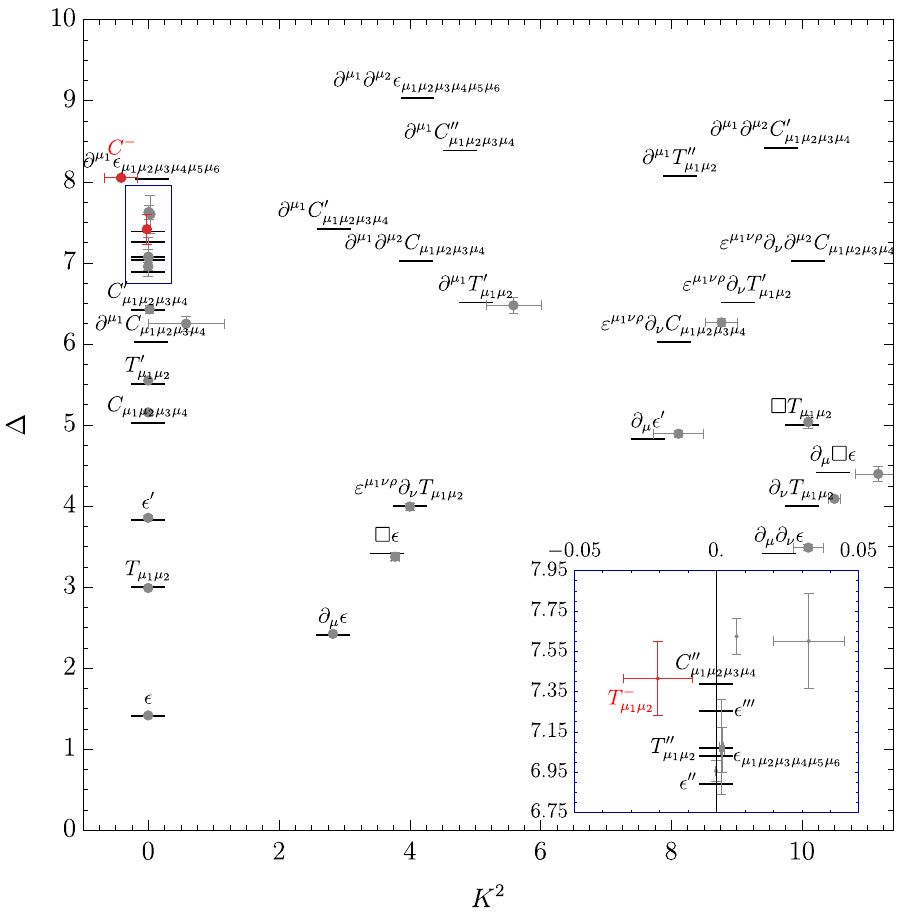}
    \end{minipage}
\caption{Plot of the $|K|^2$ and $\Delta$ values of the states obtained with our construction. Left is $\mathbb{Z}_2$ odd and right is $\mathbb{Z}_2$ even.  Solid horizontal lines indicate the location of states predicted by the bootstrap.  Previously unknown primaries are shown in red.  The right plot has an inset blowing up the region at $6.75 < \Delta < 8$ near $|K|^2=0$, where many primaries lie close together.  Error bars are estimated as described in the text.  }\label{Fig:Ksquared}
\end{figure}
On the fuzzy sphere, the main errors in $|K|^2$ come from the presence of irrelevant operators.  Schematically,
\begin{equation}
K = K_{\rm CFT} + O
\end{equation}
where $O$ is the integral of a local irrelevant operator.  Therefore, 
\begin{equation}
|K|^2 = |K_{\rm CFT}|^2 + K_{\rm CFT}^\dagger O + O^\dagger K_{\rm CFT} + O^\dagger O.
\end{equation}
A nice property of $|K|^2$ is that, acting within the space of primaries, the first three terms vanish and therefore the errors are quadratic in the deformation.\footnote{Mixing between the space of primaries and descendants is linear in $O$, so the correction to the eigenvalues of $|K|^2$ is still quadratic in $O$.  }  This statement should also hold approximately for the `almost-primaries', where $K_{\rm CFT}$ is small but nonzero.  If we work with states below a fixed upper limit $\Delta_{\rm max}$ in dimension, then as we increase $N$ in the fuzzy sphere the corrections from $O$ all decrease.  Consequently, at sufficiently large $N$, we should see that there are no $|K|^2$ eigenvalues in some window between $\sim 0.0726$ and $1$, as discussed in the previous subsection. Therefore, we should be able to select primaries and almost-primaries simply taking all states with $|K|^2$ less than any threshold that lies in this window. On the other hand, at fixed $N$ as the dimension $\Delta_{\rm max}$ increases, the size of corrections from the irrelevant operators  increases on the states near $\Delta_{\rm max}$.  Because of this,  we do see some $|K|^2$  eigenvalues in the middle of the window $0.0726 < |K|^2 < 1$, especially at small $N$ or at larger dimensions. 
  In practice, we take our threshold for primaries to be $|K|^2 \le 0.17$.\footnote{At $N=16$, only three states below our $\Delta_{\rm max}$ are in the range $0.17 < |K|^2 < 1$, so this threshold for $|K|^2$ picks out the primaries that appear to become reliable by $N=16$. } The corresponding conformal multiplets are generated acting with $P_z$,  for instance for a spin-$\ell_0$ primary $\CO$
\eqna{
\langle \CO |  |K|^2 |\CO\rangle \leq 0.17 \, ,  \qquad
 |\partial \CO\rangle \propto P_z  |\CO \rangle\, .
}[]

In Fig.~\ref{Fig:Ksquared}, we show the expectation values of $|K|^2$ against the corresponding operator dimensions in the $\mathbb{Z}_2$ even and odd sectors.\footnote{For primaries we show the exact eigenvalues, while for the descendants, obtained acting with $P$, their expectation values. } We compare the values for different $N$ with the CFT predictions in~\eqref{Ksqrdvalues}. Notice that some descendants also exhibit a small value of $|K|^2$, but this is fully consistent with the CFT expectations from~\eqref{Ksqrdvalues} for descendants of almost conserved currents.  After identifying and excluding these special cases, we can safely assume that all the remaining states with small $|K|^2$ are genuine primaries. Using this criterion we were  able to identify all primary operators predicted by the numerical bootstrap up to dimension 8.7, as well as  additional previously unidentified ones\footnote{After this work was completed, we learned that work in progress on the stress tensor bootstrap contains many new previously unknown operators as well. Comparing our results with theirs \cite{PolandInPrep}, all our previously unidentified primaries are also seen in the stress tensor bootstrap.  We thank the authors of \cite{PolandInPrep} for sharing their preliminary results with us. } --- see Table \ref{tab:NewPrimaries}.\footnote{ The convention used in the table is that errors in parentheses indicate the uncertainty in the final digits - for instance, $7.53(6)$ means $7.53 \pm 0.06$ whereas $7.41(18)$ means $7.41 \pm 0.18$.  Errors are estimated by taking half the difference between $N=\infty$ and $N=16$.} 
In Fig.~\ref{Fig:conformalMultiplets} and~\ref{Fig:conformalMultipletsEven} we show the $|K|^2$ eigenvalues as well as the dimensions computed as  $\langle \CO | H| \CO \rangle$ for all the identified primaries as  functions of $N$.   For low-lying operators,
$|K|^2$  is extremely close to zero for every $N$,  while for higher-energy states closer to the cutoff, $|K|^2$ only approaches zero as a function of $N$.  In cases where bootstrap data are available, our extrapolated dimensions show good agreement.  The states shown in Fig.~\ref{Fig:epsilonPrimes} and~\ref{Fig:Cprimes} deserve special mention. In both cases there is clear evidence of mixing between two nearby states. To address this, we consider the two-dimensional subspace spanned by the corresponding small-$|K|^2$ eigenvectors and re-diagonalize $H$ within this subsector. This procedure improves the the extracted dimensions as functions of $N,$ and generally leads to more reliable extrapolations of OPE coefficients.\footnote{We performed the same procedure for the two $\mathbb{Z}_2$-odd spin-6 operators for extracting OPE coefficient. We do not show this case explicitly since there was no significant change in either $|K|^2$ or the extracted dimensions.}
In Fig.~\ref{Fig:conformalMultiplets} and \ref{Fig:conformalMultipletsEven}, we show the conformal multiplets obtained for various spin sectors and for different values of $N$,  constructed by acting with $P_z$ on the primaries defined as above.  An 
advantage of this construction is that the spectra exhibit no trace of the avoided level crossings emphasized in~\cite{Lauchli:2025fii}.  Here the authors observe that when the energy levels were plotted as a function of $N$, certain states --- primarily $\square \epsilon$ and $\epsilon^\prime$ and their spin 1 descendants --- repel each other as a $N$ varies. In our approach, these mixings are automatically resolved and each operator can be tracked unambiguously as a smooth function of $N$.  Finally, we note that higher-dimensional descendants, obtained through repeated action of $P_z$, become increasingly degraded, as errors accumulate at each step. 
\begin{table}
\begin{center}
\begin{tabular}{|c|c|c|c|c|}
%New & primaries \\ 
\hline
$\CO$ & $\Delta$ & $\ell$ & $\mathbb{Z}_2$ & $\mathcal{P}$ \\
\hline
$\sigma'_{\mu_1 \mu_2 \mu_3} $ & 7.53(6)   & 3 & - & + \\
\hline
$\sigma^{-}_{\mu_1 \mu_2 \mu_3 \mu_4}$ &  6.53(5)   & 4 & - & - \\
\hline
$\sigma'_{\mu_1 \mu_2 \mu_3 \mu_4 \mu_5}$ & 8.37(10)  & 5 & - & + \\
\hline
$\sigma'_{\mu_1 \mu_2 \mu_3 \mu_4 \mu_5 \mu_6}$ & 7.73(5)  & 6 & - & + \\
\hline
$T^{-}_{\mu_1 \mu_2} $ & 7.41(18)  & 2 & + & - \\
\hline
$C^{-}_{\mu_1 \mu_2 \mu_3 \mu_4} $ & 8.04(2) & 4 & + & - \\
\hline
\end{tabular}
\end{center}
\caption{New primaries obtained from our analysis, along with their dimension $\Delta$, spin $\ell$, internal $\mathbb{Z}_2$ spin $\sigma^z \rightarrow - \sigma^z$ and parity $\CP$ quantum numbers.}
\label{tab:NewPrimaries}
\end{table}%

%Starting from the $\mathbb{Z}_2$ odd in order of spin
\begin{figure}
  \centering
  \subfloat[$\sigma$]{
    \includegraphics[width=0.29\textwidth]{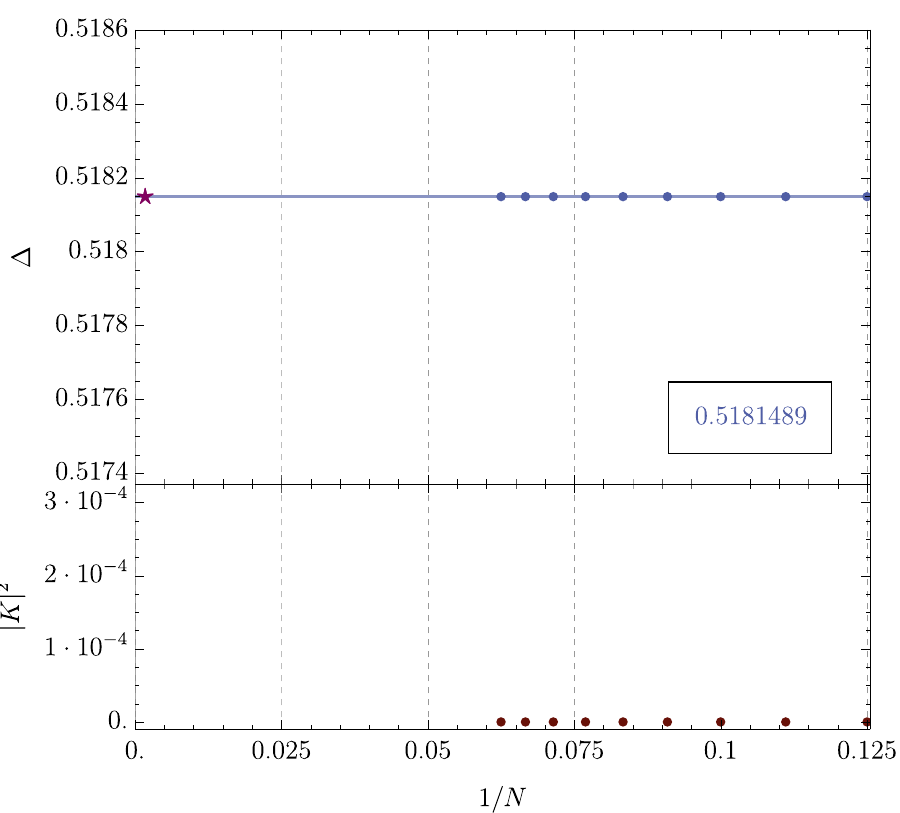}
  }
  \hfill
  \subfloat[$\sigma^\prime$]{
    \includegraphics[width=0.29\textwidth]{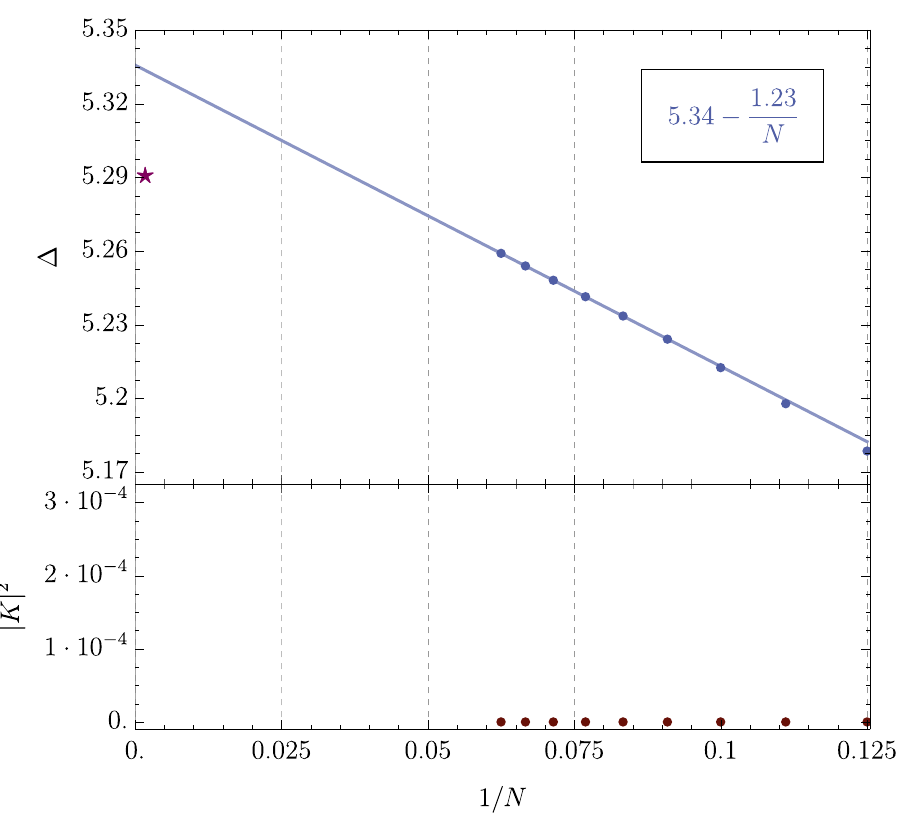}
  }
    \hfill
  \subfloat[$\sigma_{\mu_1\mu_2}$]{
    \includegraphics[width=0.29\textwidth]{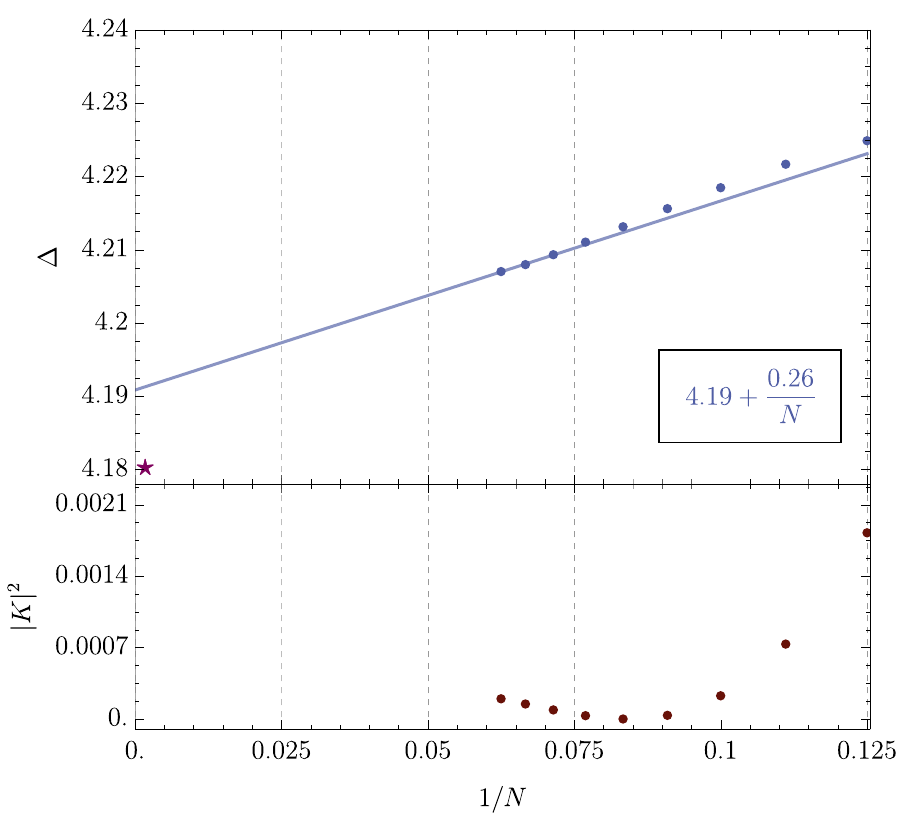}
  }
  \vspace{0.1cm}
    \subfloat[$\sigma_{\mu_1\mu_2}^\prime$]{
    \includegraphics[width=0.29\textwidth]{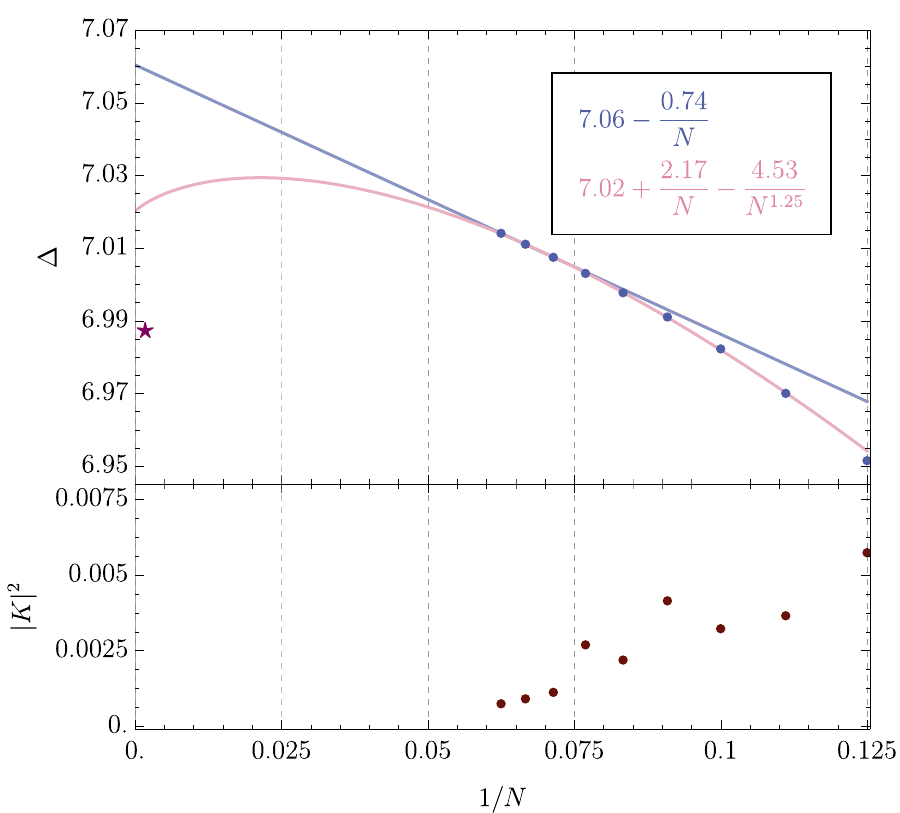}
  }
  \hfill
  \subfloat[$\sigma_{\mu_1\mu_2\mu_3}$]{
    \includegraphics[width=0.29\textwidth]{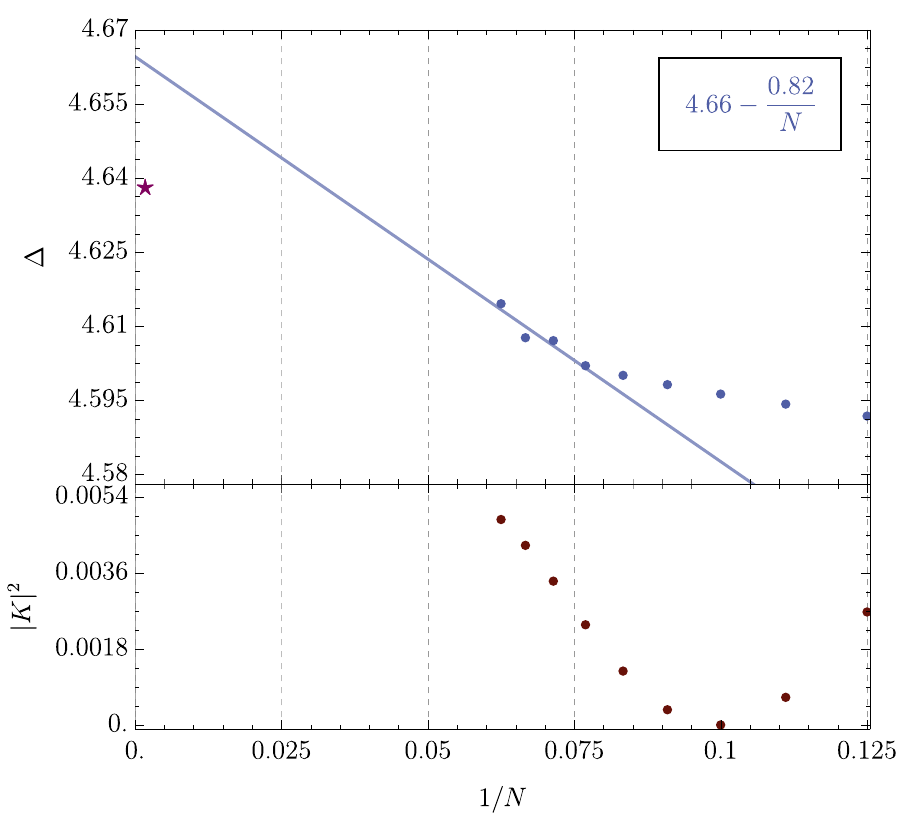}
  }
    \hfill
  \subfloat[$\sigma_{\mu_1\mu_2\mu_3}^\prime$]{
    \includegraphics[width=0.29\textwidth]{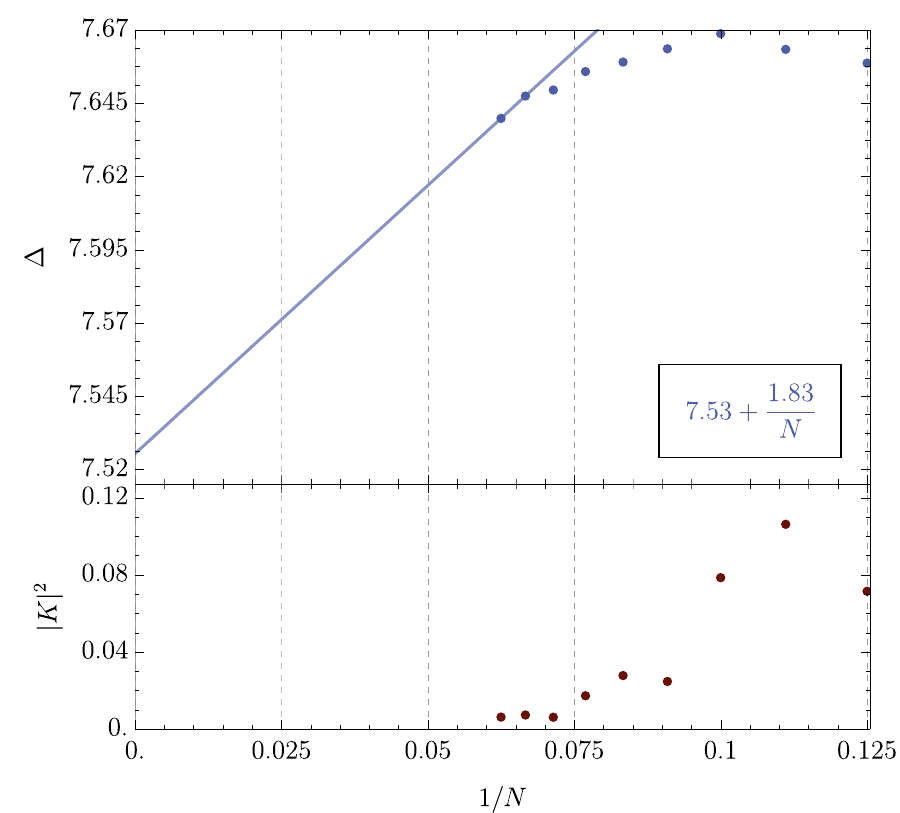}
  }
  \vspace{0.1cm}
    \subfloat[$\sigma_{\mu_1\mu_2\mu_3\mu_4}$]{
    \includegraphics[width=0.29\textwidth]{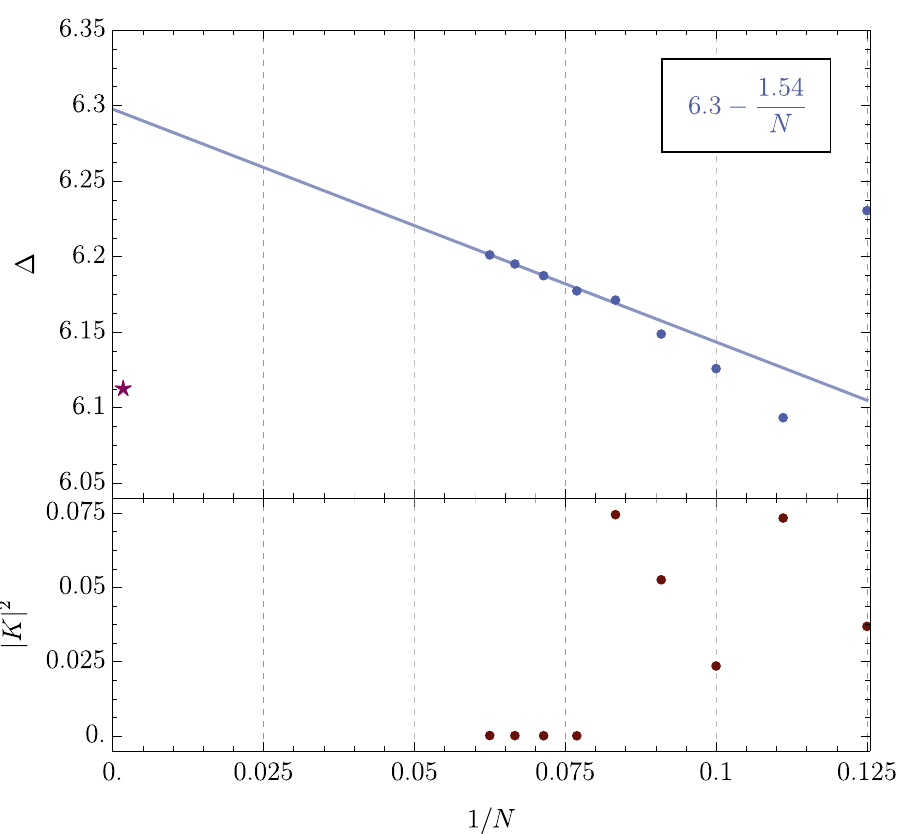}
  }
  \hfill
  \subfloat[$\sigma_{\mu_1\mu_2\mu_3\mu_4}^{-}$]{
    \includegraphics[width=0.29\textwidth]{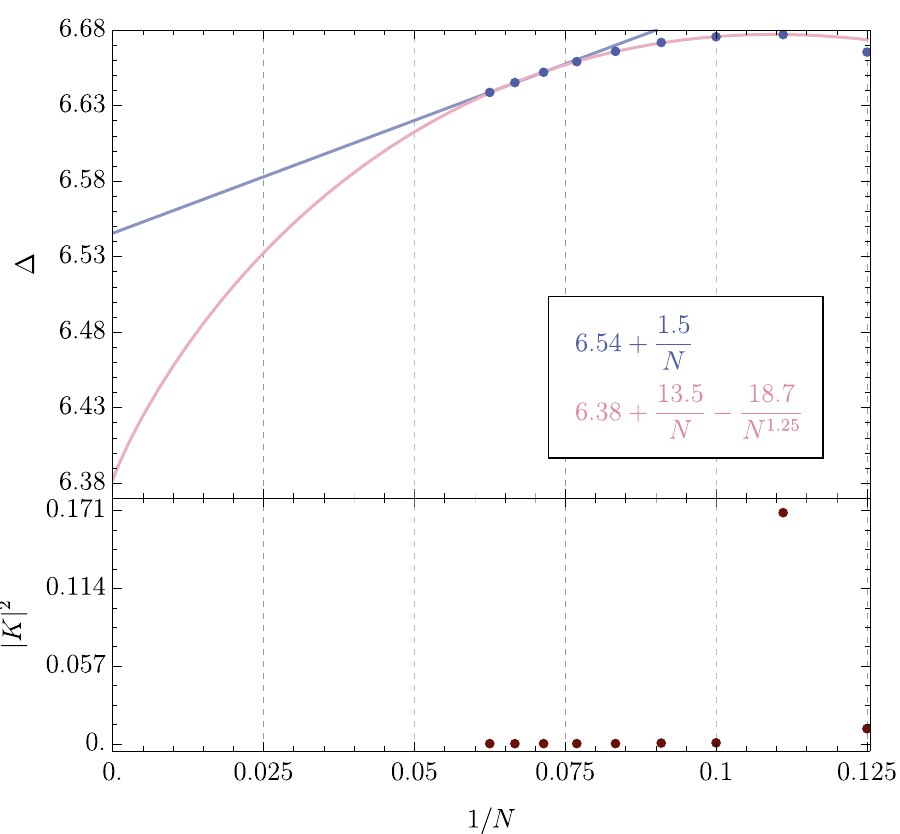}
  }
    \hfill
  \subfloat[$\sigma_{\mu_1\mu_2\mu_3\mu_4\mu_5}$]{
    \includegraphics[width=0.29\textwidth]{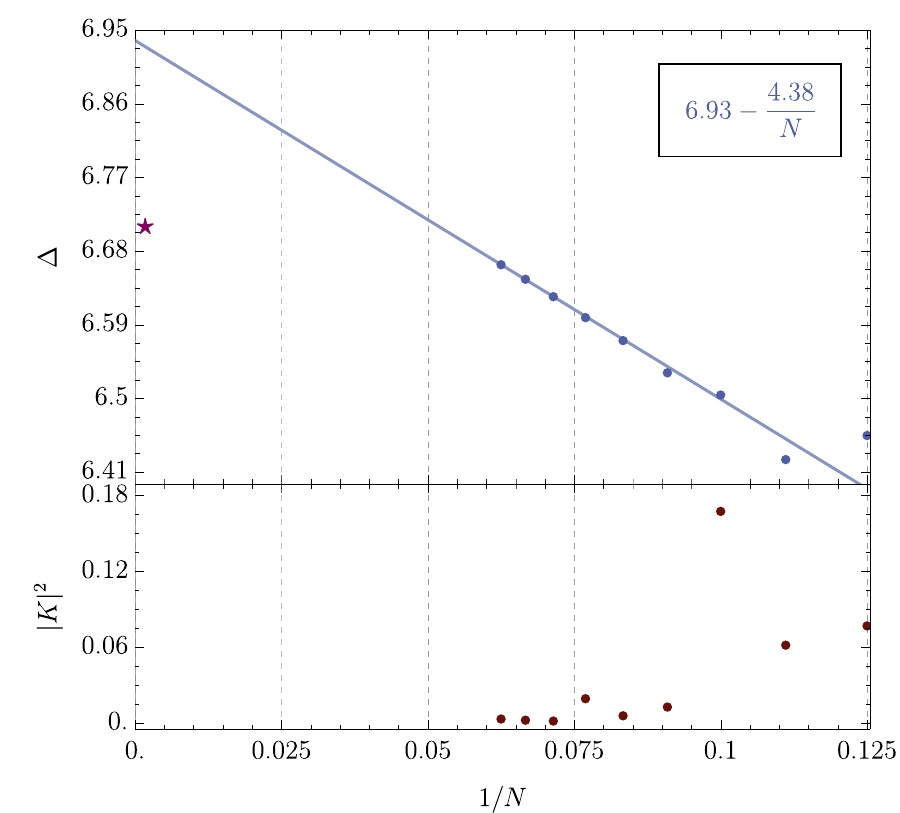}
  }
    \vspace{0.1cm}
    \subfloat[$\sigma_{\mu_1\mu_2\mu_3\mu_4\mu_5}^\prime$]{
    \includegraphics[width=0.29\textwidth]{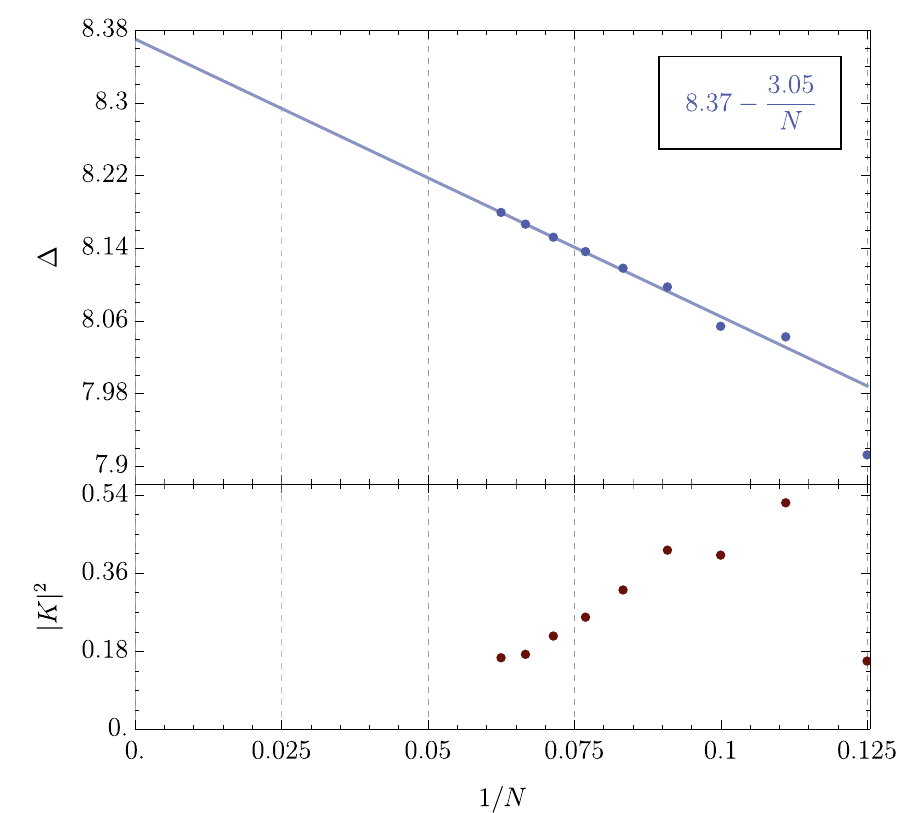}
  }
  \hfill
  \subfloat[$\sigma_{\mu_1\mu_2\mu_3\mu_4\mu_5\mu_6}^\prime$]{
    \includegraphics[width=0.29\textwidth]{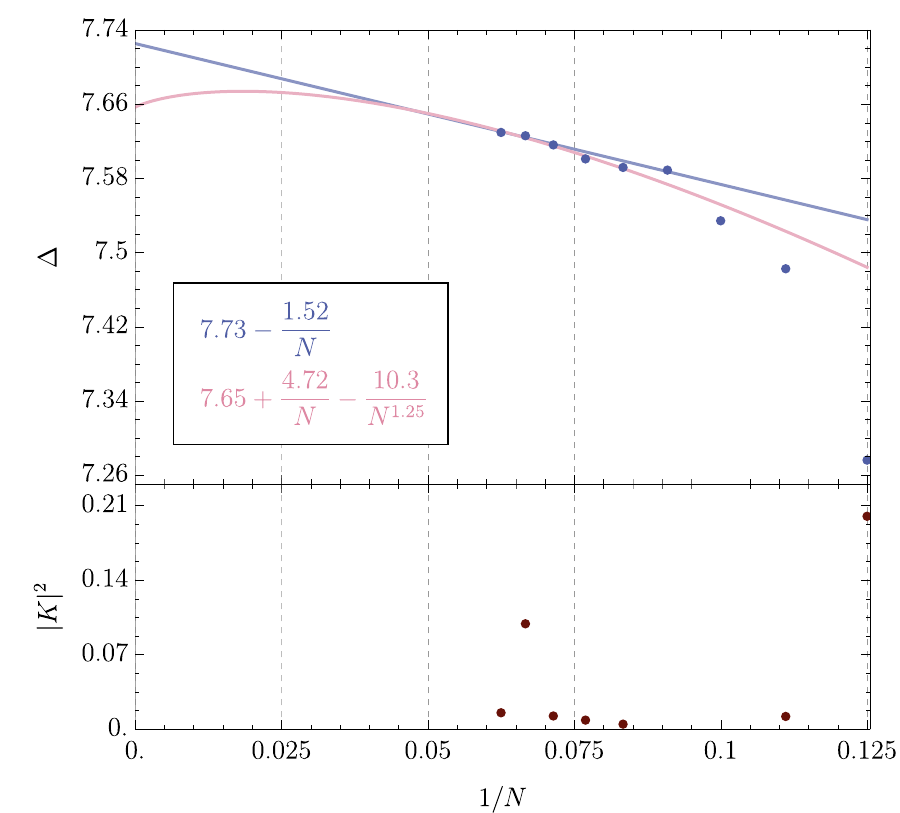}
  }
    \hfill
  \subfloat[$\sigma_{\mu_1\mu_2\mu_3\mu_4\mu_5\mu_6}$]{
    \includegraphics[width=0.29\textwidth]{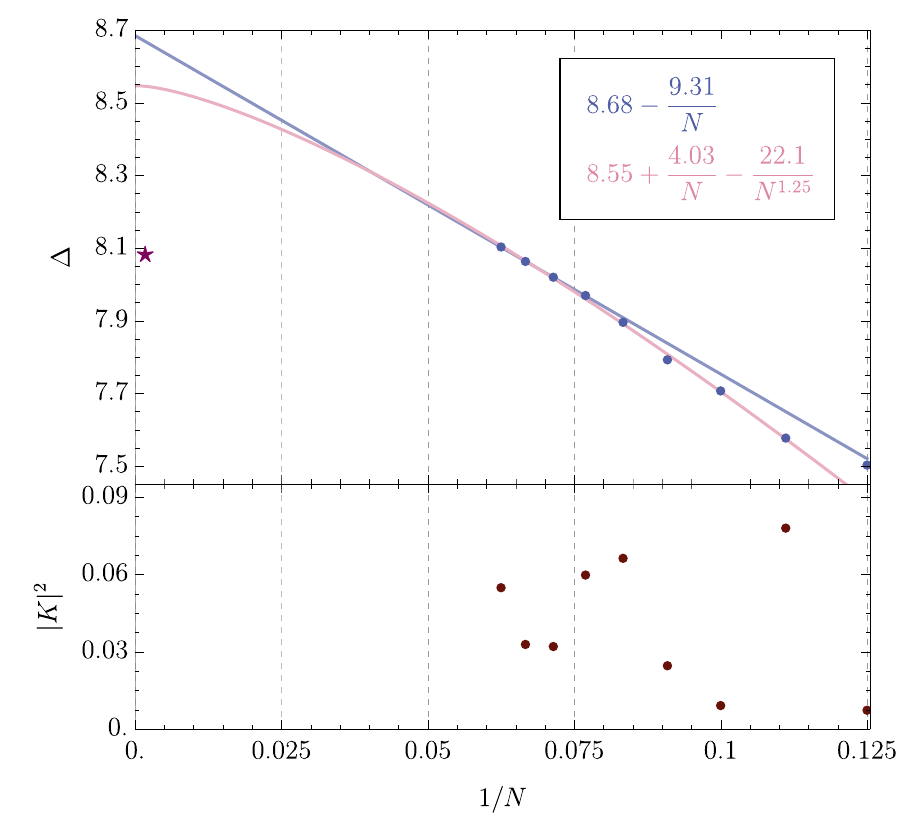}
  }
  \caption{$|K|^2$ eigenvalues, and dimension expectation values $\Delta \equiv \< H \>$,   as a function of $N$, and extrapolations to large $N$, for $\mathbb{Z}_2$-odd operators.  Stars represent known results from numerical conformal bootstrap.  In the language of~\eqref{WC}, the $1/N$ correction corresponds to the $\gamma_{\CR}$ term, while the $1/N^{1.25}$ scaling arises from $\CO = T^\prime$. We assume $g_{\epsilon}=g_{\CR\epsilon}=g_{\epsilon^\prime}\simeq 0$ and approximate $g_C$ by the $1/N$ term, since $(3-\Delta_C)/2 \simeq -1.01$. }
  \label{fig:Z2oddPrimaries}
\end{figure}
%Now for the $\mathbb{Z}_2$ even in order of spin (excluding the vacuum)
\begin{figure}
  \centering
  \subfloat[$\epsilon$]{
    \includegraphics[width=0.29\textwidth]{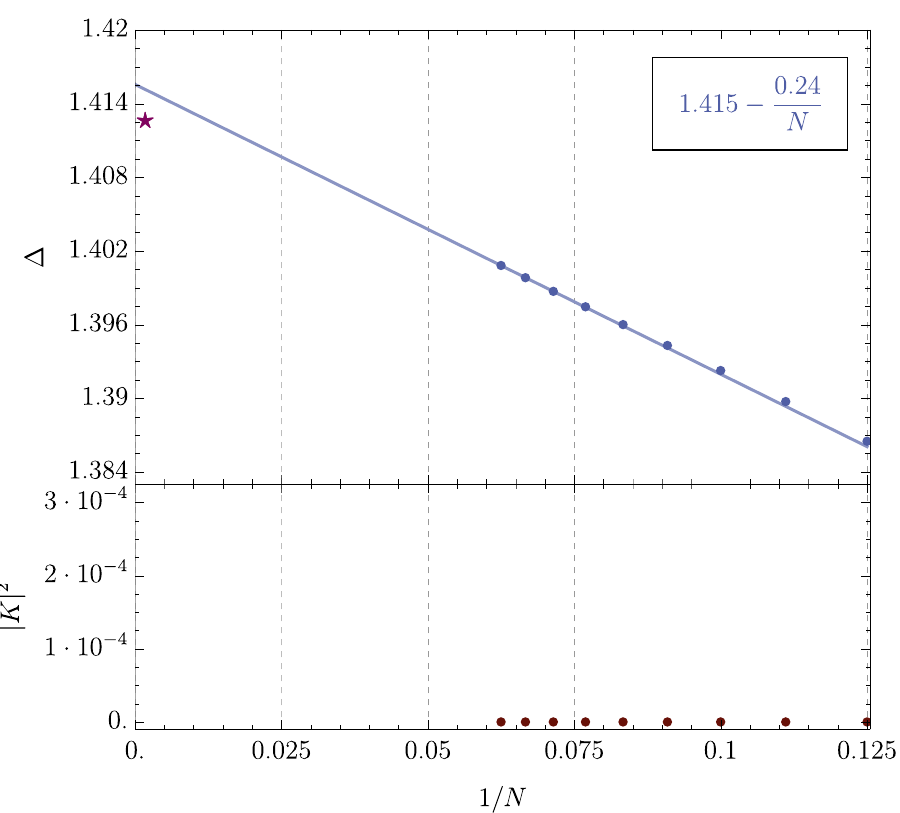}
  }
  \hfill
  \subfloat[$\epsilon^\prime$]{
    \includegraphics[width=0.29\textwidth]{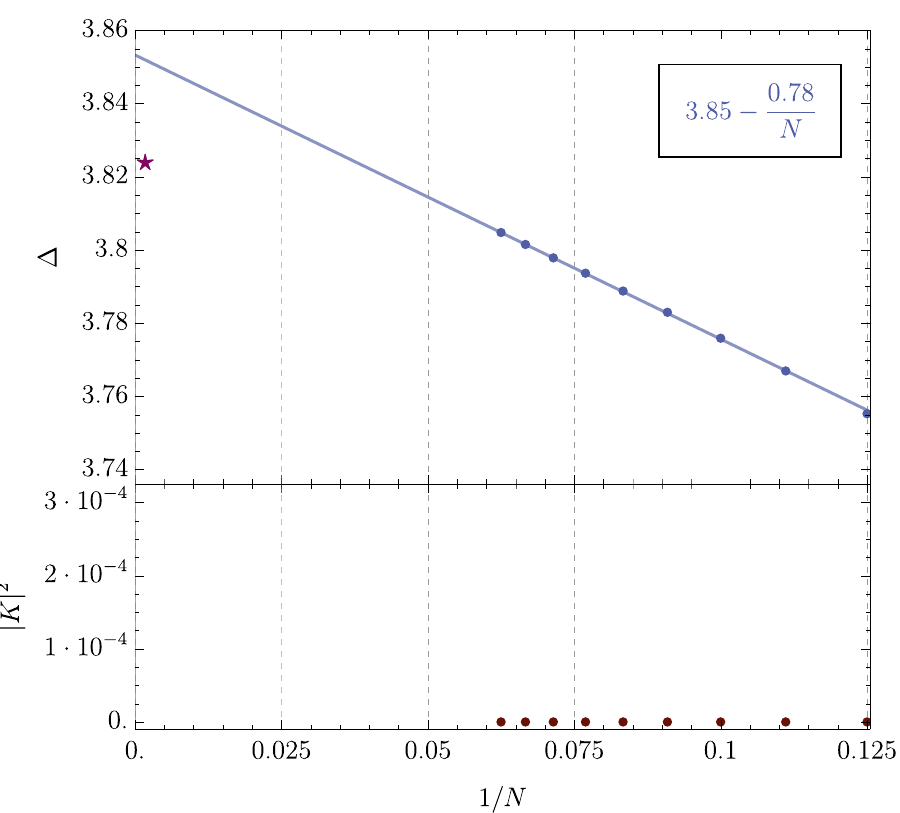}
  }
  \hfill
  \subfloat[$T_{\mu_1\mu_2}$]{
    \includegraphics[width=0.29\textwidth]{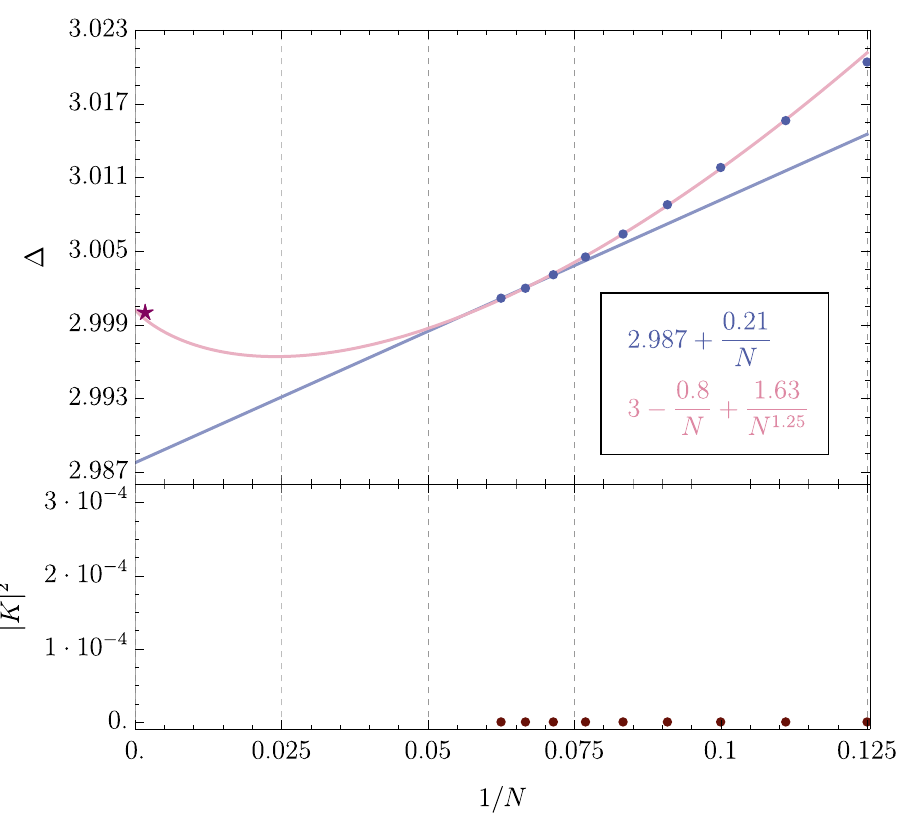}
  }
  \vspace{0.1cm}
  \subfloat[$T^\prime_{\mu_1\mu_2}$]{
    \includegraphics[width=0.29\textwidth]{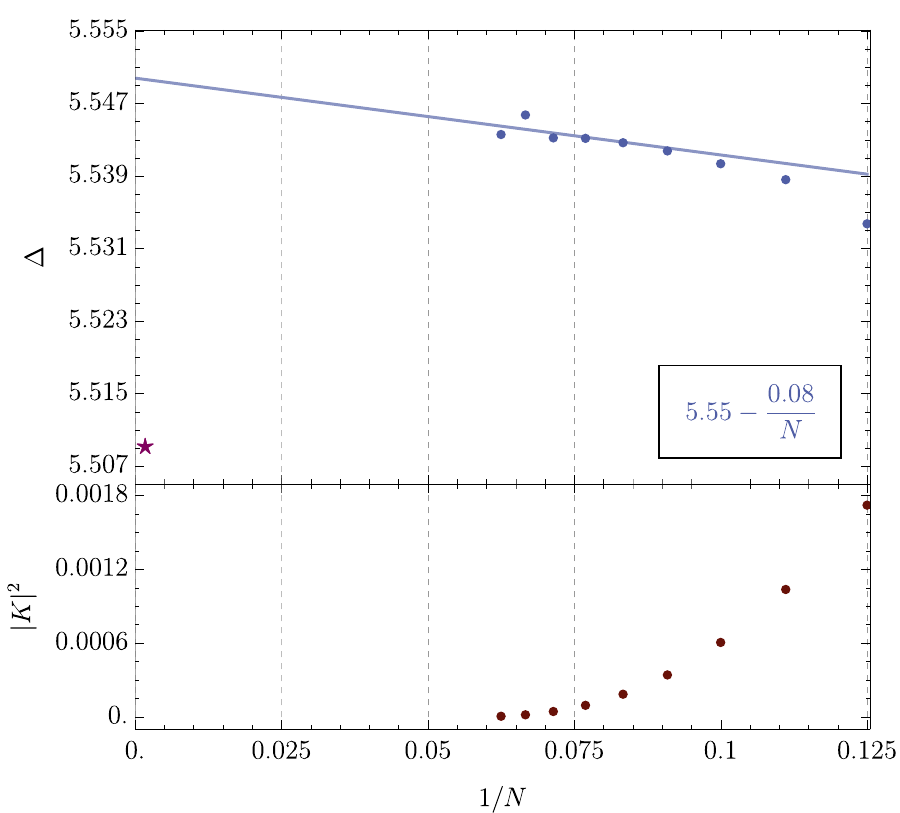}
  }
  \hfill
    \subfloat[$T^{\prime\prime}_{\mu_1\mu_2}$]{
    \includegraphics[width=0.29\textwidth]{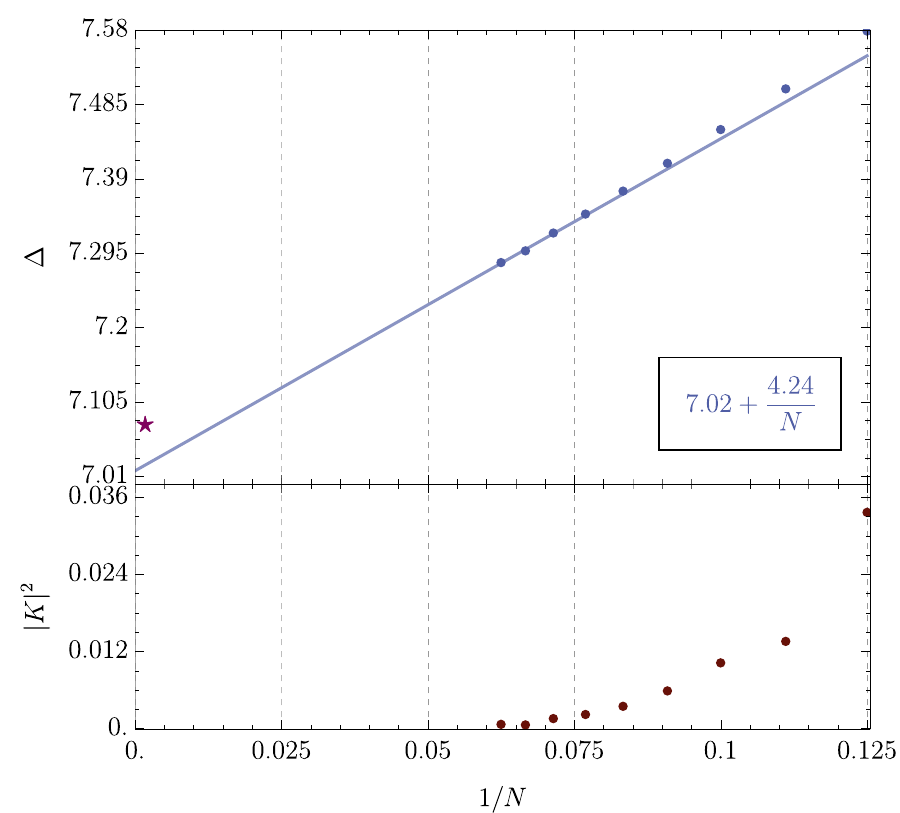}
  }
  \hfill
  \subfloat[$T^{-}_{\mu_1\mu_2}$]{
    \includegraphics[width=0.29\textwidth]{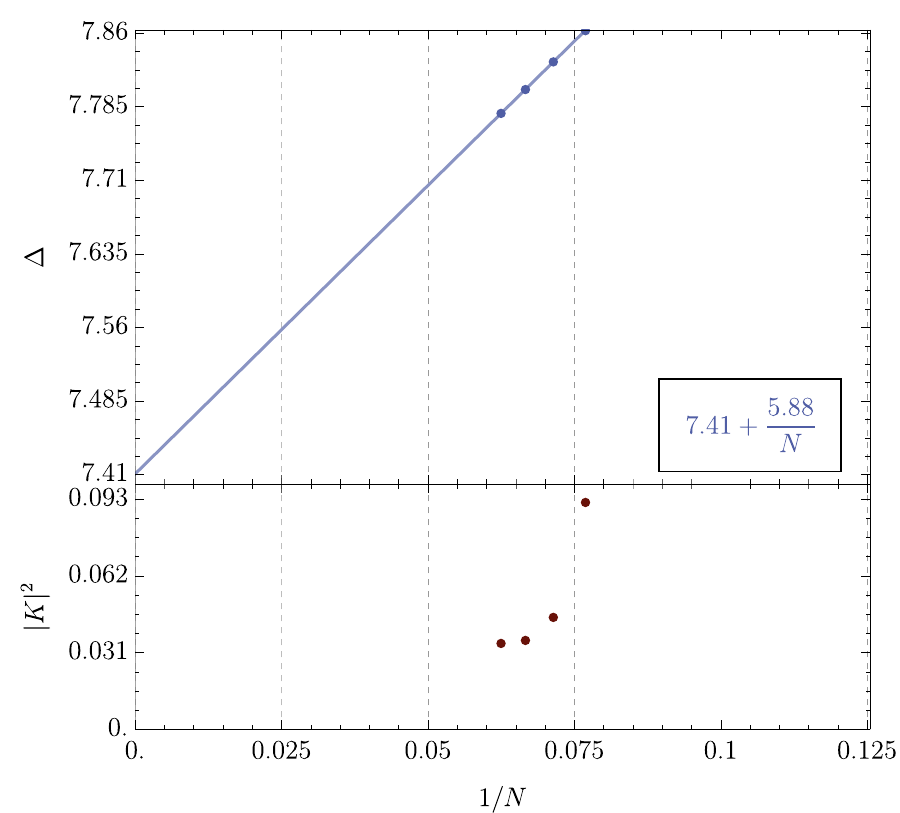}
  }
  \vspace{0.1cm}
  \subfloat[$C_{\mu_1\mu_2\mu_3\mu_4}$]{
    \includegraphics[width=0.29\textwidth]{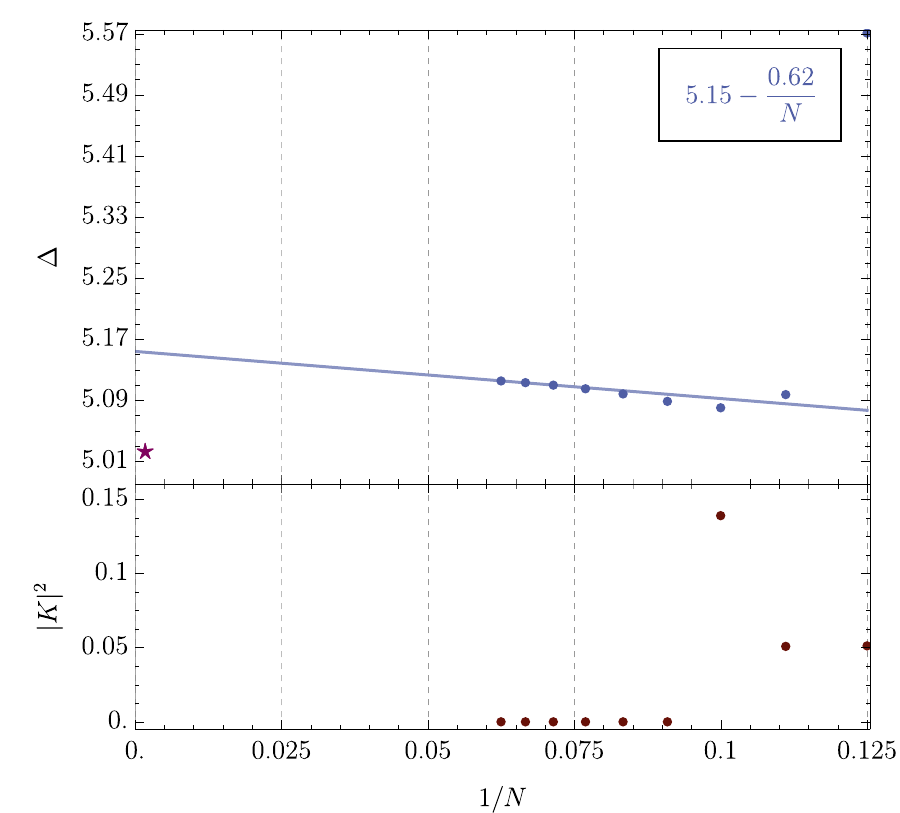}
  }
  \hfill
  \subfloat[$C^-_{\mu_1\mu_2\mu_3\mu_4}$]{
    \includegraphics[width=0.29\textwidth]{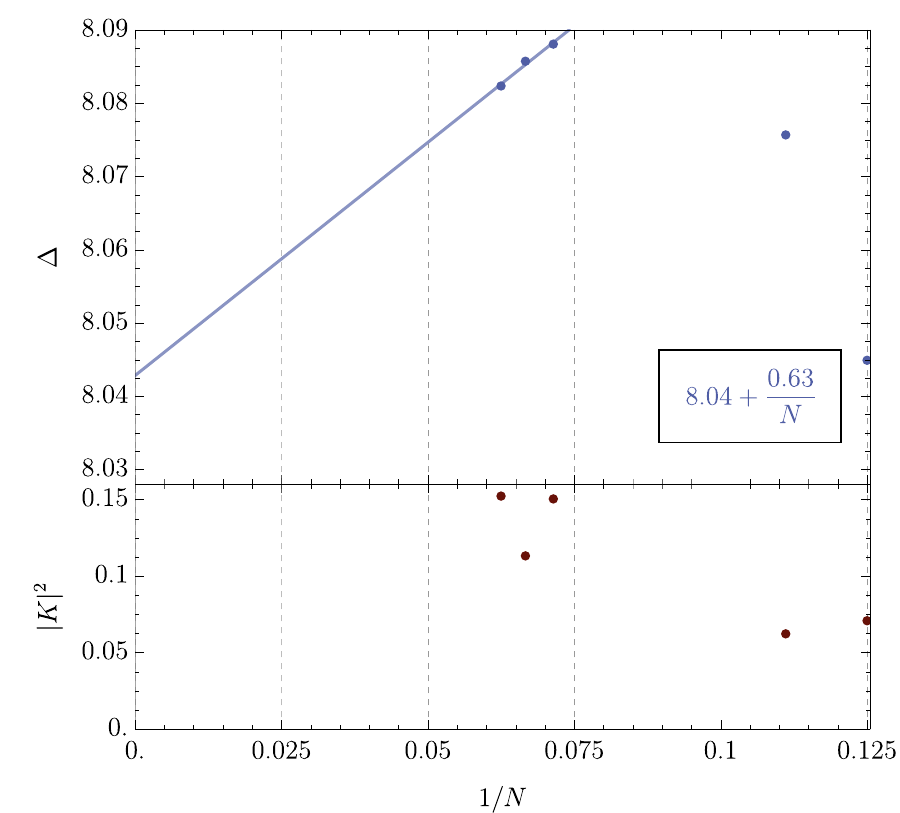}
  }
    \hfill
  \subfloat[$\epsilon_{\mu_1\mu_2\mu_3\mu_4\mu_5\mu_6}$]{
    \includegraphics[width=0.29\textwidth]{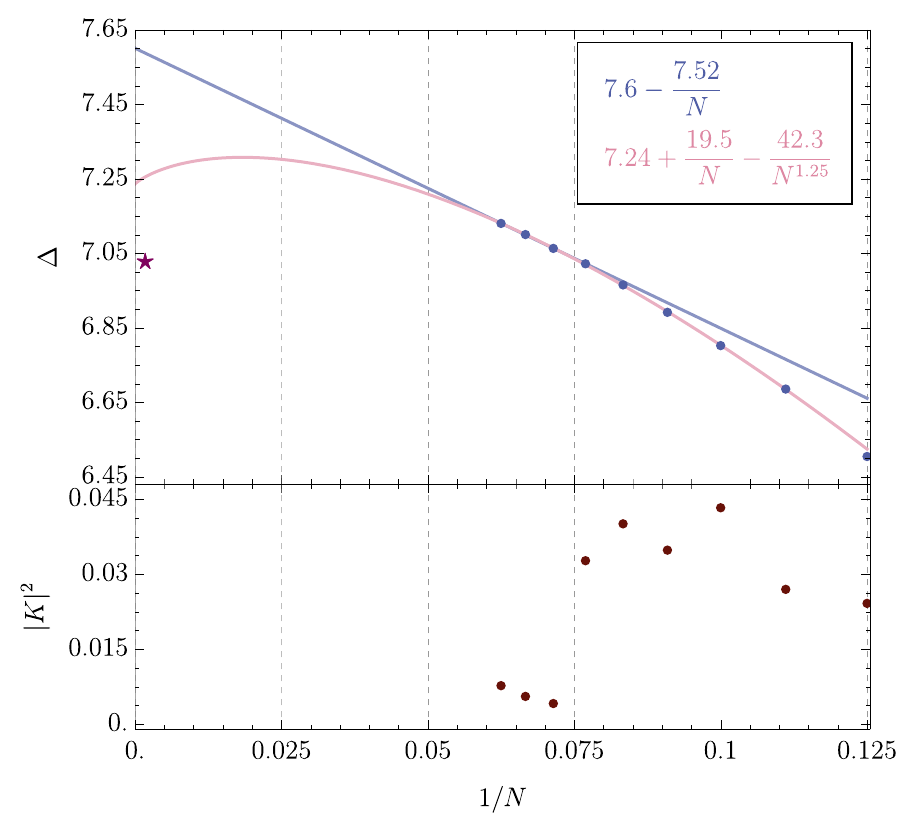}
  }
  \caption{Same as in Fig~\ref{fig:Z2oddPrimaries} for $\mathbb{Z}_2$-even operators.}
  \label{fig:Z2evenPrimaries}
\end{figure}

\begin{figure}
\centering
\begin{minipage}{0.41\textwidth}
  \centering
  \includegraphics[width=\linewidth]{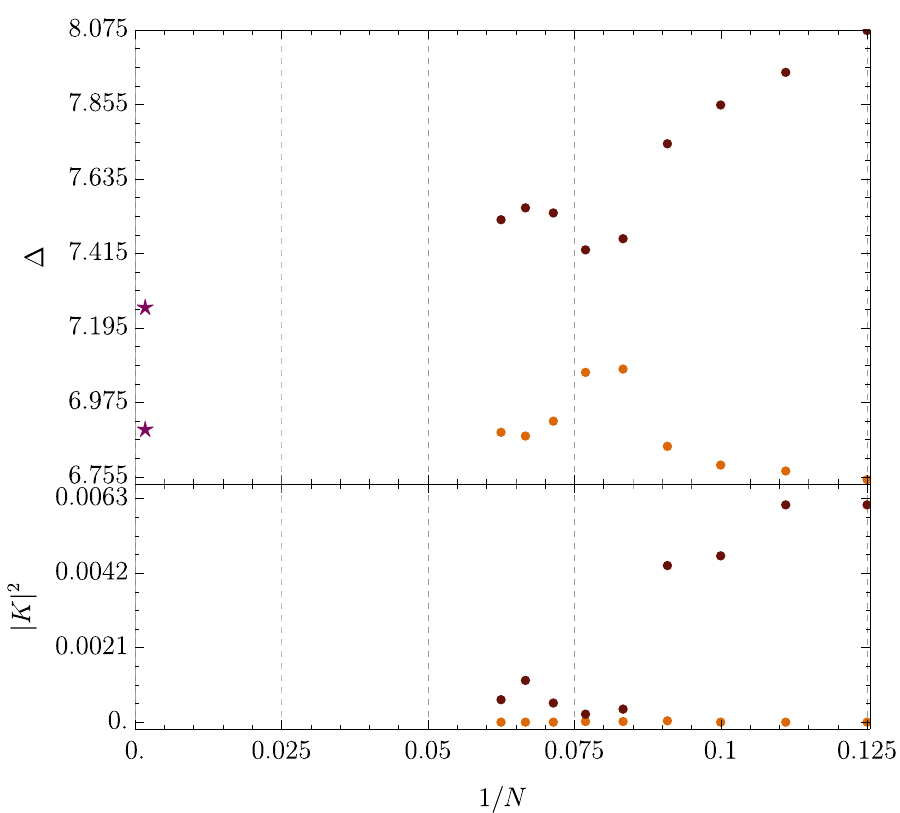}
\end{minipage}
\begin{minipage}{0.4\textwidth}
  \centering
  \includegraphics[width=\linewidth]{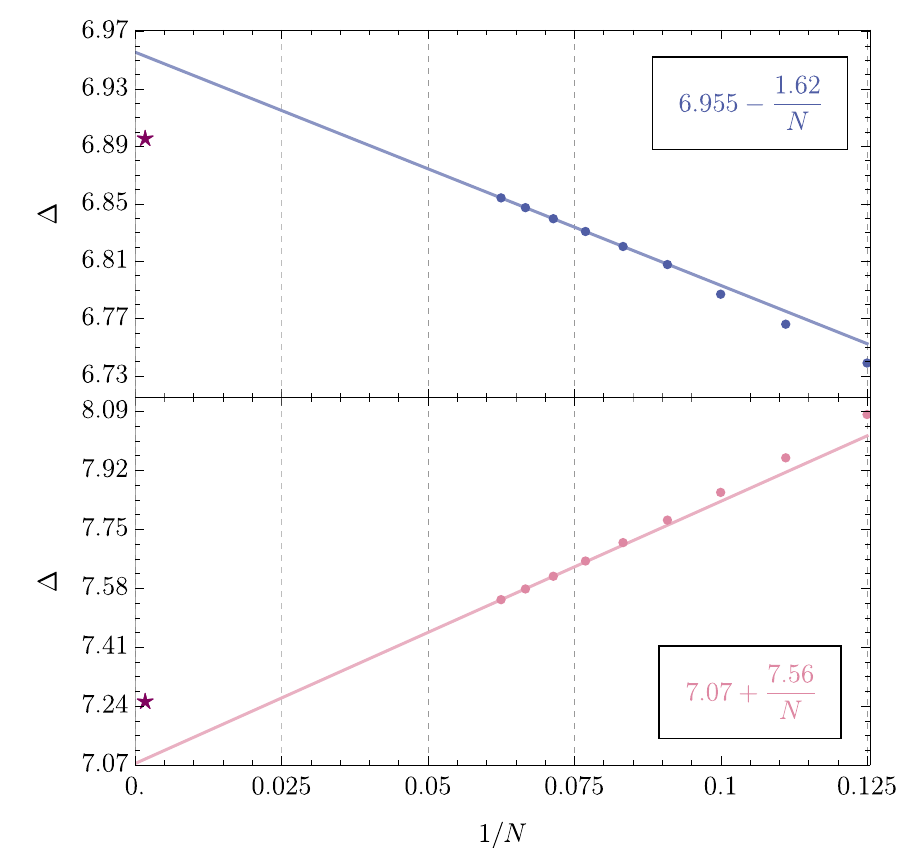}
\end{minipage}
\caption{Results from rediagonalizing the $\epsilon^{\prime \prime}$ and $\epsilon^{\prime \prime\prime}$ subsector.  ({\it Left}): Raw results for $|K|^2$ eigenvalues and $\Delta \equiv \< H \>$ expectation values. ({\it Right}):  We first take the subspace scanned by the two small-eigenvalues eigenvectors in the left panel, and then rediagonalize $H$ within this subspace to get the two eigenvalues of $\Delta$ shown.  }
\label{Fig:epsilonPrimes}
\end{figure}

\begin{figure}
\centering
\begin{minipage}{0.41\textwidth}
  \centering
  \includegraphics[width=\linewidth]{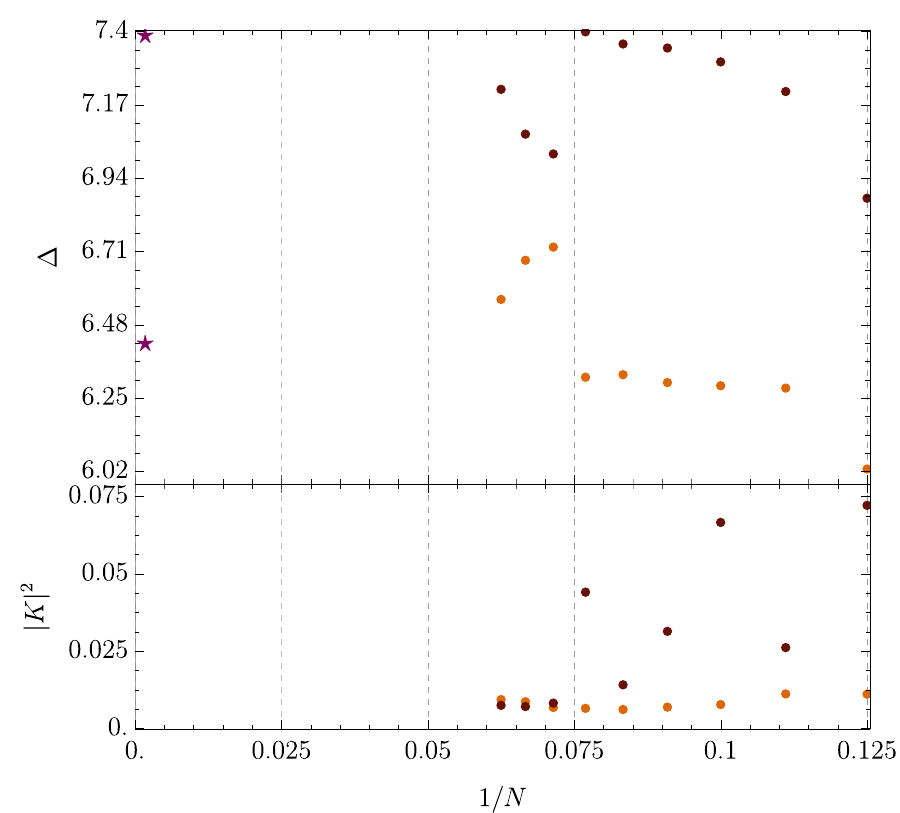}
\end{minipage}
\begin{minipage}{0.4\textwidth}
  \centering
  \includegraphics[width=\linewidth]{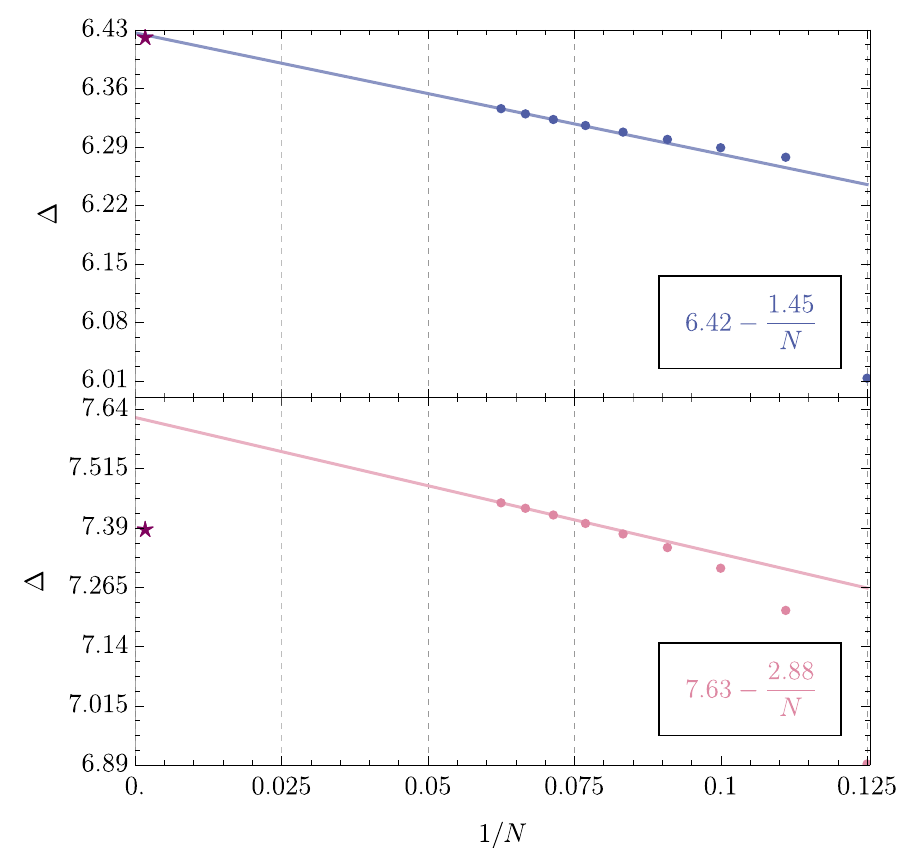}
\end{minipage}
\caption{Same as Fig.~\ref{Fig:epsilonPrimes} but for the operators $C^{\prime }$ and $C^{\prime \prime}$. }
\label{Fig:Cprimes}
\end{figure}
\subsection*{New Primares}
Next we turn to a brief discussion of the new primaries obtained by our analysis, listed in Table \ref{tab:NewPrimaries}.
As noted in~\cite{Zhu:2022gjc}, it is particularly interesting to look for new parity-odd states, since these would not appear in most bootstrap analyses and therefore remain largely unconstrained.\footnote{The stress tensor bootstrap~\cite{Chang:2024whx}  is a notable exception, as is the energy correlator bootstrap \cite{Mecaj:2025ecl}.}  Among our states with small values of $|K|^2$, we find three clear parity-odd primaries.\footnote{As an example of how $|K|^2$ reduces operator mixing for primaries, we can look at the descendant states that would most strongly mix with $\sigma^-_{\mu_1 \mu_2 \mu_3 \mu_4}$ when we diagonalize the Hamiltonian without using $|K|^2$.  There are three descendant states with the same quantum numbers $(4, -, -)$ in the range $6.5 \lesssim \Delta \lesssim 7.3$.  These are a level-two descendant of $\sigma_{\mu_1 \mu_2}$ ($\Delta_{\rm desc} = 7.18$), a level-three descendant of $\sigma_{\mu_1 \mu_2 \mu_3}$  ($\Delta_{\rm desc} = 6.6$) and a level-one descendant of $\sigma_{\mu_1 \mu_2 \mu_3 \mu_4}$  ($\Delta_{\rm desc} = 7.1$); the $|K|^2$ values of these descendants in the CFT limit is 34.9, 26.5, and 6.8, respectively.}  We label them $\sigma_{\mu_1 \mu_2 \mu_3 \mu_4}^-, T^-_{\mu_1 \mu_2}$ and $C^-_{\mu_1 \mu_2 \mu_3 \mu_4}$, indicating that they have $(\ell, \mathbb{Z}_2, \CP)$ quantum numbers $(4, -, -), (2, +, -)$ and $(4, +, -)$, respectively.

One way to get a sense of whether or not these new primaries should be expected in 3d Ising is to compare them with the spectrum of parity-odd primaries in the 3d free scalar CFT.  At a qualitative level, the free theory operators should flow to the operators in the 3d Ising model under RG flow, and so provide a rough guide as to where one might expect to find new parity-odd operators at each spin and $\mathbb{Z}_2$ sector. In Appendix~\ref{app:Counting}, we review how to efficiently count the number of primary operators at a given dimension with fixed charge, spin, parity, and particle number.  The result for the lowest-lying primaries is shown in Fig. \ref{fig:ParityOdd}.  In particular, while the first spin-0 parity odd state arises at $\Delta=11$, at higher spins the parity odd states begin at much lower $\Delta$, with the lowest occurring at $\Delta=6.5, \ell=4$, and is $\mathbb{Z}_2$ odd. The next two-lowest parity odd primaries are at $\Delta=7$, with spins $\ell=2$ and $\ell=4$.  Thus the lowest-dimension parity-odd operators in the free theory have precisely the same quantum numbers as our new primaries.  

It is also interesting to consider how the new primaries would be interpreted in the framework of large spin.  This framework treats operators as Generalized Free Fields (GFFs) built from $\sigma$, together with small interactions suppressed by powers of spin.  The way multi-twist operators are built out of $\phi$ in the free theory is structurally the same as how they are built out of $\sigma$ in the GFF theory, and so our comparison with the free field states in Fig.~\ref{fig:ParityOdd} suggests how many $\sigma$s each of our operators should be built out of.  For instance, $\sigma_{\mu_1 \mu_2 \mu_3 \mu_4}^-$ is close to the $n_\phi=3, \ell=4$ free theory primary, suggesting it should be a triple-twist operator in the large spin expansion.  And indeed, $\Delta(\sigma_{\mu_1 \mu_2 \mu_3 \mu_4}^-) = 6.53$ is extremely close to the value $3 \Delta_\sigma + \ell + 1 = 6.554$ one would get for a triple-twist subleading-twist  parity-odd operator ($\Delta-\ell- n \Delta_\phi=1$) in a GFF.   By contrast, the new primaries $C^-_{\mu_1 \mu_2 \mu_3 \mu_4}$ and $T^-_{\mu_1 \mu_2}$ have dimensions that are not as close to that expected for quadruple-twist subleading-twist operators in GFF, 
but this is natural since  they have lower spin-per-particle.  It would be extremely interesting to calculate the leading large-spin anomalous dimensions for these operators in 3d Ising.

Next, we turn to the new parity-even operators. These are all $\mathbb{Z}_2$ odd, and should in principle be accessible to the conformal bootstrap through the $\sigma \times \epsilon$ OPE (or the $\sigma \times T$ OPE in the stress-tensor bootstrap).   Generally, at high dimensions, lower-twist operators are more accessible than higher-twists.  From this perspective, the new primary $\sigma'_{\mu_1 \dots \mu_6}$ is particularly interesting, because it is a leading-twist triple twist operator ($\Delta \approx \ell+3 \Delta_\sigma = 7.554$).  In the GFF limit, there are two degenerate triple-twist leading-twist operators at $\ell=6$. After their energies are split by interactions, the more energetic state is dominantly the operator $[\sigma, \epsilon]_6$ which can be thought of as a double-twist operator built from $\epsilon$ and  $\sigma$.  This operator has a large OPE coefficient in the $\sigma \times \epsilon$ OPE and therefore is clearly visible in previous bootstrap studies.  On the other hand, the lower-dimension one is the one we call the new primary $\sigma'_{\mu_1 \dots \mu_6}$. In the large spin framework, it should have a positive binding energy (due to $\epsilon$ bulk exchange)\footnote{See~\cite{Fardelli:2024heb} for a similar analysis in the 3d O(2) model.} and therefore have dimension slightly below $3\Delta_\sigma+6 = 7.554$.  Comparing to Fig.~\ref{fig:Z2oddPrimaries}(k), our large $N$ extrapolation $7.73$ is a bit high compared to this expectation, but close enough that it seems reasonable for the difference to be due to small corrections from additional irrelevant operators in the fit. 

Finally, we comment that while the quality of our primaries appears to degrade when the dimension of the state gets too high, the transition is not completely sharp and taking a more liberal approach one could also keep the first few primaries that we discard.  In Appendix~\ref{app:discarded}, we show the next four primaries that we have discarded.  
\newpage
\begin{figure}[h!]
\centering
\begin{minipage}{0.32\textwidth}
  \centering
  \includegraphics[width=\linewidth]{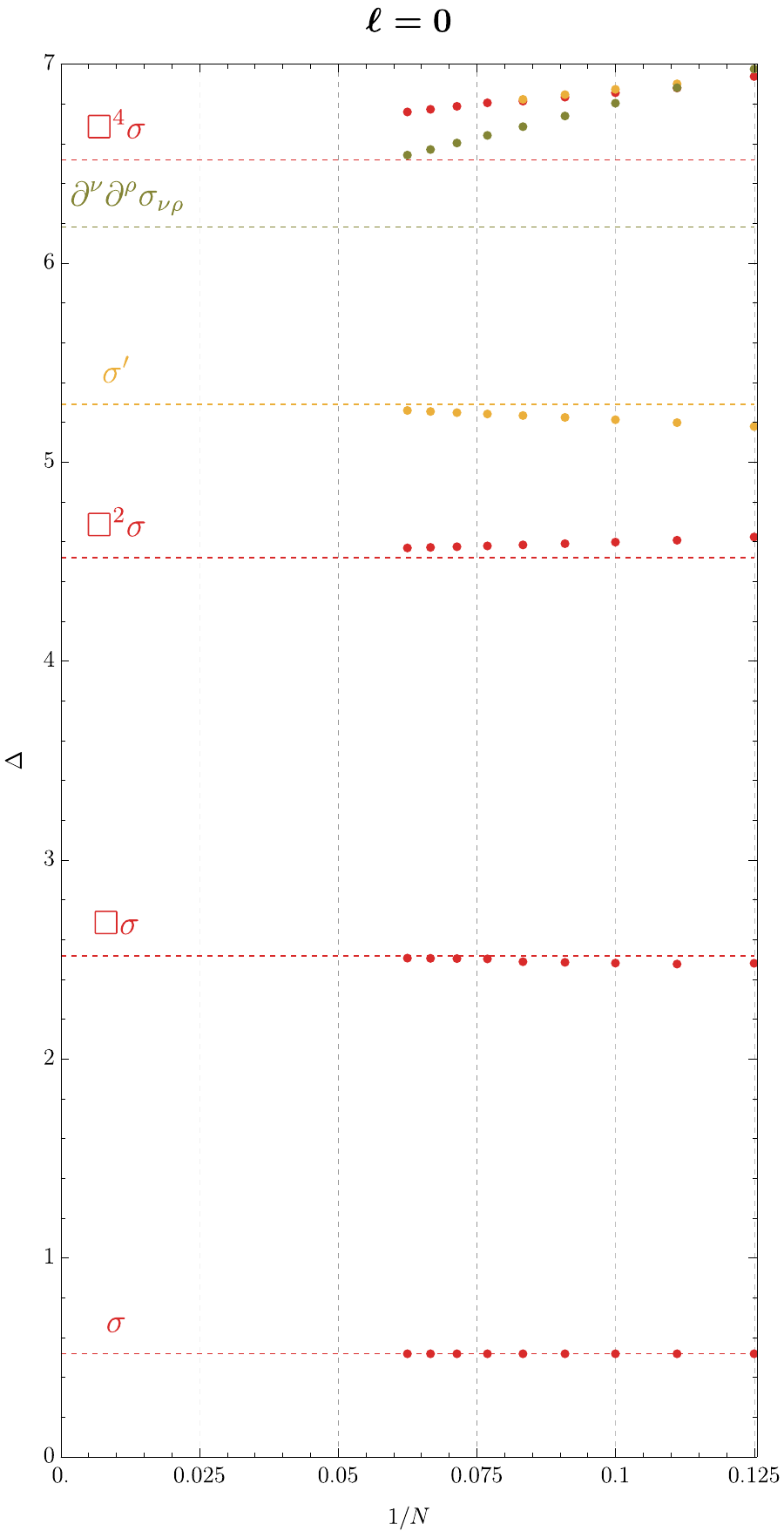}
\end{minipage}
\begin{minipage}{0.32\textwidth}
  \centering
  \includegraphics[width=\linewidth]{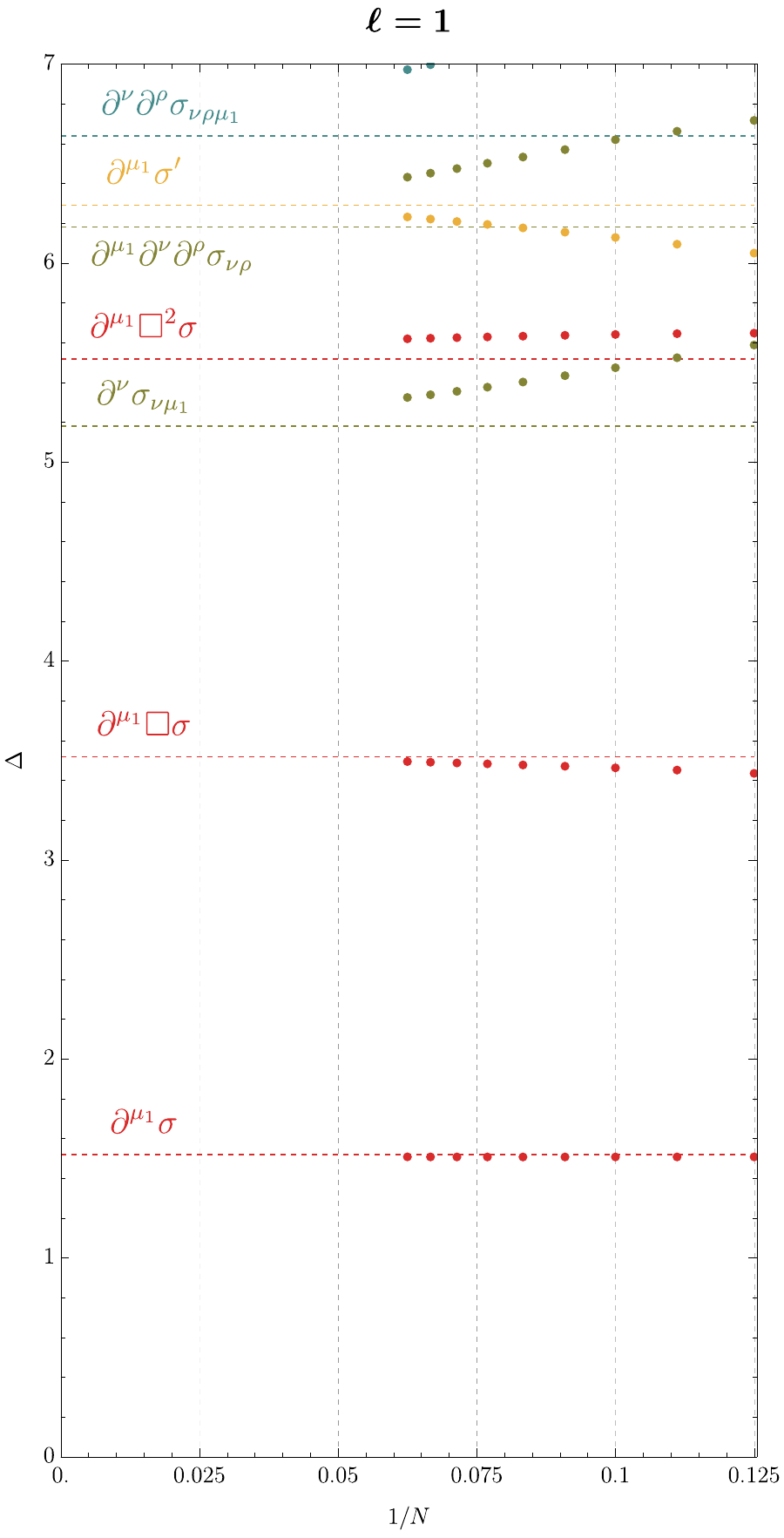}
\end{minipage}
\begin{minipage}{0.32\textwidth}
  \centering
  \includegraphics[width=\linewidth]{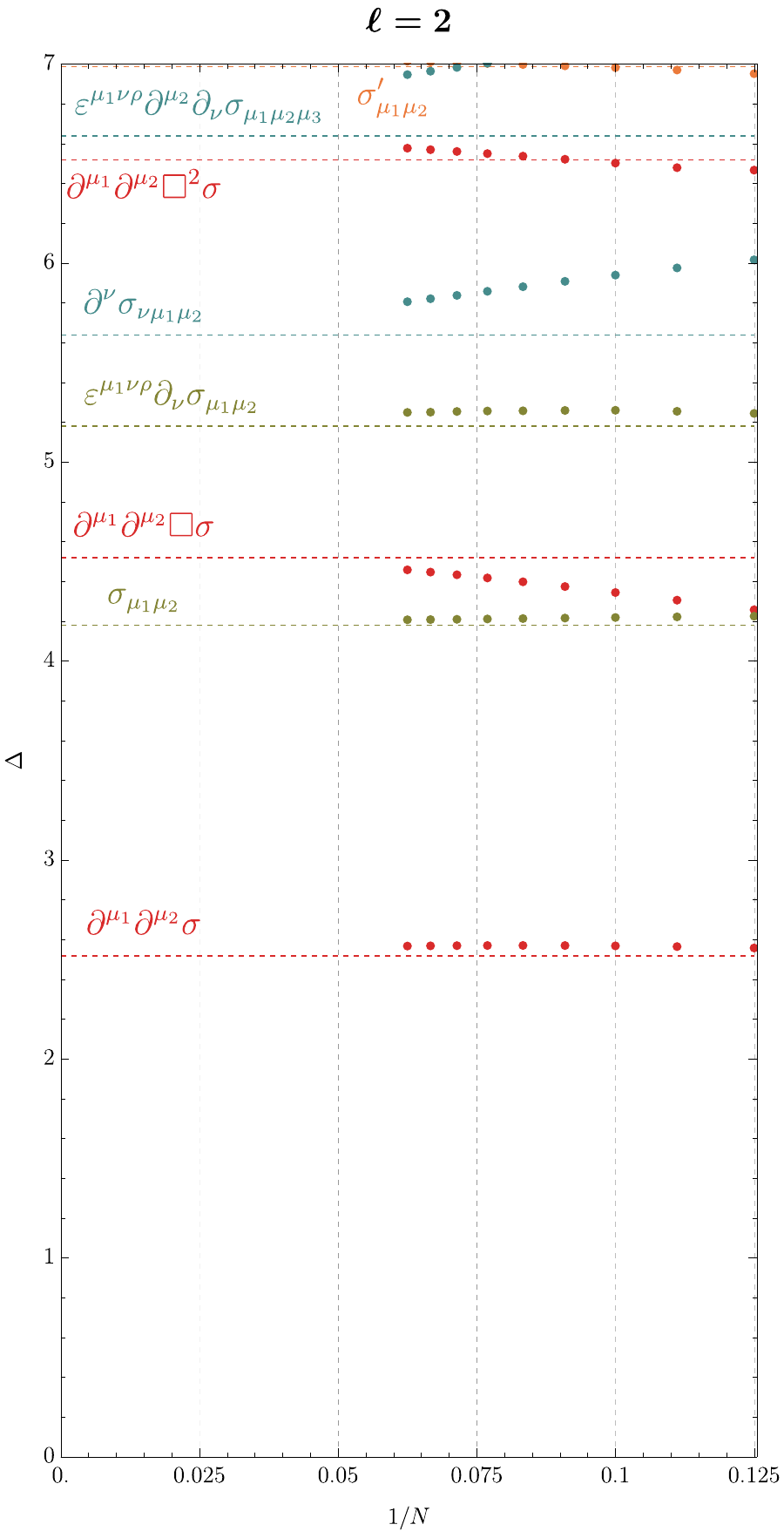}
\end{minipage}
\begin{minipage}{0.24\textwidth}
  \centering
  \includegraphics[width=\linewidth]{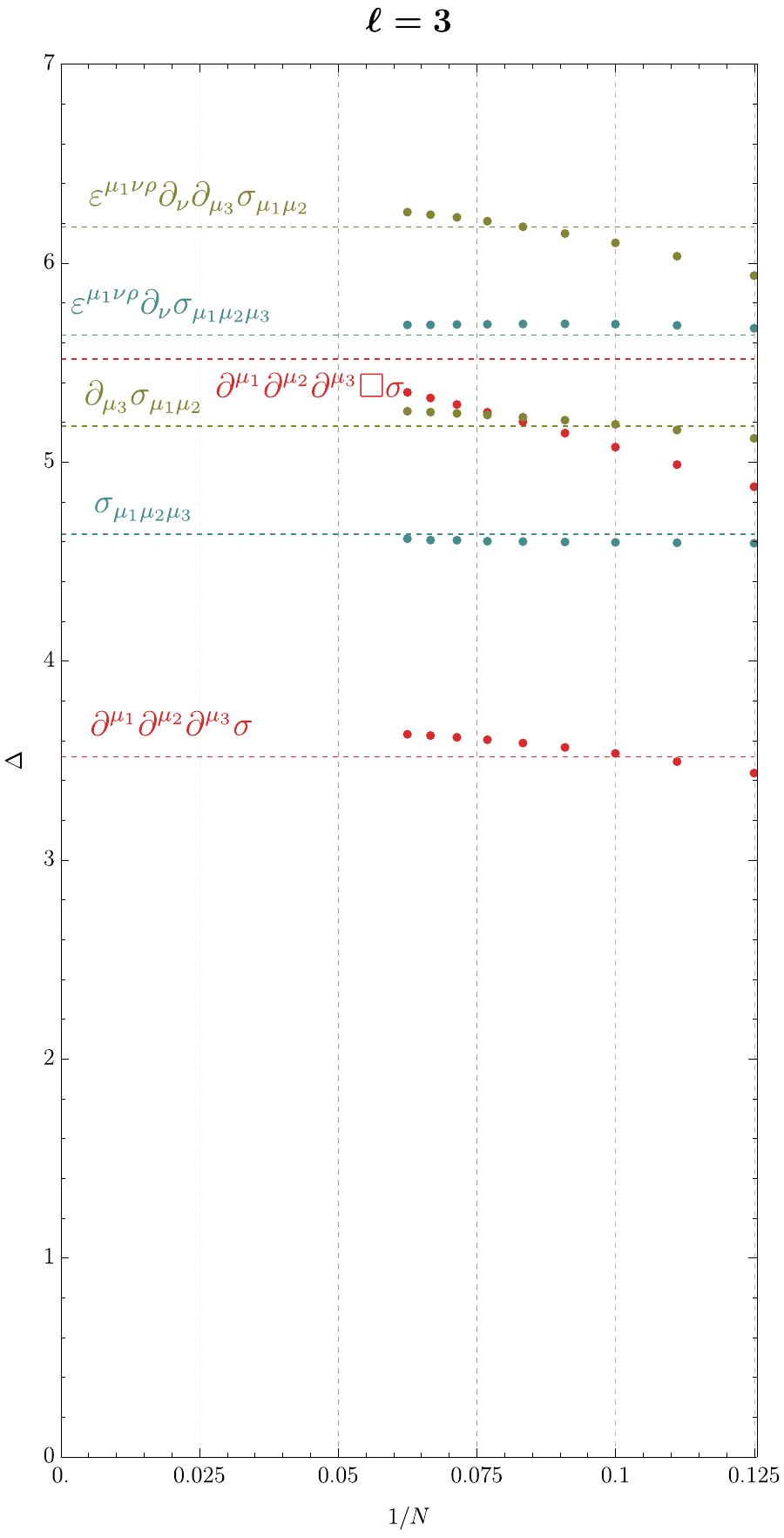}
\end{minipage}
\begin{minipage}{0.24\textwidth}
  \centering
  \includegraphics[width=\linewidth]{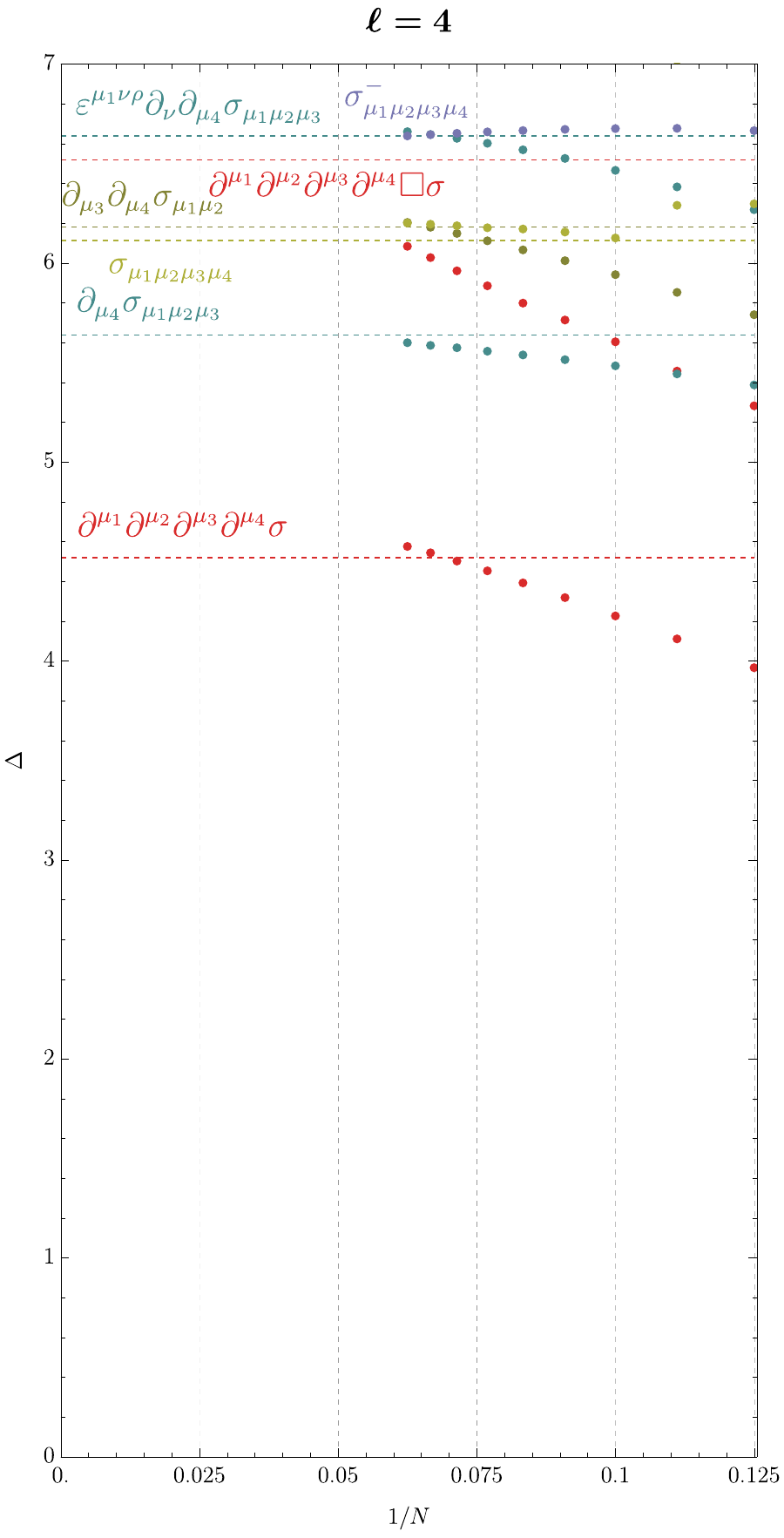}
\end{minipage}
\begin{minipage}{0.24\textwidth}
  \centering
  \includegraphics[width=\linewidth]{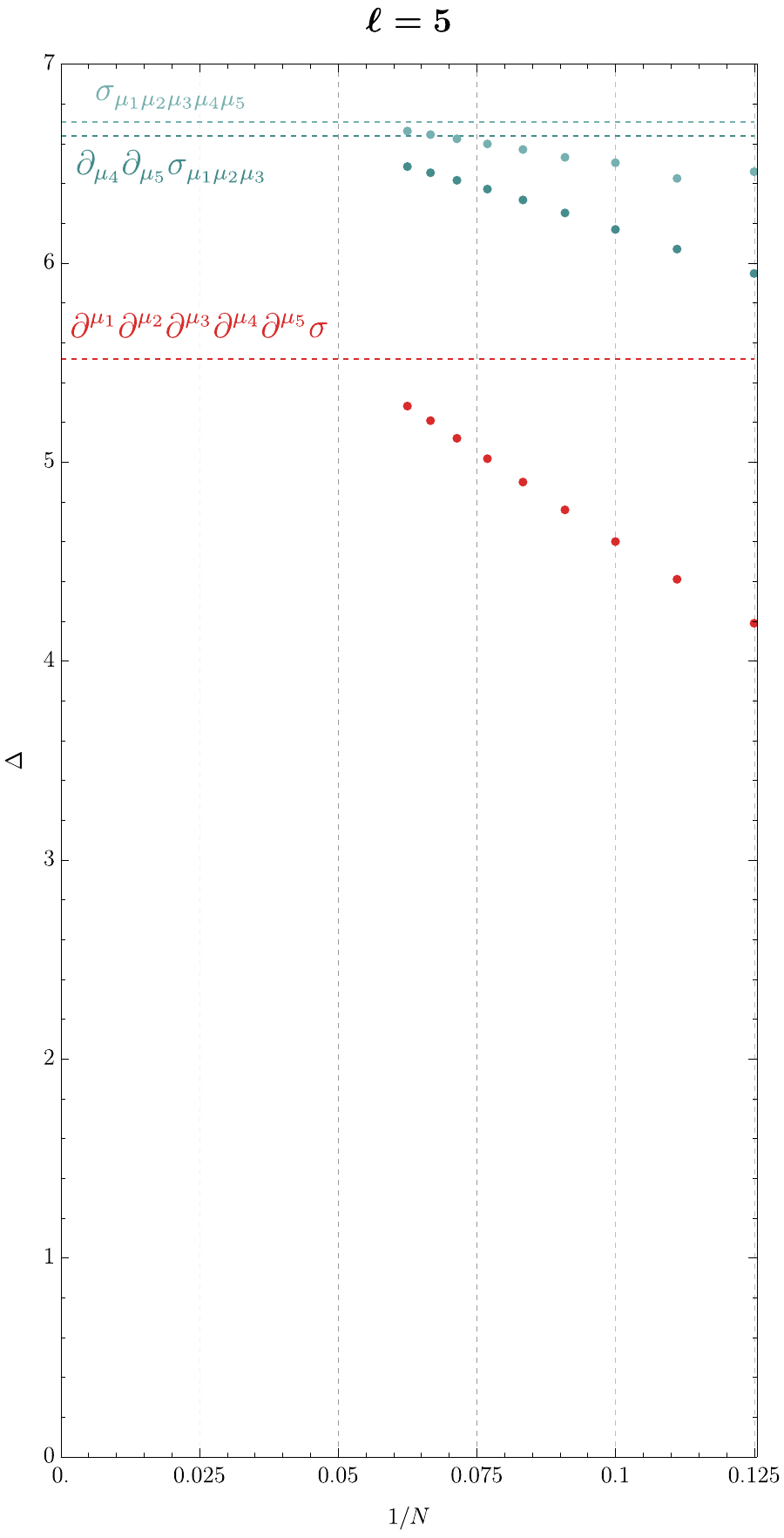}
\end{minipage}
\begin{minipage}{0.24\textwidth}
  \centering
  \includegraphics[width=\linewidth]{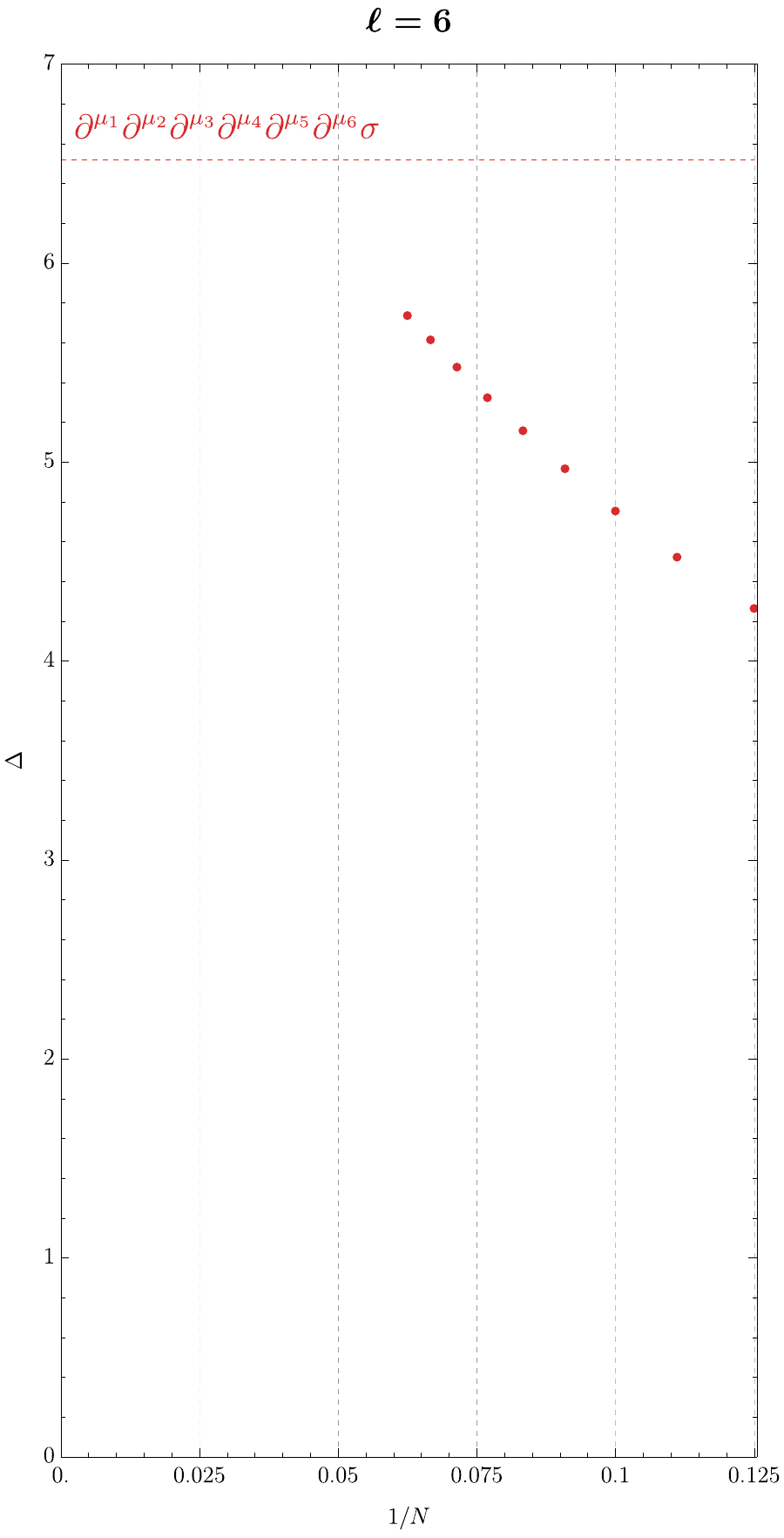}
\end{minipage}
\caption{Conformal multiplets in the $\mathbb{Z}_2$ odd sector obtained for various spin sectors and for different values of $N$. Descendants are constructed by acting with $P_z$ on the primaries.}
\label{Fig:conformalMultiplets}
\end{figure}
\newpage
\begin{figure}[h!]
\centering
\begin{minipage}{0.32\textwidth}
  \centering
  \includegraphics[width=\linewidth]{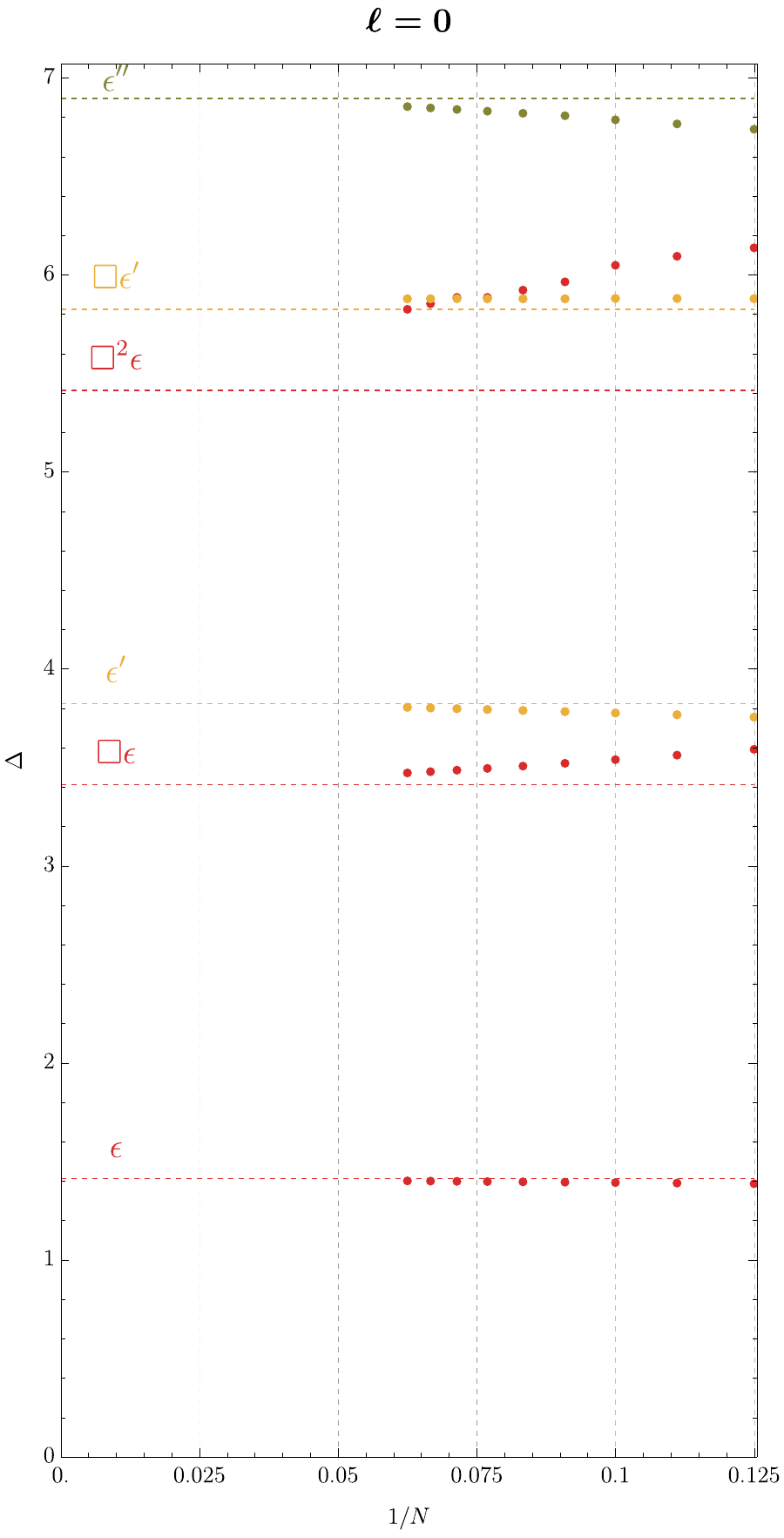}
\end{minipage}
\begin{minipage}{0.32\textwidth}
  \centering
  \includegraphics[width=\linewidth]{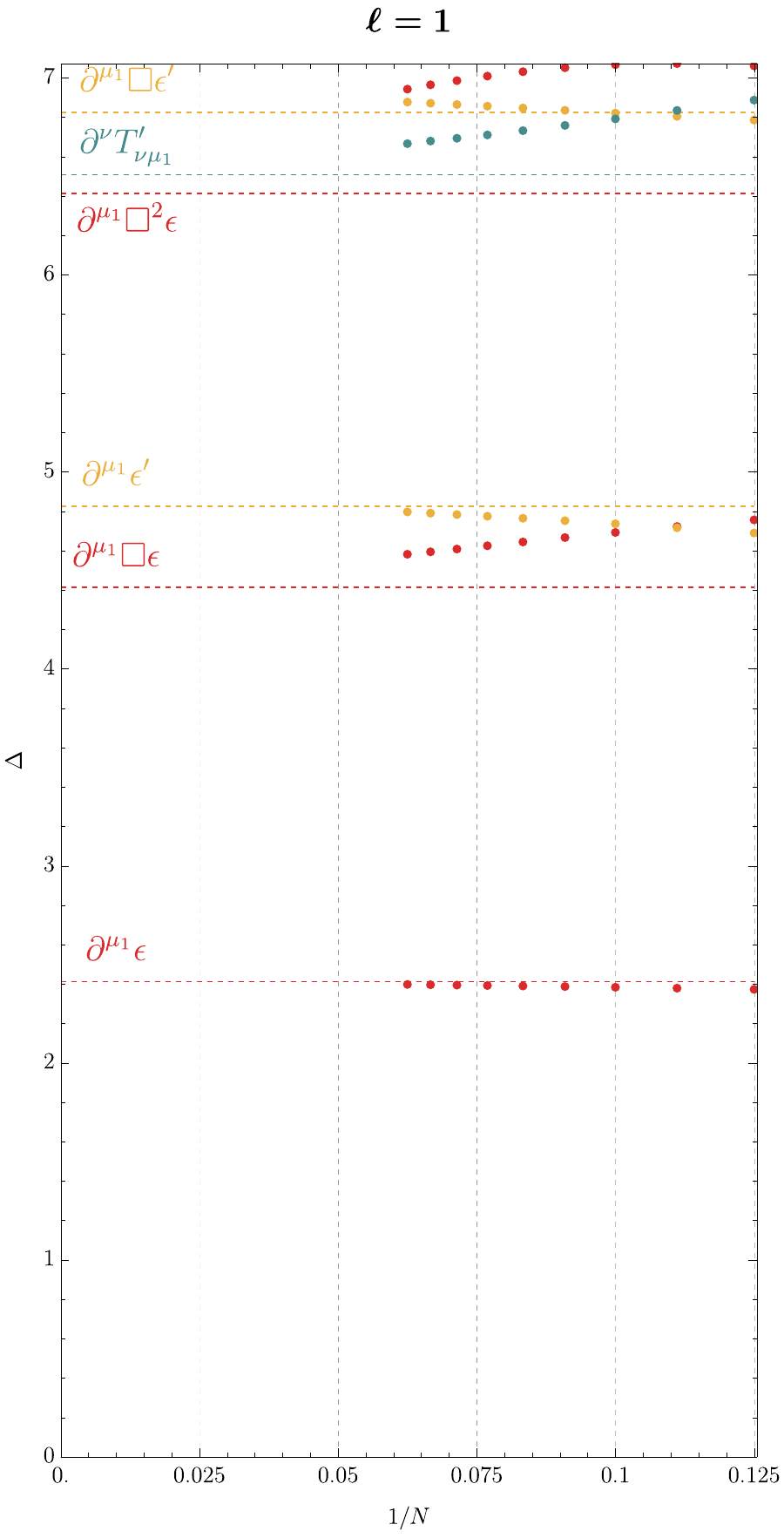}
\end{minipage}
\begin{minipage}{0.32\textwidth}
  \centering
  \includegraphics[width=\linewidth]{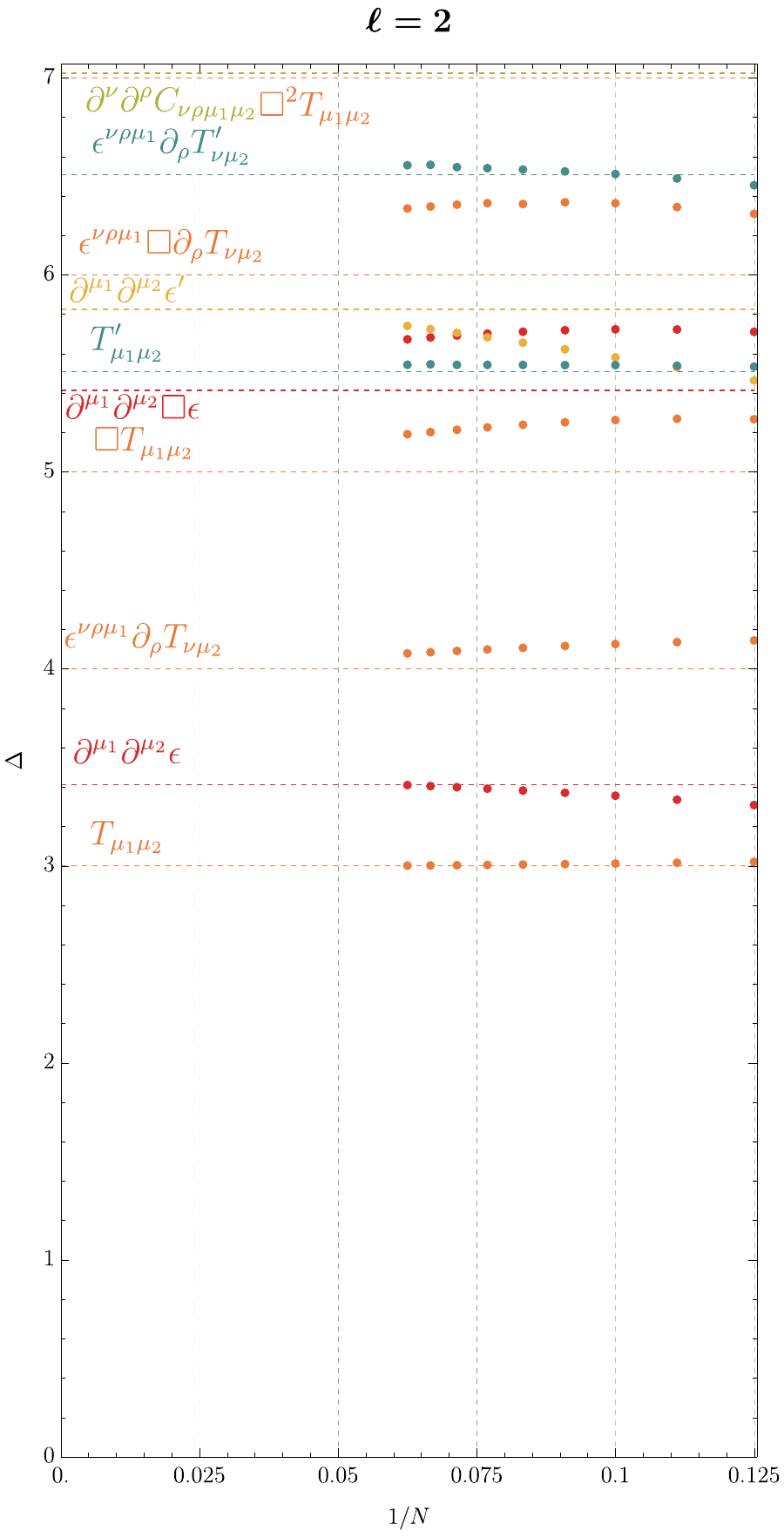}
\end{minipage}
\begin{minipage}{0.24\textwidth}
  \centering
  \includegraphics[width=\linewidth]{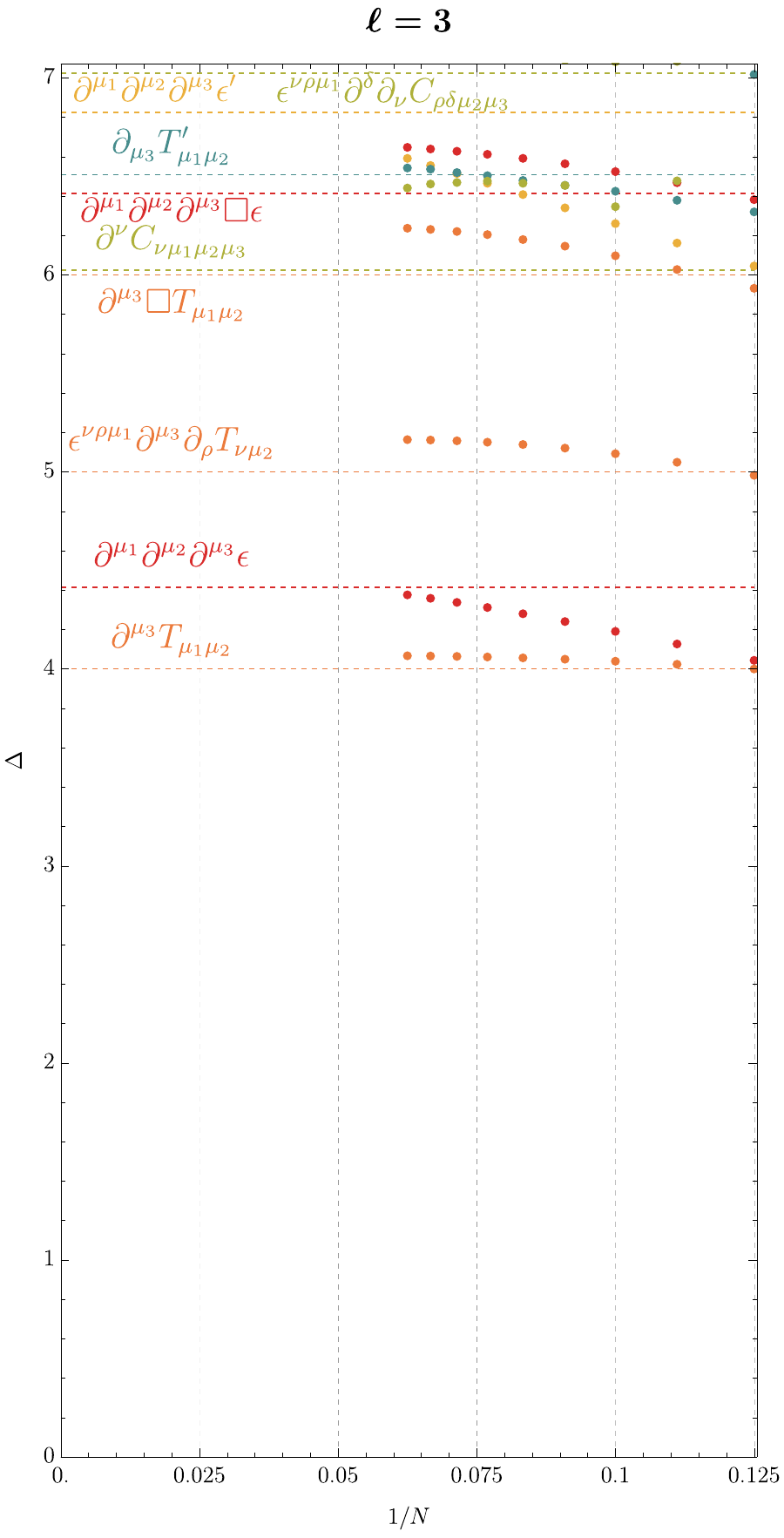}
\end{minipage}
\begin{minipage}{0.24\textwidth}
  \centering
  \includegraphics[width=\linewidth]{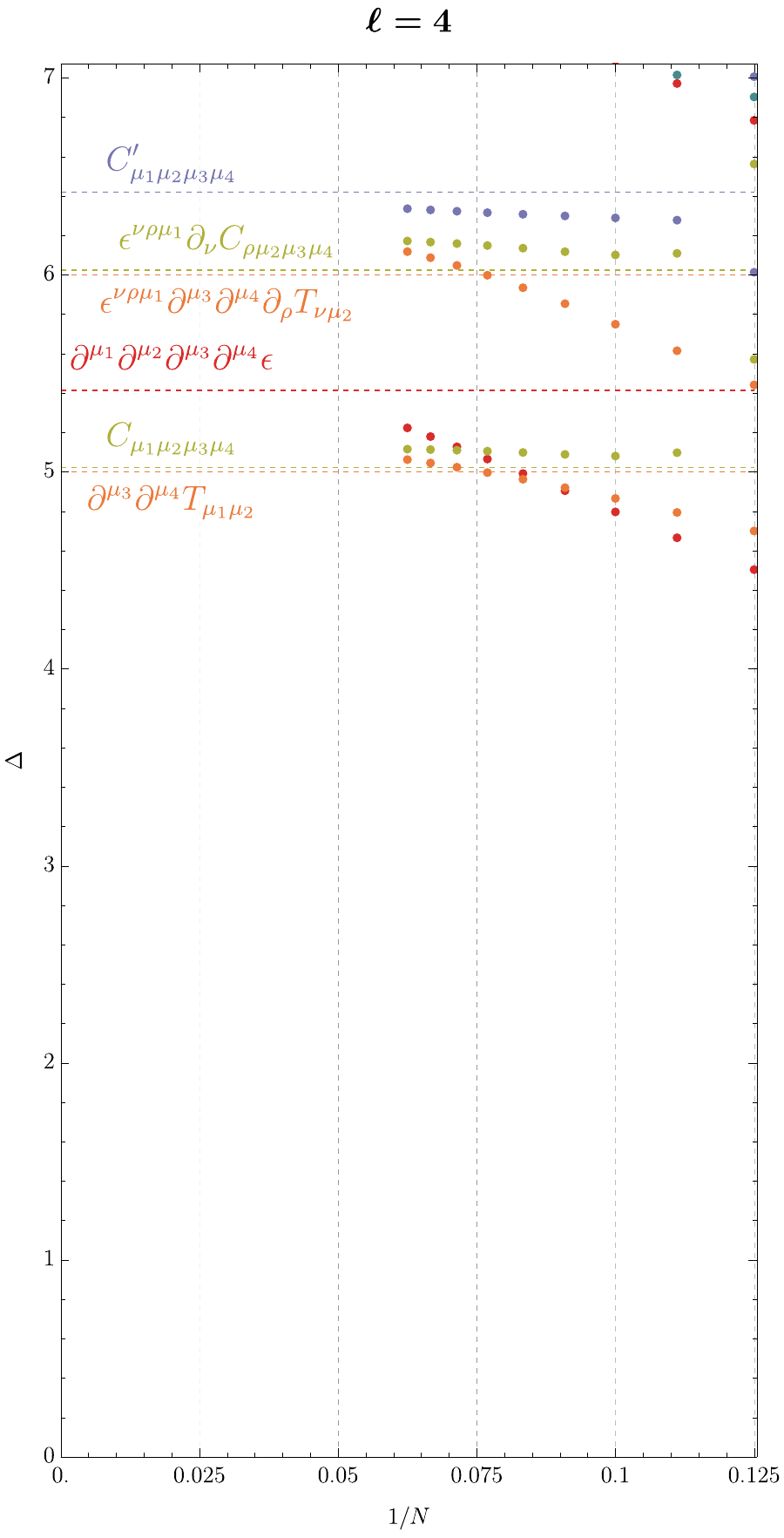}
\end{minipage}
\begin{minipage}{0.24\textwidth}
  \centering
  \includegraphics[width=\linewidth]{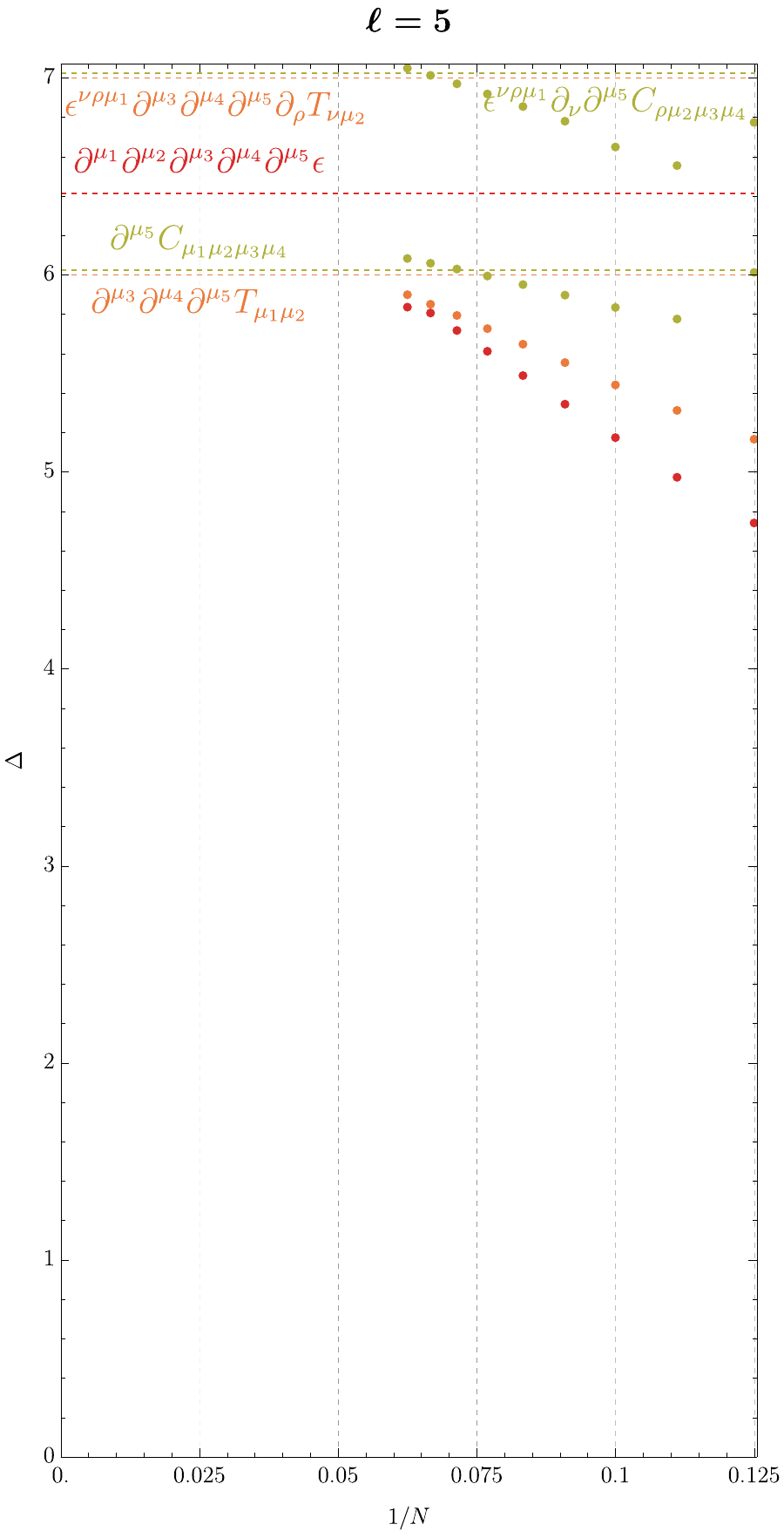}
\end{minipage}
\begin{minipage}{0.24\textwidth}
  \centering
  \includegraphics[width=\linewidth]{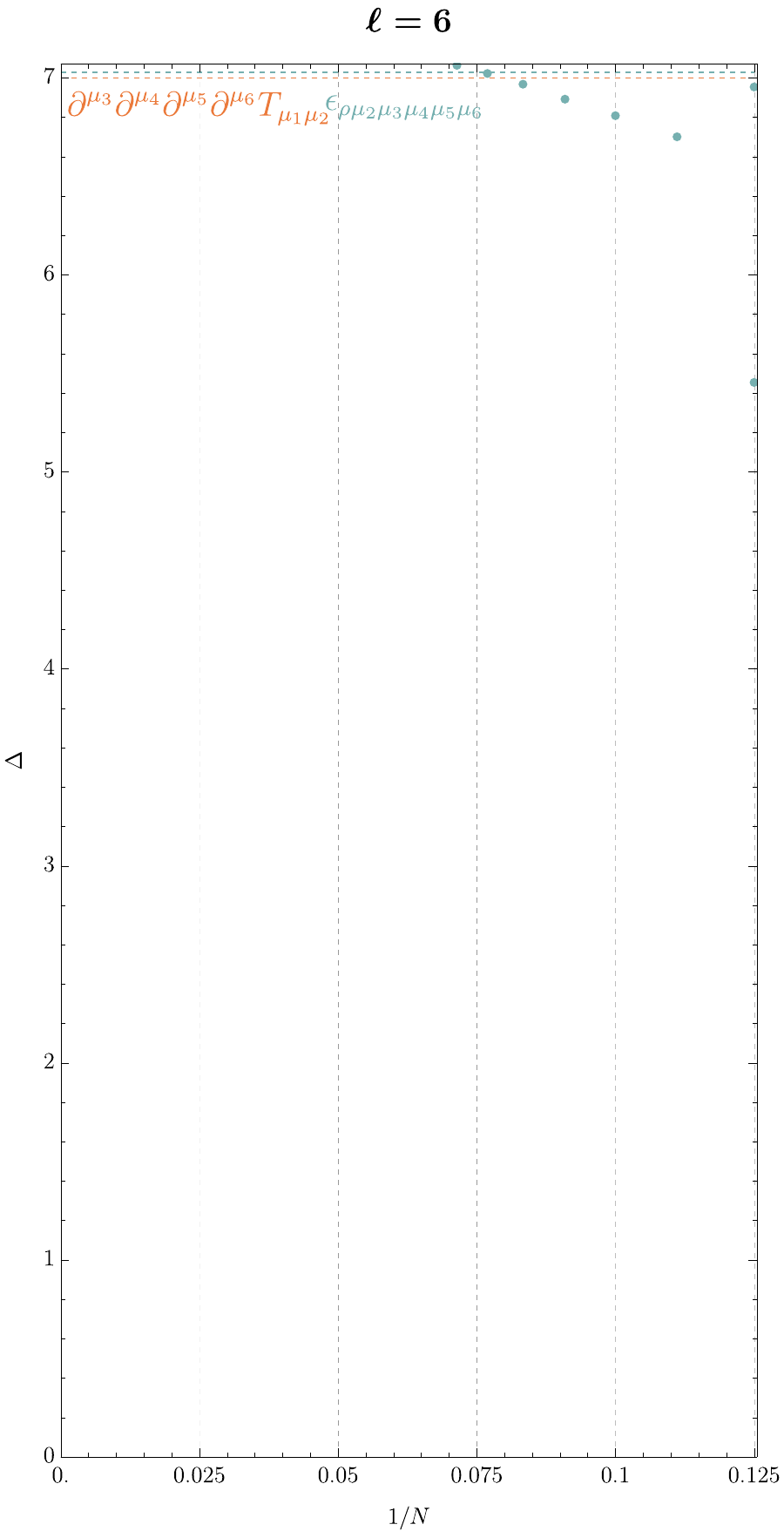}
\end{minipage}
\caption{Conformal multiplets in the $\mathbb{Z}_2$ even sector obtained for various spin sectors and for different values of $N$. Descendants are constructed by acting with $P_z$ on the primaries. }
\label{Fig:conformalMultipletsEven}
\end{figure}
\newpage
\section{OPE Coefficients from the Fuzzy Sphere}
\label{sec:OPE}
\begin{figure}
\begin{center}
\includegraphics[width=0.6\textwidth]{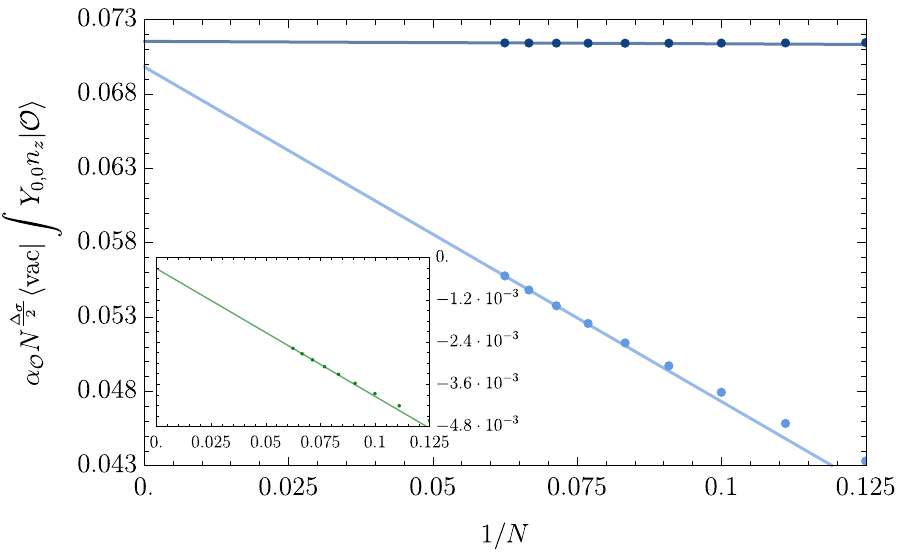}
\caption{Integrated matrix elements of $n_z$ between the vacuum and  $\sigma$ (dark blue) and $\square\sigma $ (light blue) as defined in~\eqref{sigmaOverlaps} with $\alpha_\sigma\equiv (4\pi)^{-1/2}$ and $\alpha_{\square \sigma}\equiv \sqrt{\frac{3}{4\pi \Delta_{\sigma}(2\Delta_\sigma-1)}}$.  The inset shows $\frac{N^{\frac{\Delta_{\sigma}}{2}}}{-4( \Delta_\sigma +1)}\langle \text{vac}| \int Y_{0,0}n_z(\alpha_{\sigma} \ket{\sigma}-\alpha_{\square \sigma }\ket{\sigma}) \approx \frac{a_{\partial_0^2\sigma}}{N}$ as a function of $1/N$.
Solid lines represent linear fits. }
\label{fig:sigmaOverlaps}
\end{center}
\end{figure}
In order to extract OPE coefficients directly on the fuzzy sphere,  we must represent continuum CFT operators in terms of microscopic fermionic variables. Within this framework, good candidates for the $\mathbb{Z}_2$-even and $\mathbb{Z}_2$-odd lowest dimension operators are~\cite{Hu:2023xak}   $n_x$ and $n_z$  introduced in~\eqref{niDef}.
More explicitly,  we can think of the fuzzy sphere operator as a linear combination of CFT ones as follows
\twoseqn{
n_z(\Omega)&=a_{\sigma}(R)\sigma+a_{\partial_0^2 \sigma }(R)\partial_0^2  \sigma+a_{\nabla^2 \sigma }(R)\nabla^2  \sigma+a_{\sigma_2}(R)\sigma^{00}+\cdots \,,
}[]
{
n_x(\Omega)&=a_{\mathds{1}}(R)\mathds{1}+a_{\epsilon}(R)\epsilon +a_{T}(R)T^{00}+a_{\partial_0^2 \epsilon }(R)\partial_0^2  \epsilon+a_{\nabla^2 \epsilon }(R)\nabla^2  \epsilon+\cdots\, ,
}[][]
where the $a_{\CO}$ coefficients are a function of $R$, to take into account possible curvature corrections in the sense of~\eqref{eq:RFuncN},  and they scale as
\eqna{
a_{\CO} \sim \left( \frac{1}{\Lambda_{\rm UV}}\right)^{\Delta_{\CO}}\, .
}[]
To gain intuition about which CFT operators dominate inside $n_{z,x}$ , or in another words which coefficients $a_{\CO}$ are largest, we can study overlaps of the form
\eqna{
\bra{\rm vac} \int d\Omega \lsp  Y_{\ell,0} \lsp n_{z,x}(\Omega)\ket{\CO}\, , 
}[]
where both  bra and ket states are taken to be primary states constructed via the procedure described above and in the $j_z=0$ sector.  Assuming that these states are ``perfect’’ in the sense of conformal perturbation theory, i.e.  that all relevant and slightly irrelevant deformations have been tuned to zero, we can analyze these overlaps using  CFT two-point functions. Focusing on the $\mathbb{Z}_2$ odd sector and  on $\sigma$ and its first scalar descendant state, we obtain
\twoseqn{
&\phantom{\square}\bra{\rm vac} \int d\Omega \lsp Y_{0,0}\lsp  n_{z}(\Omega)\ket{\sigma}=\sqrt{4\pi }\left( \frac{a_{\sigma}(R)}{R^{\Delta_{\sigma}}}+\frac{a_{\partial_0^2 \sigma}(R)\Delta_{\sigma}^2}{R^{\Delta_{\sigma}+2}} \right)+\CO\left(\frac{1}{R^{\Delta_\sigma+4}}\right)\,,
}[]
{ &\begin{aligned}
\bra{\rm vac} \int d\Omega \lsp Y_{0,0}\lsp  n_{z}(\Omega)\ket{\square \sigma}&=\sqrt{4\pi }\sqrt{\frac{\Delta_\sigma(2\Delta_\sigma-1)}{3}} \Bigg( \frac{a_{\sigma}(R)}{R^{\Delta_{\sigma}}} \\
&\quad\, +\frac{a_{\partial_0^2 \sigma}(R)(\Delta_{\sigma}+2)^2}{R^{\Delta_{\sigma}+2}}\Bigg)+\CO\left(\frac{1}{R^{\Delta_\sigma+4}}\right)
\end{aligned}
}[][sigmaOverlaps]
Fig.~\ref{fig:sigmaOverlaps} shows  these quantities as a function of $1/N$ as well as  their combination that isolates $a_{\partial_0^2\sigma}$. The results confirm that $n_z$ couples dominantly with $\ket{\sigma}$.\footnote{We have checked that the overlaps with $\sigma^{00}$ and other higher-dimensional operators are even smaller. } Therefore,  for the purpose of extracting three-point functions, $n_z$ can be treated as a good approximation for the $\sigma$ operator. Accordingly,  we extract OPE coefficients involving $\sigma$ from the ratios
\eqna{
\frac{\bra{\CO_1}\int d\Omega \lsp Y_{\ell, 0} \lsp n_z(\Omega) \ket{\CO_3}}{\bra{\rm vac}\int d\Omega \lsp Y_{0, 0} \lsp n_z(\Omega) \ket{\sigma}}\, ,
}[intnz]
where $\CO_1$ and $\CO_3$ are  primary operators with spin $\ell_1$ and $\ell_3$ in the $j_z=0$ sector\footnote{For the cases where we rediagolize $H$ in the small $|K|^2$ sector (namely $\mathbb{Z}_2$- even spin 0 and 4 and $\mathbb{Z}_2$-odd spin 6), these are the primary operators that we consider.} and they are non vanishing for $\ell=|\ell_1-\ell_3|, |\ell_1-\ell_3|+2, \cdots, \ell_1+\ell_3$.   Explicitly  in the case of two scalar primaries
\eqna{
\frac{\bra{\CO_1}\int d\Omega \lsp Y_{0, 0} \lsp n_z(\Omega) \ket{\CO_3}}{\bra{\rm vac}\int d\Omega \lsp Y_{0, 0} \lsp n_z(\Omega) \ket{\sigma}}&=\frac{\sqrt{4\pi}\lambda^{(0)}_{\CO_1\sigma\CO_3}\left( \frac{a_\sigma}{R^{\Delta_\sigma}}+\frac{a_{\partial_0^2 \sigma}(\Delta_{1}-\Delta_3)^2}{R^{\Delta_{\sigma}+2}}+\cdots\right)}{\sqrt{4\pi }\left( \frac{a_{\sigma}}{R^{\Delta_{\sigma}}}+\frac{a_{\partial_0^2 \sigma}\Delta_{\sigma}^2}{R^{\Delta_{\sigma}+2}} +\cdots \right)}\\&=\lambda^{(0)}_{\CO_1\sigma\CO_3}+\CO(N^{-1})\,.
}[]
For spinning operators, the precise relation between~\eqref{intnz} and the corresponding CFT OPE coefficients is more involved and it is discussed  in Appendix~\ref{app:CFTtoFS}.
\begin{figure}
\centering
\begin{minipage}{0.32\textwidth}
  \centering
  \includegraphics[width=\linewidth]{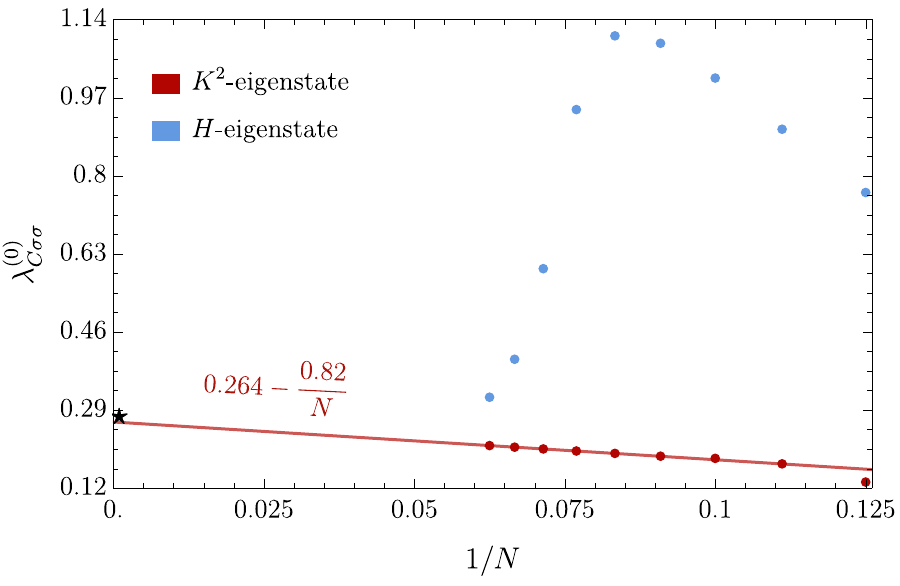}
\end{minipage}
\hfill
\begin{minipage}{0.32\textwidth}
  \centering
  \includegraphics[width=\linewidth]{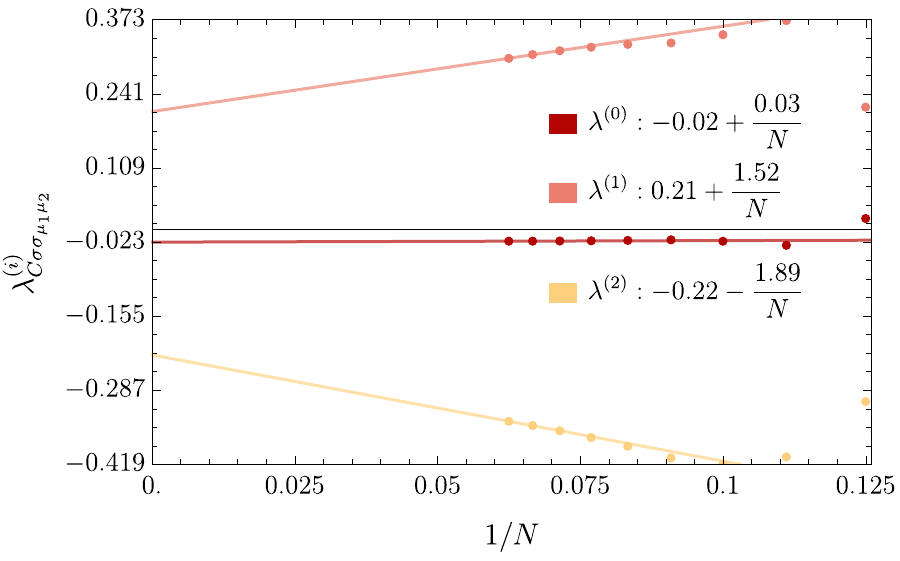}
\end{minipage}
\hfill
\begin{minipage}{0.32\textwidth}
  \centering
  \includegraphics[width=\linewidth]{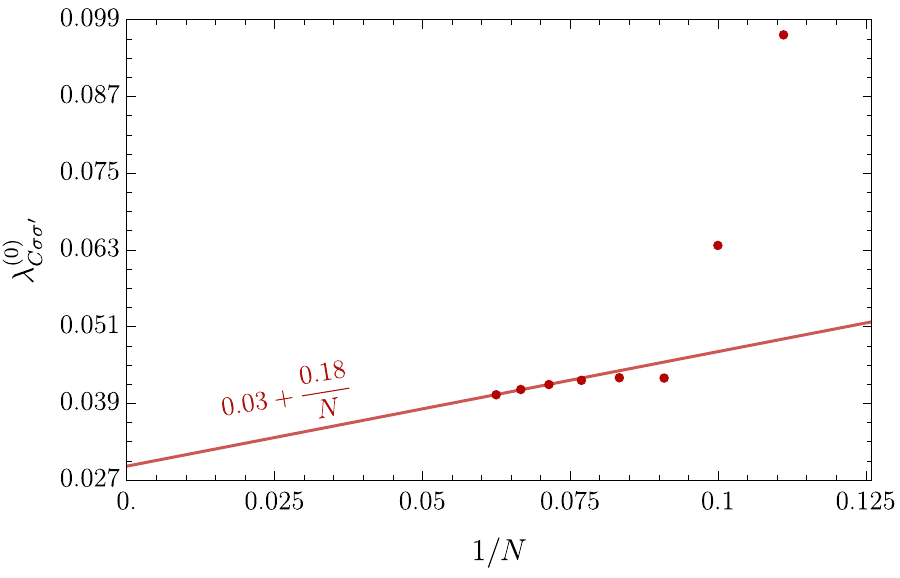}
\end{minipage}
\begin{minipage}{0.4\textwidth}
  \centering
  \includegraphics[width=\linewidth]{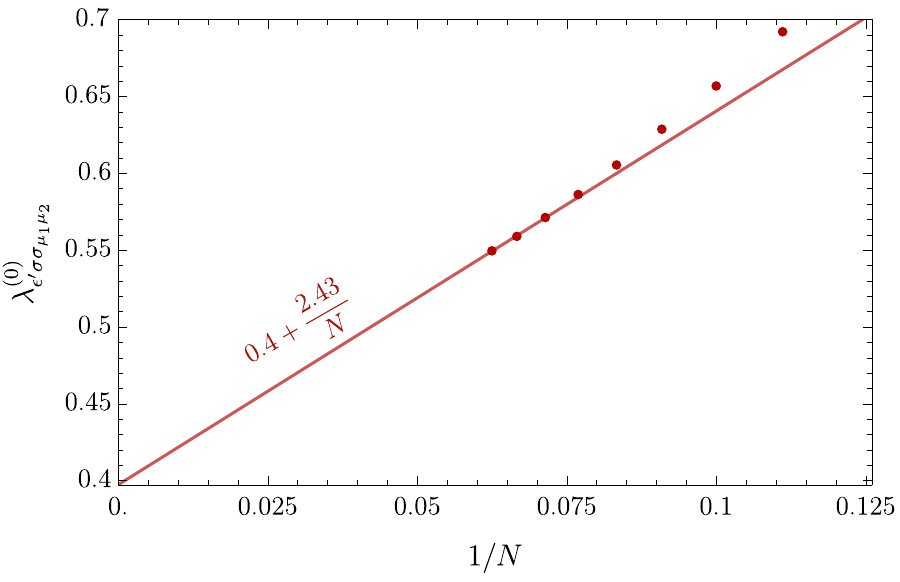}
\end{minipage}
\hspace{1cm}
\begin{minipage}{0.4\textwidth}
  \centering
  \includegraphics[width=\linewidth]{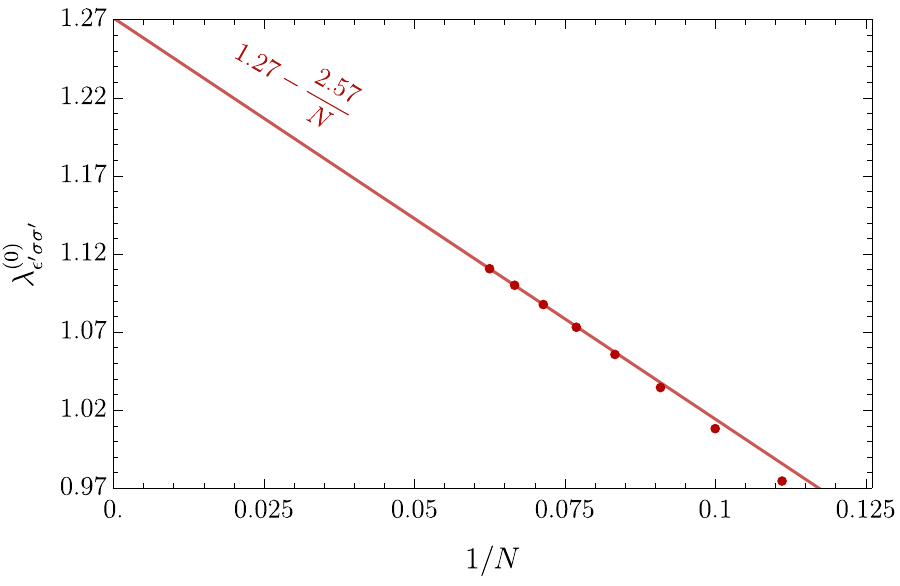}
\end{minipage}
\begin{minipage}{0.41\textwidth}
  \centering
  \includegraphics[width=\linewidth]{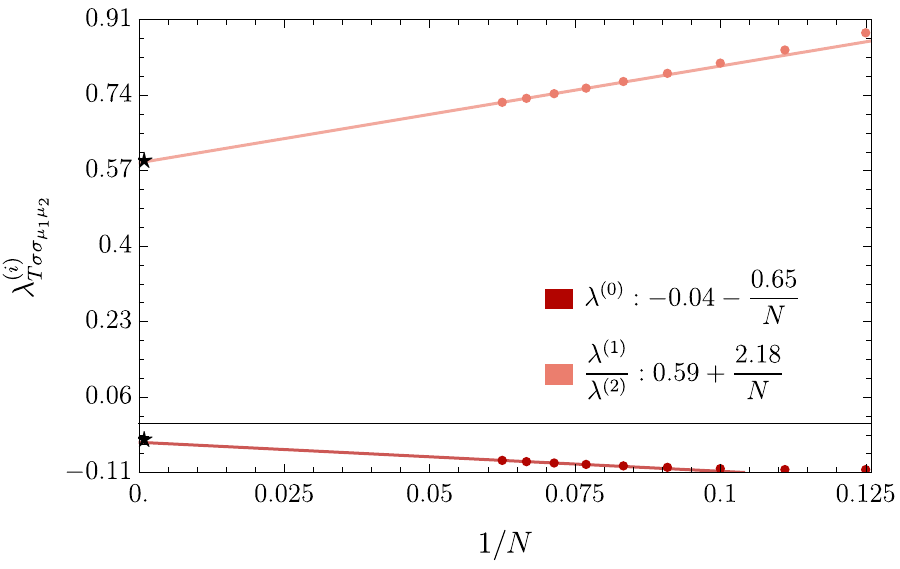}
\end{minipage}
\hspace{1cm}
\begin{minipage}{0.41\textwidth}
  \centering
  \includegraphics[width=\linewidth]{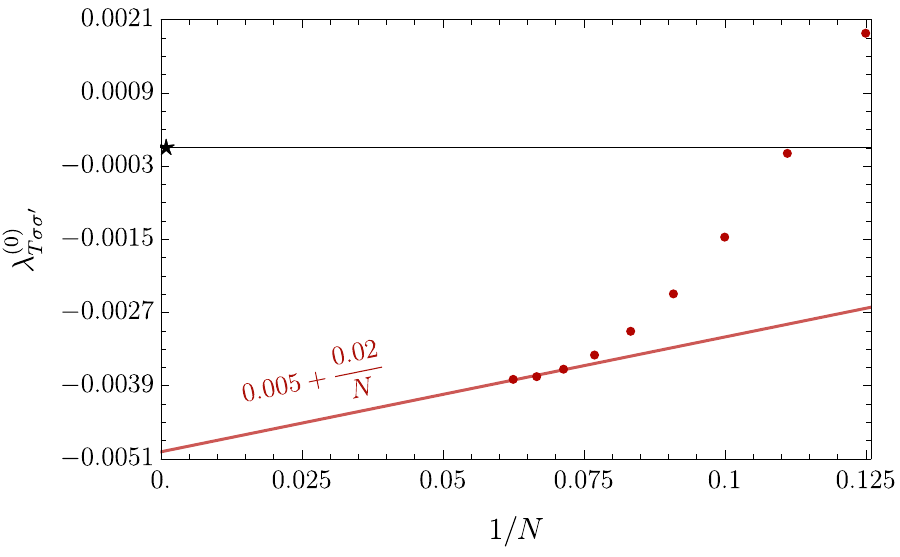}
\end{minipage}
\caption{ ({\it Upper left}): Comparison of OPE coefficient $\lambda_{C \sigma \sigma}$  obtained using $n_z$ between primary states,  constructed by diagonalizing $|K|^2$, vs using pure energy eigenstates.
({\it Upper center}): An example ($\lambda_{C \sigma \sigma_{\mu_1 \mu_2}}$) of an OPE coefficient involving multiple tensor structures.
({\it Upper right}, {\it middle row}): Representative examples of OPE coefficients with a single tensor structure.
({\it Lower row}): Check that OPE coefficients involving the stress tensor $T$ satisfy the Ward identities that fix relations among the different $\lambda^{(i)}$s.  The bottom right OPE coefficient should vanish by the Ward identities; one can see by eye that the $N$-dependence has some curvature, so that including higher order terms in the fit will bring the extrapolated value closer to zero.}
\label{Fig:OPEwithnz}
\end{figure}
In Fig.~\ref{Fig:OPEwithnz} we present several examples of OPE coefficients involving $\sigma$ extracted using this method.  In the first panel, we compare results obtained using primary states,  constructed by diagonalizing $|K|^2$, with those obtained using pure energy eigenstates. The primary-state construction leads to a much clearer and more systematic dependence on $N$. This effect becomes increasingly important for higher-dimensional operators, since a larger number of energy eigenstates must be combined to form a primary.
For three-point functions involving spinning operators, we present results in the standard CFT tensor-structure basis. In the case of OPE coefficients involving the stress tensor $T$, Ward identities fix relations among the different $\lambda^{(i)}$s, as in~\eqref{WIconstraint}, or for the OPE coefficient itself. In the last row of Fig.~\ref{Fig:OPEwithnz}, we show that our extracted coefficients are consistent with these constraints. In particular, for $\lambda_{T\sigma \sigma^\prime}^{(0)}$ it is reasonable to expect that the result can be pushed closer to zero by using more refined fits (including higher powers of $1/N$) and by further improving the definition of $n_z$ to remove residual contamination from $\sigma^\prime$.  Table~\ref{Tab: sigma} report all the remaining  OPE coefficients involving $\sigma$ extracted using a linear extrapolation in $1/N$, using data at $N=14, 15, 16$.

\begin{figure}
\begin{center}
\includegraphics[width=0.6\textwidth]{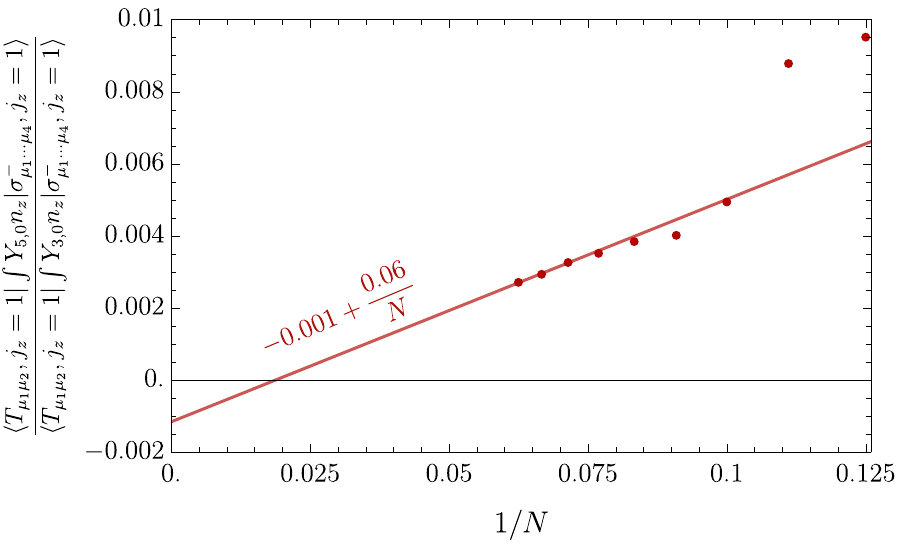}
\caption{Ratios of the two parity odd tensor structures between $j_z=1$ states, in the spherical harmonics basis, appearing in the three-point function of $T_{\mu_1\mu_2}$, $\sigma$ and $\sigma_{\mu_1\mu_2\mu_3\mu_4}^-$.  }
\label{fig:TsS4M}
\end{center}
\end{figure}
So far, all extracted OPE coefficients have corresponded to parity-even tensor structures. When a parity-odd operator is involved, however, parity-odd structures must also be taken into account, as explained in Appendix~\ref{app:CFTtoFS}. These matrix elements vanish identically when the states are restricted to the $j_z=0$ sector.  As an illustrative example, Fig.~\ref{fig:TsS4M} shows the analysis of the OPE coefficient involving the stress tensor and the new parity-odd operator 
$\sigma_{\mu_1\mu_2\mu_3\mu_4}^{-}$,  computed using data in the $j_z = 1$ sector. In this case, two parity-odd tensor structures contribute, and their ratio is fixed by Ward identities to be
\eqna{
\frac{\langle T_{\mu_1\mu_2}, j_z=1|\int Y_{5,0} n_z|\sigma_{\mu_1 \cdots \mu_4}^{-}, j_z=1\rangle}{\langle T_{\mu_1\mu_2}, j_z=1|\int Y_{3,0} n_z|\sigma_{\mu_1 \cdots \mu_4}^{-}, j_z=1\rangle}=-\frac{\sqrt{\frac{7}{11}} (\Delta_{\sigma_4^-}-\Delta_\sigma -6)}{\Delta_{\sigma_4^-}-\Delta_\sigma +3}\, .
}[TNzSigM]
This relation provides an alternative way to extract the dimension of $\sigma_{\mu_1\mu_2\mu_3\mu_4}^{-}$
\eqna{
\Delta_{\sigma_4^{-}}=6.531\, .
}[]
In fact, because the matrix element (\ref{TNzSigM}) turns out to be very small, it is quite useful for constraining $\Delta_{\sigma_4^{-}}$ even though the relative error on (\ref{TNzSigM}) is fairly large.  The point is that just knowing that (\ref{TNzSigM}) is very small tells us that $\Delta_{\sigma_4^{-}}$ is very close to $\Delta_\sigma + 6$.

Now moving on to  the $\mathbb{Z}_2$-even operator, we consider the following matrix elements in the scalar sector
\threeseqn{
\bra{\rm vac} \int \!d\Omega \lsp Y_{0,0}\lsp  n_{x}\ket{\rm vac}&=\sqrt{4\pi} a_{\mathds{1}}(R) +\cdots\, ,
}[]
{
\bra{\rm vac} \int \!d\Omega \lsp Y_{0,0}\lsp  n_{x}\ket{\rm \epsilon}&=\frac{\sqrt{4\pi}}{R^{\Delta_\epsilon}}\left( a_{\epsilon}(R)+\frac{a_{\partial_{0}^2\epsilon}(R)\Delta_\epsilon^2}{R^2}\right)+\cdots\, ,
}[]
{
\bra{\rm vac} \int \!d\Omega \lsp Y_{0,0}\lsp  n_{x}\ket{\rm \square \epsilon}&=\frac{\sqrt{4\pi}}{R^{\Delta_\epsilon}}\sqrt{\frac{\Delta_\epsilon(2\Delta_\epsilon-1)}{3}}\left(a_{\epsilon}(R)+ \frac{a_{\partial_{0}^2\epsilon}(R)(\Delta_{\epsilon}+2)^2}{R^2}\right)+\cdots\, ,
}[][scalarEpsOv]
and, in the spin-2 sector,
\eqna{
\bra{\rm vac} \int d\Omega \lsp Y_{2,0}\lsp  n_{x}\ket{T_{\mu_1\mu_2}}&=\bra{\rm vac} \int d\Omega \lsp Y_{2,0}\lsp  T^{00}(\Omega)\ket{T_{\mu_1\mu_2}}+\cdots \\
&=\sqrt{\frac{8\pi}{15}}\frac{a_T}{R^3}+\cdots\, , 
}[spinEpsOv]
where we have used the (slightly unconventional) normalization $\langle T T \rangle=\frac{H_{12}^2}{(x_{12}^2)^5}$  for the stress-tensor two-point function.   
\begin{figure}
\centering
\begin{minipage}{0.48\textwidth}
        \centering
        \includegraphics[width=1\textwidth]{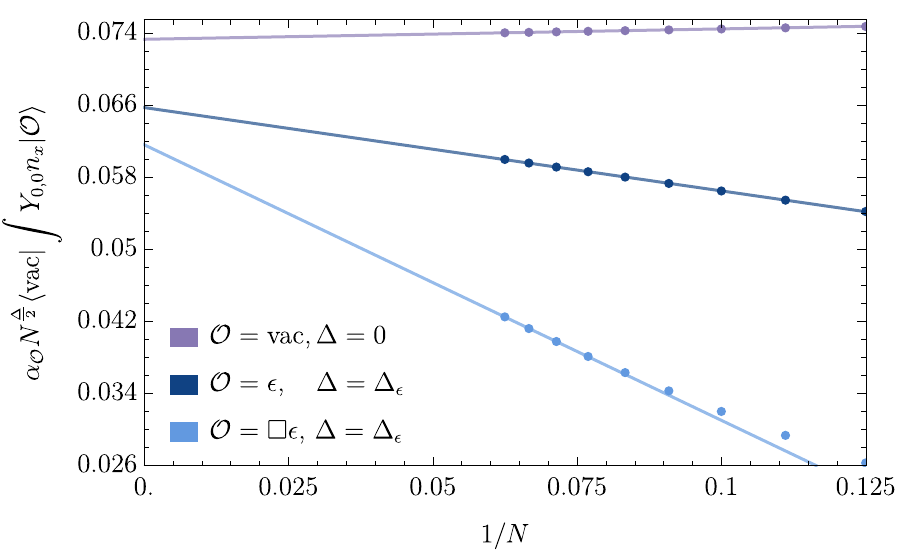}
    \end{minipage}
    \hfill
    % Second figure
    \begin{minipage}{0.48\textwidth}
        \centering
        \includegraphics[width=1\textwidth]{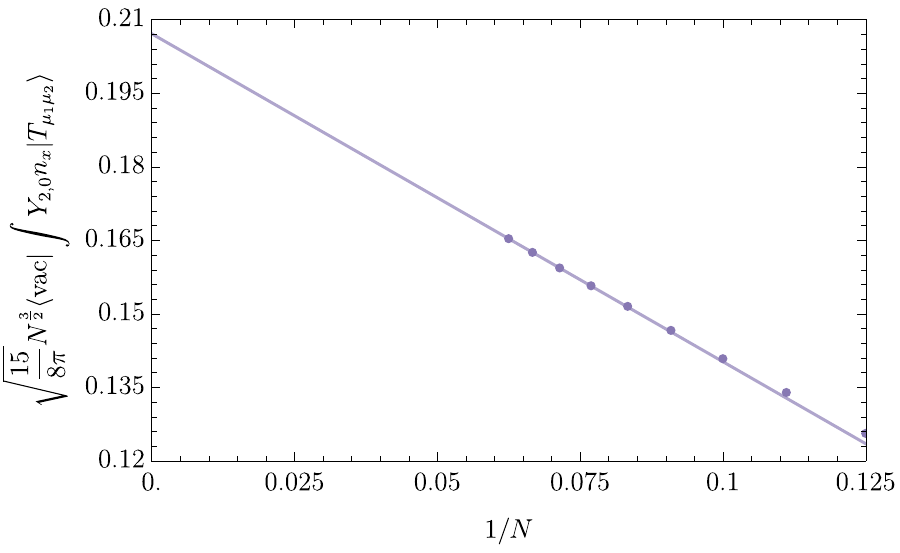}
    \end{minipage}
\caption{Integrated matrix elements of $n_x$ between the vacuum and $\CO=\text{vacuum}, \epsilon, \square\epsilon$ (\textit{left}) and the stress tensor (\textit{right}) as defined in~\eqref{scalarEpsOv} and~\eqref{spinEpsOv} with $\alpha_{\rm vac}=\alpha_\epsilon\equiv (4\pi)^{-1/2}$ and $\alpha_{\square \epsilon}\equiv \sqrt{\frac{3}{4\pi \Delta_{\epsilon}(2\Delta_\epsilon-1)}}$.  
Solid lines represent linear fits. }\label{Fig:epsilonOverlaps}
\end{figure}
Fig.~\ref{Fig:epsilonOverlaps} shows these matrix elements as a function of $1/N$ together with their extrapolation, which provides an estimate of the coefficients $a_\CO$  and confirms that $n_x$ receives dominant contributions from  $\mathds{1}$, $\epsilon$ and $T$.  Hence, when using  $n_x$ to extract three-point functions involving $\epsilon$,  we must first remove the identity contribution. More explicitly, to extract the OPE coefficients we use the ratio
\eqna{
&\frac{\bra{\CO_1}\int d\Omega \lsp Y_{\ell, 0} \lsp n_x(\Omega) \ket{\CO_3}}{\bra{\rm vac}\int d\Omega \lsp Y_{0, 0} \lsp n_x(\Omega) \ket{\epsilon}} &&\qquad \text{for }\CO_1\neq\CO_3\, ,  \\
&\frac{\bra{\CO_1}\int d\Omega \lsp Y_{\ell, 0} \lsp n_x(\Omega) \ket{\CO_3}-\bra{\rm vac}\int d\Omega \lsp Y_{0, 0} \lsp n_x(\Omega) \ket{\rm vac}}{\bra{\rm vac}\int d\Omega \lsp Y_{0, 0} \lsp n_x(\Omega) \ket{\epsilon}} &&\qquad \text{for }\CO_1=\CO_3\text{ and }\ell=0\, . \\
}[intnx]
As an explicit example, let us consider the three-point function between $\epsilon$ and two identical scalar operators
\eqna{
&\frac{\bra{\CO}\int d\Omega \lsp Y_{0, 0} \lsp n_x(\Omega) \ket{\CO}-\bra{\rm vac}\int d\Omega \lsp Y_{0, 0} \lsp n_x(\Omega) \ket{\rm vac}}{\bra{\rm vac}\int d\Omega \lsp Y_{0, 0} \lsp n_x(\Omega) \ket{\epsilon}}=\frac{R^{\Delta_\epsilon}}{{a_\epsilon}\left(1+\frac{a_{\partial_0^2\epsilon}\Delta_\epsilon^2}{a_\epsilon R^2}+O\left(\frac{1}{R^4}\right)\right)}\\
&\qquad \times\left[  \left(a_\mathds{1}+a_\epsilon \frac{\lambda_{\CO\epsilon\CO}^{(0)}}{R^{\Delta_{\epsilon}}}+a_T \frac{d-1}{d} \frac{\lambda_{\CO T \CO}^{(0)}}{R^3}+\cdots \right)-\left(a_\mathds{1}+\cdots \right)\right]\\
&\qquad=\lambda_{\CO\epsilon\CO}^{(0)}+\frac{2 a_T}{3 a_\epsilon} \frac{\lambda_{\CO T \CO}^{(0)}}{N^{\frac{3-\Delta_{\epsilon}}{2}}}+\CO(N^{-1}) . 
}[nxMatrices]
The contribution  of the  $T^{00}$ component inside $n_x$ can either be removed by fitting it as a subleading term in $1/N$ or  subtracted explicitly using the value of $a_T$ extracted from Fig.~\ref{Fig:epsilonOverlaps}.  In the case of two external scalar states,\footnote{In the case of spinning operators,  removing the contribution from $T^{00}$ inside $n_x$ is more involved due to the presence of multiple tensor structures, and in that case we will simply treat the $T^{00}$ contamination as a subleading effect and remove it via a fit in $1/N$ when extracting OPE coefficients.} the OPE coefficient $\lambda_{\CO T \CO}^{(0)}$ is fixed by the Ward identities
\eqna{
\langle \CO(P_1) T(Z_2, P_2) \CO(P_3)\rangle &=\lambda_{\CO T \CO}^{(0)}\frac{V_{2,13}^2}{(2P_1\cdot P_2)^{\frac{2\Delta_{\CO}-5}{2}}(2 P_1\cdot P_3)^{\frac{5}{2}}(2 P_2\cdot P_3)^{\frac{5}{2}}} \, ,\\
\lambda_{\CO T \CO}^{(0)}&=-\frac{\Delta}{\sqrt{c_T}} \frac{d}{(d-1)S_d}\, , 
}[]
where we have introduced the central charge $c_T$~\cite{Chang:2024whx}
\eqna{
\frac{c_T}{c_B}=0.946538675(42)\, , \qquad c_B=\frac{d}{d-1}\frac{1}{S_d^2}\, ,  \qquad S_d=\frac{2\pi^{d/2}}{\Gamma\mleft(\frac{d}{2} \mright)}\,.
}[]

\begin{figure}
\centering
\begin{minipage}{0.4\textwidth}
  \centering
  \includegraphics[width=\linewidth]{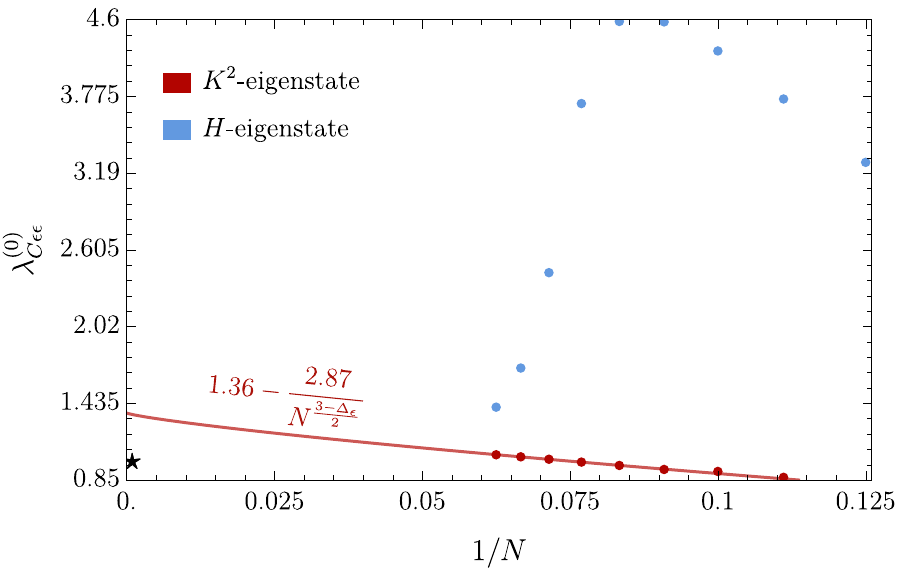}
\end{minipage}
\hspace{1cm}
\begin{minipage}{0.4\textwidth}
  \centering
  \includegraphics[width=\linewidth]{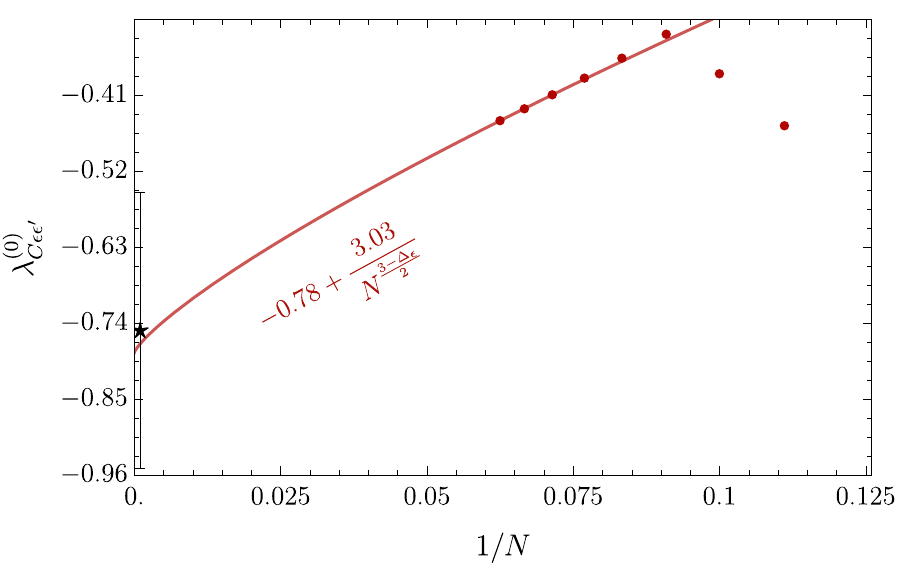}
\end{minipage}
\begin{minipage}{0.4\textwidth}
  \centering
  \includegraphics[width=\linewidth]{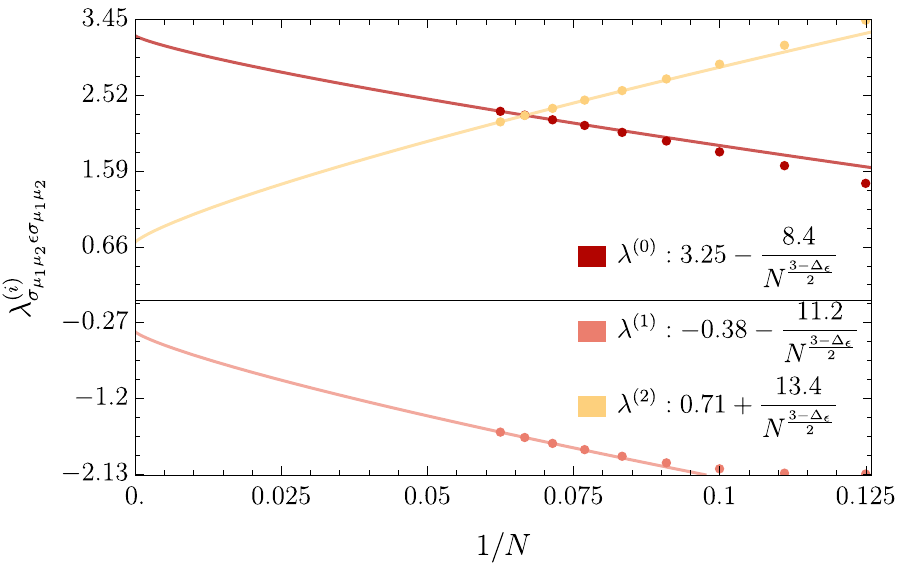}
\end{minipage}
\hspace{1cm}
\begin{minipage}{0.4\textwidth}
  \centering
  \includegraphics[width=\linewidth]{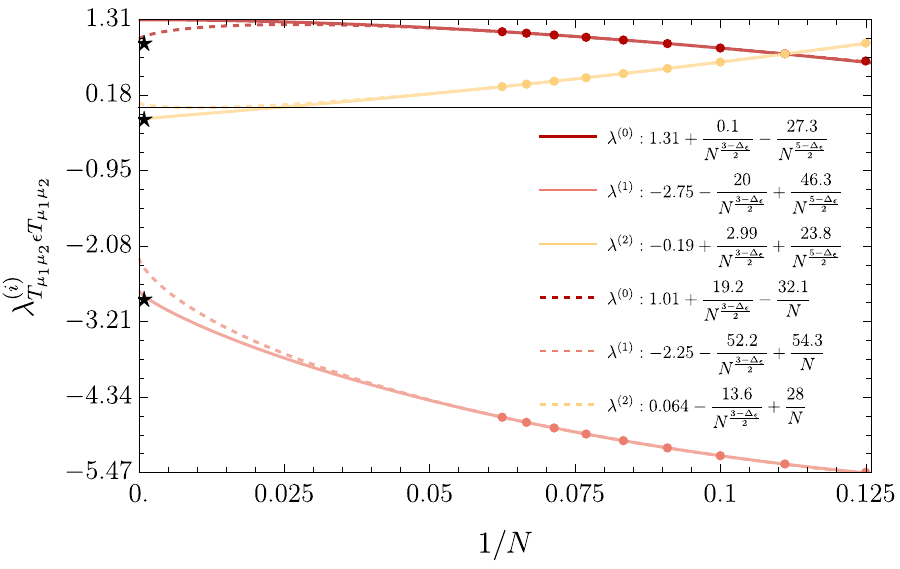}
\end{minipage}
\caption{Analogous to Fig.~\ref{Fig:OPEwithnz}, for OPE coefficients using $n_x$. ({\it Upper left}): Comparison of OPE coefficient $\lambda_{C \epsilon \epsilon}$  obtained using $n_x$ between primary states,  constructed by diagonalizing $|K|^2$, vs using pure energy eigenstates.
({\it Upper right}):  Representative example of an OPE coefficients with a single tensor structure.
({\it Lower right}):  Representative example of an OPE coefficient involving multiple tensor structures.
({\it Lower left}): Comparison of $\lambda_{T \epsilon T}$ with  results from bootstrap and Ward identities.  Different fits are shown, where the main subleading correction to $n_x$ is assumed to be a descendant of $T$ vs a descendant of $\epsilon$. }
\label{Fig:OPEwithnx}
\end{figure} 
Fig.~\ref{Fig:OPEwithnx} shows several examples of OPE coefficients involving $\epsilon$ extracted from matrix elements with $n_x$, without explicitly removing $T^{00}$, also in this case results obtained using primaries are in general better than those obtained using pure energy eigenstates.  For the $\langle T \epsilon T\rangle$ three-point function,  we present multiple fitting procedures. The star symbols denote bootstrap results obtained from~\cite{Chang:2024whx}.
\begin{figure}
\centering
\begin{minipage}{0.45\textwidth}
  \centering
  \includegraphics[width=\linewidth]{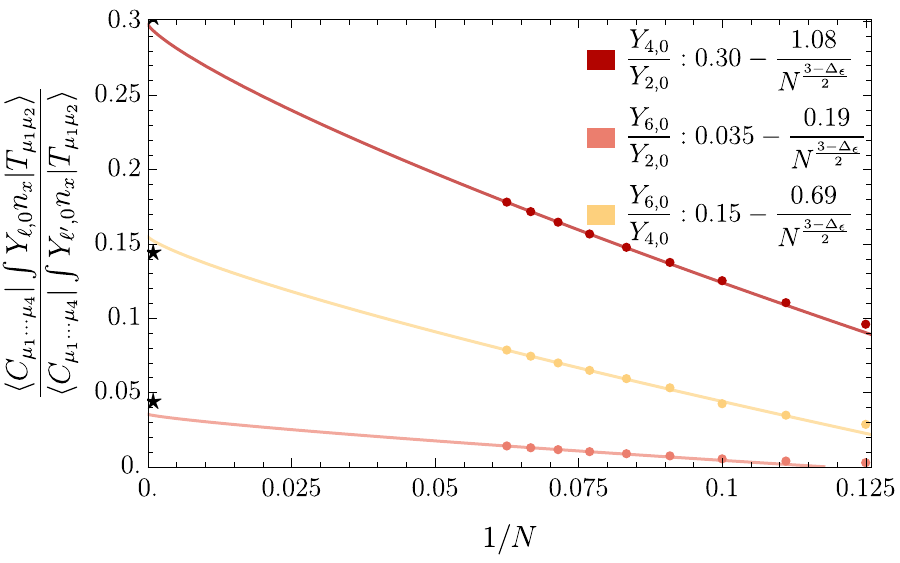}
\end{minipage}
\hfill
\begin{minipage}{0.47\textwidth}
  \centering
  \includegraphics[width=\linewidth]{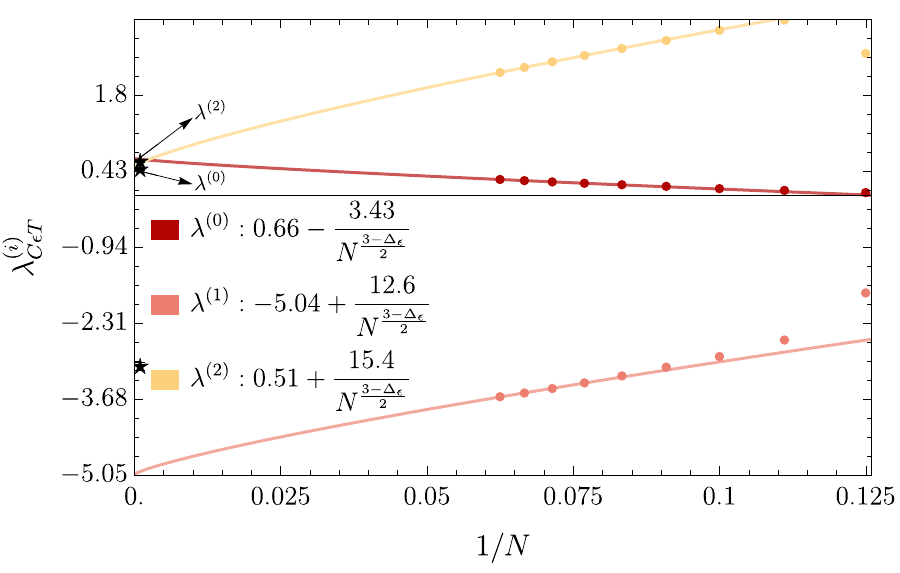}
\end{minipage}
\caption{Analysis of the three-point function involving $C_{\mu_1 \cdots \mu_4}$, $\epsilon$ and $T_{\mu_1\mu_2}$.  \textit{(Left)}: ratios of integrations over different harmonics and comparison to the value imposed by Ward identities. \textit{(Right)}: data for $\lambda^{(i)}$s, fits assuming that the leading correction to $n_x$ is coming from $T$ and comparison to the bootstrap data (stars).}
\label{Fig:CET}
\end{figure}
Fig.~\ref{Fig:CET}  analyzes the three-point function involving the stress tensor, $C_{\mu_1\mu_2\mu_3\mu_4}$ and $\epsilon$.  In this case there are three possible tensor structures, or equivalently three nonvanishing integrals involving spherical harmonics. However, only one of these structures is independent due to Ward identities.  In the left panel, we show the ratios of the different harmonic integrals. The star denotes the prediction from the Ward identities, which is well reproduced by the data. In the right panel, we display the coefficients of the corresponding tensor structures and compare the extrapolated values with predictions from the five-point bootstrap analysis~\cite{Poland:2025ide} (shown as stars).  Notably, the harmonic ratios satisfy the Ward identity constraints more accurately than the individual coefficients $\lambda^{(i)}$.  It is therefore reasonable to expect that cancellations occur at the level of the harmonic integrals, but are partially lost when these are combined to construct the $\lambda^{(i)}$s. The remaining OPE coefficients involving $\epsilon$ can be found in Table~\ref{Tab: epsilonOdd} between $\mathbb{Z}_2$ odd states and in Table~\ref{Tab: epsilonEven} between $\mathbb{Z}_2$ even ones.

\begin{figure}
\centering
\begin{minipage}{0.45\textwidth}
  \centering
  \includegraphics[width=\linewidth]{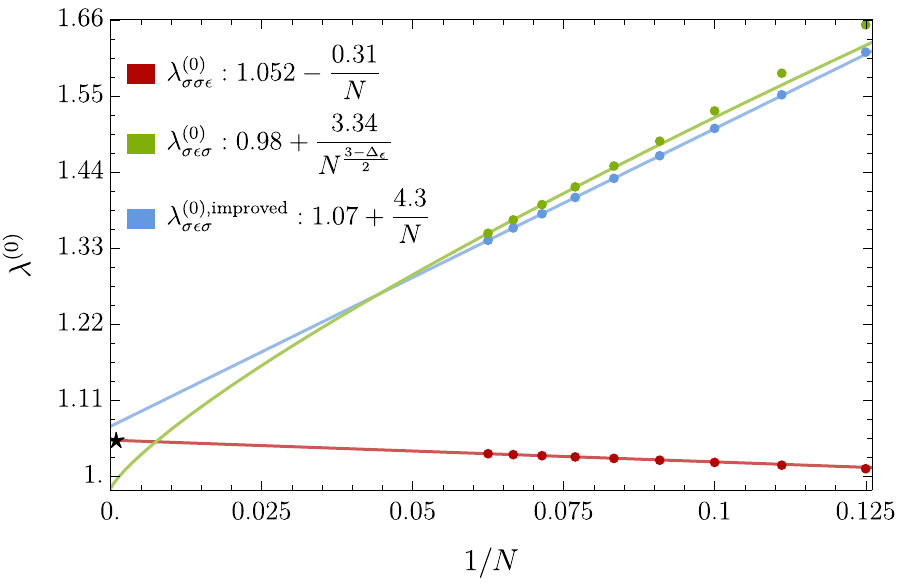}
\end{minipage}
\hfill
\begin{minipage}{0.47\textwidth}
  \centering
  \includegraphics[width=\linewidth]{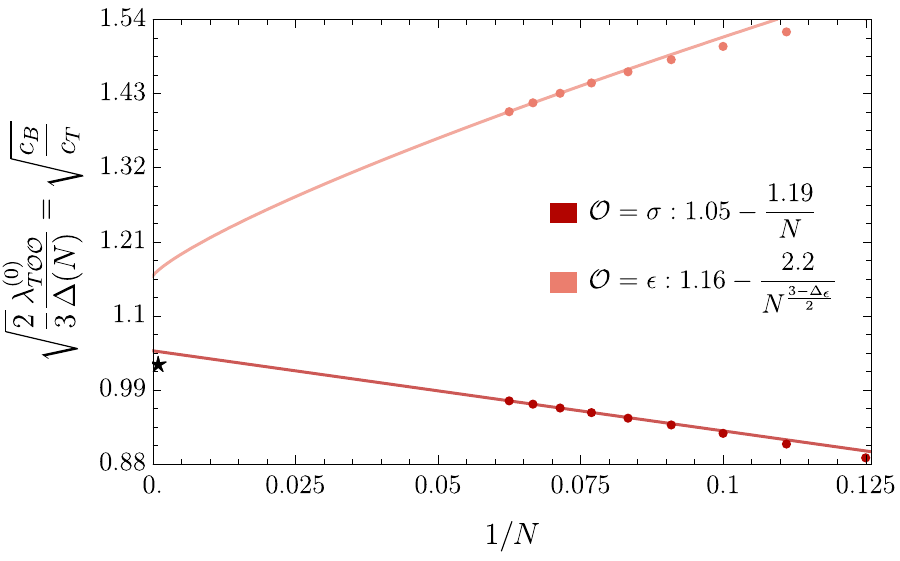}
\end{minipage}
\caption{\textit{(Left)}: $\epsilon \sigma \sigma$ OPE coefficient extracted from matrix elements of $n_z$ and $n_x$ with and without explicitly removing the contribution from $T$ inside $n_x$. \textit{(Right)}: measure of the central charge $c_T$ from $\lambda^{(0)}_{T\sigma\sigma}$ and $\lambda^{(0)}_{T\epsilon\epsilon}$, as prescribed by Ward identities.  }
\label{Fig:SSEandTOO}
\end{figure}

Finally, in Fig.~\ref{Fig:SSEandTOO} we focus on the OPE coefficients of two $\sigma$ and one $\epsilon$ extracted using both $n_z$ and $n_x$. The data obtained from $n_z$, when linearly extrapolated in $1/N$,  reproduces known bootstrap result with an accuracy of about  $0.03\%$. For the data obtained from matrix elements of $n_x$ we present two analysis.  First, we show a fit considering that the leading correction arises from contamination by $T^{00}$ in $n_x$.  Second, we consider data in which this contribution is explicitly removed using~\eqref{nxMatrices} together with the extrapolated value of $a_T$.  After subtracting this contamination, we fit the resulting data with a linear extrapolation in $1/N$. We choose a linear fit because several operators contribute corrections with powers close to $1/N$. Removing the $T^{00}$ contamination leads to a modest but noticeable improvement in the final result.
In the second panel,  we use the OPE coefficients $\lambda_{T\sigma\sigma}^{(0)}$ and  $\lambda_{T\epsilon\epsilon}^{(0)}$ to extract the value of the central charge $c_T$. Once again the result obtained from the matrix element of $n_z$ reproduces the bootstrap prediction more accurately than those obtained from $n_x$.

Overall, these results highlight the importance of the fuzzy-sphere framework for extracting OPE coefficients, especially in cases involving two external spinning operators, which are difficult to access with other bootstrap-based techniques. We also find that working with properly constructed primary states, rather than pure energy eigenstates, leads to a significant improvement in the quality and stability of the extrapolations, making the large-$N$ behavior much clearer. Finally the definition of microscopic operators,  particularly $n_x$, as already noted in~\cite{Hu:2023xak},   can be further refined to reduce residual contamination, which may allow for even more precise determinations of OPE data in future studies.

Before closing this section, we would like to make some brief comments about the signs of the OPE coefficients. These signs contain physical information that is important for various applications, e.g.~they matter when we define the Hamiltonian in the vicinity of the fixed point using Truncated Conformal Space Approach (TCSA), as well as if they are used in the boundary conformal bootstrap where they appear linearly.  However, the signs of individual OPE coefficients are unphysical, because they can be changed by flipping the sign convention of the operators (or equivalently of the states). This issue is especially troublesome for extracting `off-diagonal' OPE coefficients --- that is, OPE coefficients of the form $\< i | O |j\>$ where $|i \> \ne |j\>$ ---  from the fuzzy sphere, because the signs of the states are chosen somewhat haphazardly by the numeric diagonalization algorithms being used, and so there is no clear convention for how the signs are chosen and indeed they can even differ for a single state from one computational run to the next.  We have taken special care to use a fixed set of eigenstates for our computations so that all the signs are consistent.  Of course, many different choices of the signs of the states are allowed, and to simplify the results, we  rephased some of the states in order to  reduce the number of negative signs for the large $N$ matrix elements.   Physical signs can be diagnosed by considering products of matrix elements of the form $\< i | O | j\> \< j | O | k\> \< k | O |  \ell \> \< \ell | O | i\>$, which is manifestly invariant under any rephasing of any of the states, and therefore is convention-independent.

\section{Comparison with ETH}
\label{sec:ETH}

The Eigenstate Thermalization Hypothesis (ETH) \cite{Srednicki:1994mfb,Srednicki:1999bhx,Srednicki:1995pt} establishes a relation between expectation values $\< E | \CO_1 \CO_2 \dots \CO_n | E\>$ of operators $\CO_i$ in high energy eigenstates $|E\>$ to thermal expectation values.  In the case of one-point functions, these become predictions about the asymptotic form of OPE coefficients $\lambda^{(a)}_{\CO \phi \CO}$ when the dimension of $\CO$ is large.  In this section, we will compare our OPE coefficients for $\sigma$ and $\epsilon$ with these predictions.

More precisely, ETH predicts that matrix elements of an operator $\CO$ between eigenstates $|E\>$ and $|E'\>$ should be \cite{Srednicki:1999bhx}
\begin{equation}
\< E | \CO | E'\> = \< \CO\>_{\beta(\bar{E})} \delta_{E,E'} + e^{-S(\bar{E})/2} f_{\CO}(\omega,\bar{E}) R_{E,E'},
\label{eq:ETHPrediction}
\end{equation}
where $\bar{E} \equiv (E+ E')/2, \omega \equiv E-E'$, $\beta(E)$ is the inverse temperature at $E$, $S(E)$ is the thermodynamic entropy at $E$, and $R_{E,E'}$ is a random matrix with zero mean and unit variance. The function $f_{\CO}(\omega, \bar{E})$ is fixed by the thermal two-point function of $\CO$.  

We begin by focusing on the diagonal matrix elements.  For $\sigma$, these all vanish by symmetry, but for $\epsilon$ the ETH is a highly nontrivial prediction.  First of all,  at high temperatures $T$, observables pass over to their flat-space limit on $S^1_\beta \times \mathbb{R}^{d-1}$, where conformal invariance becomes much more constraining.  Most importantly, the free energy $F = -\beta^{-1} \log Z$ is fixed to be
\begin{equation}
F = V f \beta^{-d},
\end{equation}
where $f$ is theory-dependent numeric constant and $V$ is the volume of space; then $S = \beta^2 \partial_\beta F$ and $E=\partial_\beta(\beta F)$ as usual.  Taking $V$ to be the volume of the two-sphere, one obtains the temperature as a function of energy:
\begin{equation}
T = \left( \frac{E}{8\pi f} \right)^{1/3}.
\end{equation}
We can also make use of the fact that one-point functions of primary operators on  $S^1_\beta \times \mathbb{R}^{d-1}$ are fixed by conformal invariance up to an overall coefficient:
\begin{equation}
\< \CO^{\mu_1 \dots \mu_\ell} \>_\beta = \frac{b_\CO}{\beta^{\Delta_\CO}}\left( e^{\mu_1} \dots e^{\mu_\ell}- \textrm{traces}\right)
\end{equation}
where $e^\mu$ is the unit vector pointing in the time direction, and the constant $b_\CO$ depends on the operator. 

Due to previous numeric work \cite{vasilyev2009universal,krech1996casimir,krech1997casimir,Iliesiu:2018fao,Barrat:2025wbi}, both $f$ and $b_\epsilon$ are known to good precision in the 3d Ising model,
\begin{equation}
f = -0.153, \qquad b_\epsilon \approx 0.7 .
\end{equation}
Thus for large dimension primaries, ETH predicts the diagonal OPE coefficients of $\epsilon$.  Refinements to this prediction have been made using the thermal conformal blocks \cite{Buric:2025uqt}.

\begin{figure}[t!]
\begin{center}
\includegraphics[width=0.6\textwidth]{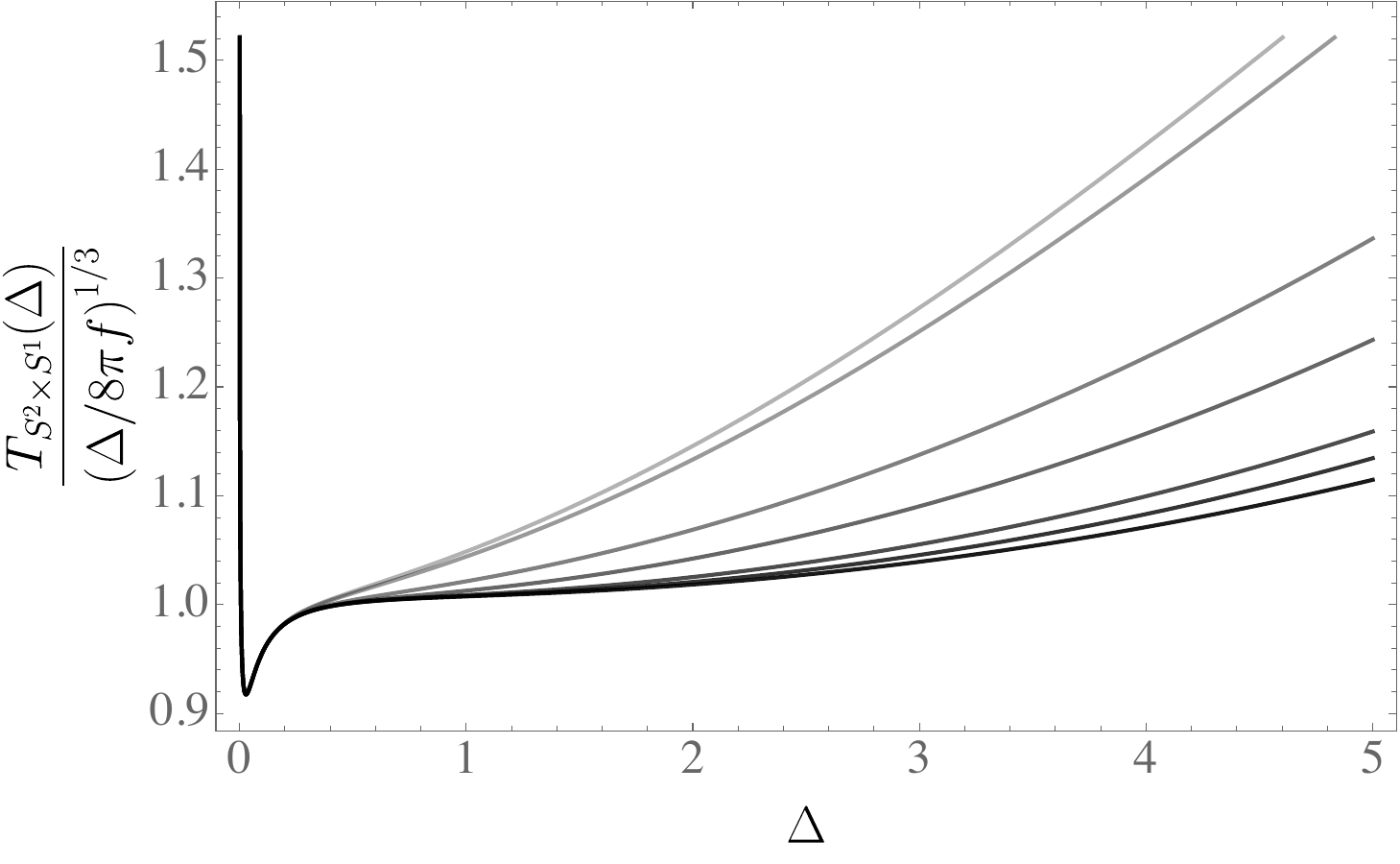}
\caption{Ratio of the thermodynamic temperature $T$ vs energy $\Delta$ relation $T(\Delta)$ on the sphere vs in infinite volume, computed by constructing the partition function $Z$ on the sphere with all known primaries in 3d Ising up to various truncation $\Delta_{\rm max}$.  Lines shown from top to bottom are $\Delta_{\rm max} = 3,4,5,6,7,8,9$, and appear to be converging to 1 to within several decimal places for all $\Delta$ above the unitarity bound.  }
\label{fig:TvsE}
\end{center}
\end{figure}

Since this prediction is an asymptotic statement about large energies, one can try to improve it to take into account finite radius effects in order to improve the accuracy at finite energies on $S^2 \times S^1$.  One improvement we can consider is to use the finite radius thermodynamic temperature-to-energy relation.  This improvement is practical because the relation depends only on the macrocanonical partition function $Z$, and therefore depends only on the spectrum of dimensions in 3d Ising.  In Fig.~\ref{fig:TvsE}, we show the ratio of $T(\Delta)$ on the sphere vs in infinite volume, using the spectrum of known operators in 3d Ising up to various truncations $\Delta_{\rm max} = 3,4,5, \dots, 9$.\footnote{Concretely, we compute $Z(\beta) =  \sum_{\Delta_i < \Delta_{\rm max}} \chi_{\Delta_i, \ell_i}(\beta)$, where $\chi_{\Delta,\ell}$ is the 3d character for primary of dimension $\Delta$ and spin $\ell$, and invert the function $E(\beta) = - \partial_\beta \log Z(\beta)$ to get $T(\Delta) = \beta^{-1}|_{E(\beta)=\Delta}$.}  Remarkably, however,  this ratio appears to converge to 1 to very high precision for all dimensions above the unitarity bound $\Delta \ge 1/2$.  For instance, at $\Delta=0.4$, the curves have converged well and the ratio is $T_{S^2 \times S^1}(0.4)/\left( \frac{0.4}{8 \pi f} \right)^{\frac{1}{3}} = 0.9985$.\footnote{In fact at $\Delta=0.4$, one reaches this value of $T(\Delta)$ including only the primaries $\{1, \sigma, \epsilon, T^{\mu\nu}, \epsilon', \sigma_{\mu \nu}, \sigma_{\mu\nu\rho}\}$ with $\Delta_i = \{0,0.5181,1.4126, 3, 3.823, 4.18, 4.64\}$ in $Z(\beta)$. } 

\begin{figure}[t!]
\begin{center}
\includegraphics[width=0.6\textwidth]{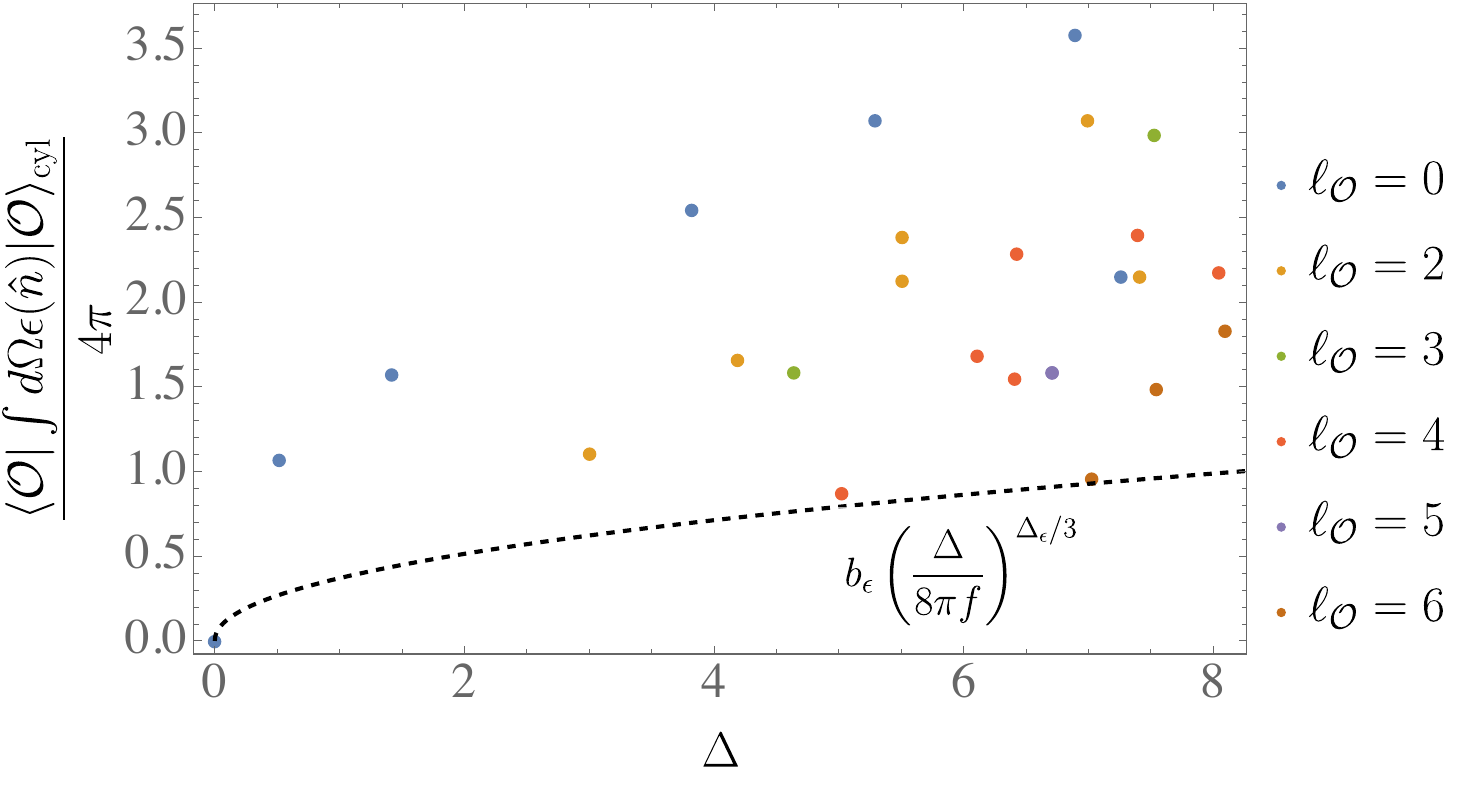}
\caption{Comparison of diagonal matrix elements of $\int d\Omega \epsilon$ with the asymptotic ETH prediction for primaries up to $\Delta \le 8.1$.  Dots indicate individual primaries  The spin of the different external states $|\Delta\>$ is indicated by the color of the dot. The black solid line shows the leading ETH prediction.}
\label{fig:ETHDiag}
\end{center}
\end{figure}

Therefore, ETH makes a clear and sharp prediction for the diagonal matrix elements of $\epsilon$ at large dimension $\Delta$:
\begin{equation}
\< \Delta | \frac{1}{4\pi} \int d \Omega \epsilon(\hat{n}) | \Delta \> = b_\epsilon \left( \frac{\Delta}{8 \pi f} \right)^{\Delta_\epsilon/3}.
\end{equation}
  This prediction has moreover been derived using bootstrap methods, purely using the decomposition of thermal correlators into thermal blocks \cite{Buric:2024kxo,Buric:2025uqt}, where it was shown that perturbative (in $1/\Delta$) corrections to this formula are suppressed by integer powers of $\Delta^{1/3}$ relative to the leading term.  In Fig.~\ref{fig:ETHDiag}, we show the result from our OPE coefficients for primaries up to $\Delta \le 8.1$. Even at the highest energies we can access, the expectation values of $\epsilon$ are significantly higher than the ETH prediction.  It is perhaps suggestive that a similar feature arises in the free theory, where even the averaged (over states with a fixed dimension) expectation values of $\phi^2$ do not begin to have a clear trend towards their asymptotic behavior until fairly large dimensions, around $\Delta \sim 10$ (see \cite{Buric:2025uqt} Fig.~6).\footnote{One can estimate the size of corrections to ETH at large $\Delta$ from the expansion in \cite{Buric:2025uqt} eq. (4.29).  Most of the terms in this expansion are unknown, but even just the known terms $\propto b_{0,0,0} = b_\epsilon$ are suppressed relative to the leading term by $\sim (7/\Delta)^{2/3}$ and therefore not small for our primary states.  Moreover, in GFF theories these subleading terms have large coefficients for $\Delta_{\epsilon}>1$.  We thank Alessandro Vichi for discussions of these points.}

\begin{figure}[t!]
\begin{center}
\includegraphics[width=0.4\textwidth]{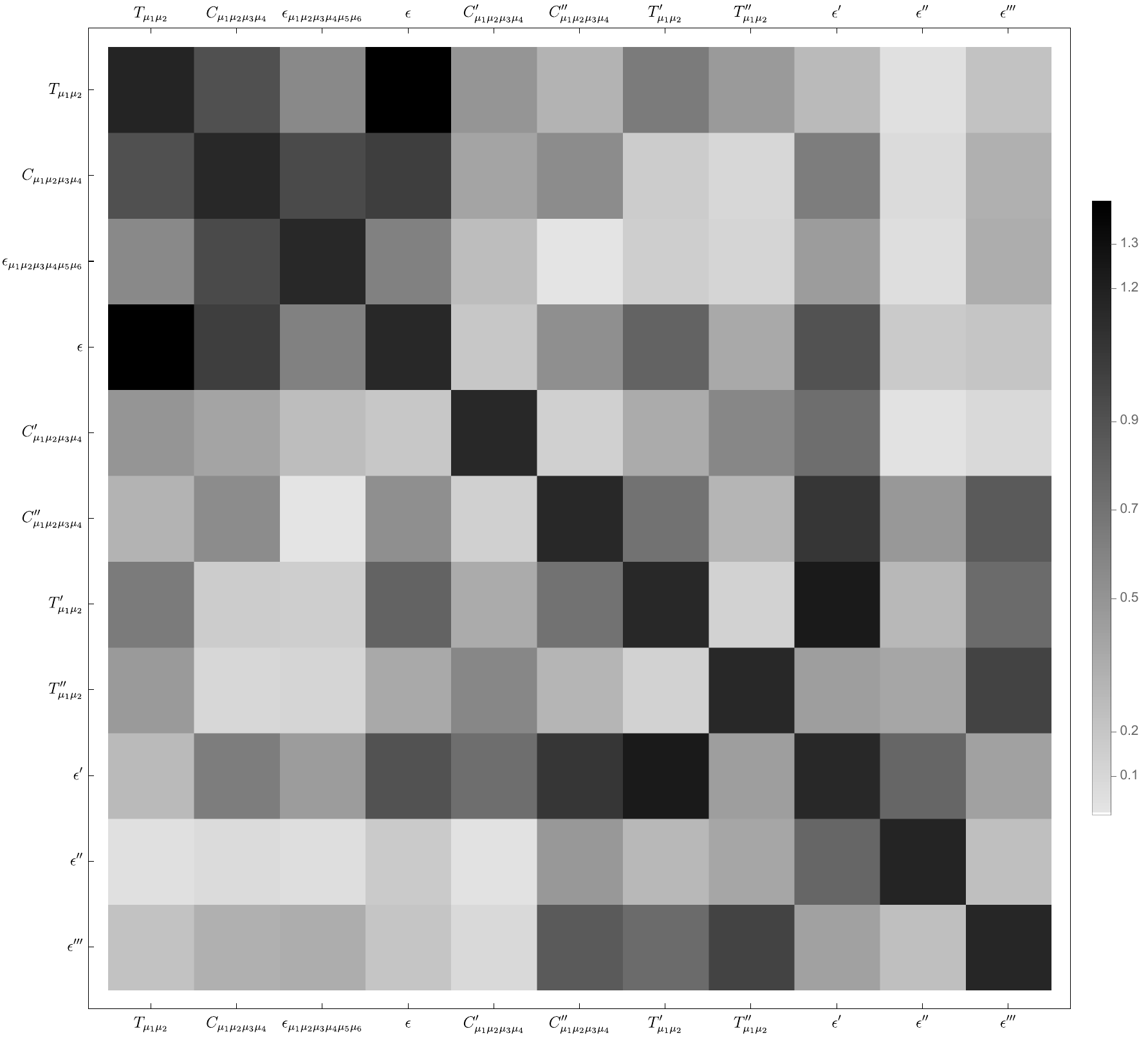}
\includegraphics[width=0.4\textwidth]{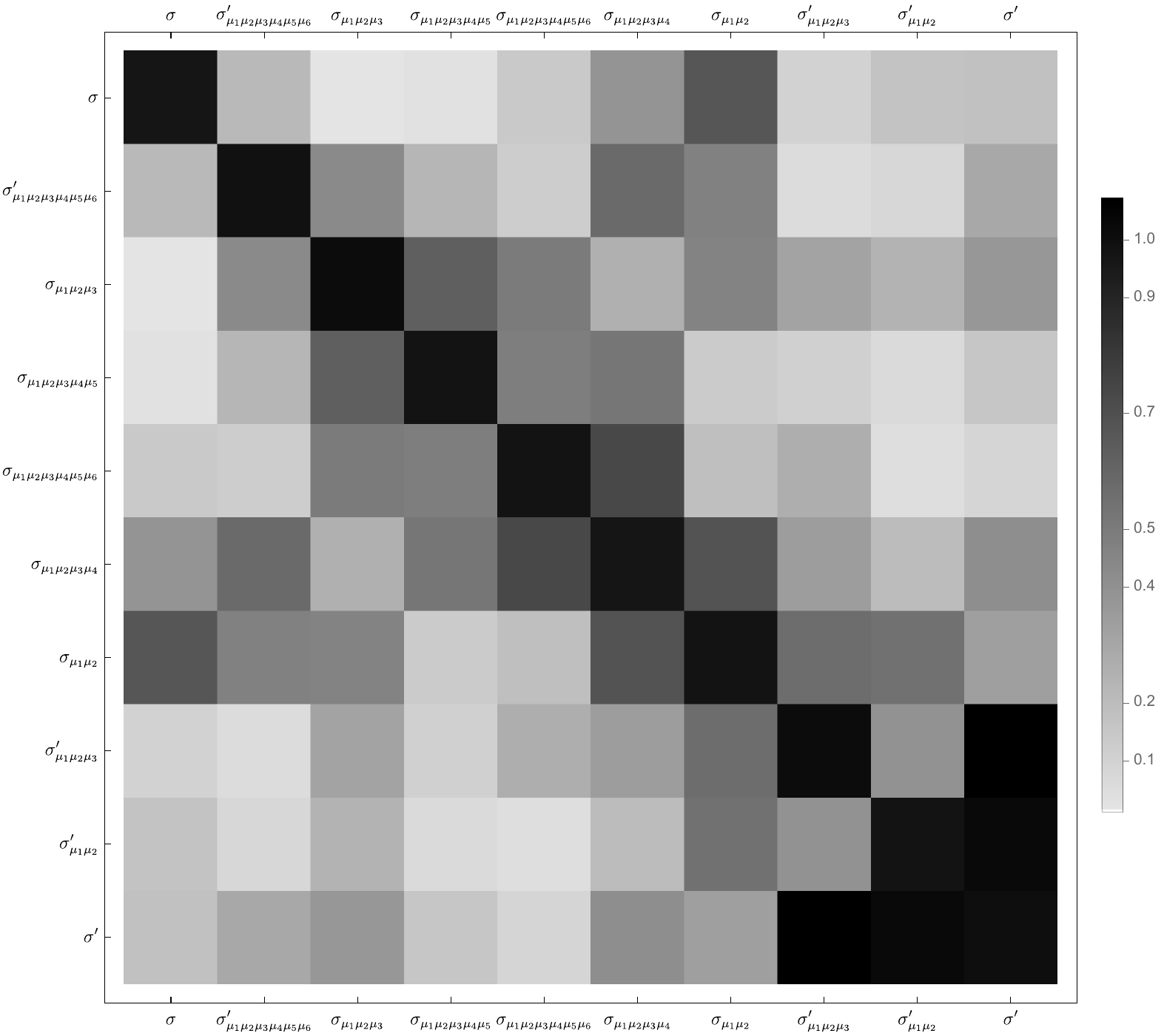}
\caption{Matrix elements of $\int d \Omega \epsilon$ in $\mathbb{Z}_2$ even ({\it left}) and odd ({\it right}) sectors, normalized by the diagonal elements (each square is $\left| \frac{\< \CO_i | \epsilon | \CO_j \>}{\sqrt{\< \CO_i | \epsilon | \CO_i \>\< \CO_j | \epsilon | \CO_j \>}}\right|$, all states are $j_z=0$). Operators are organized by increasing twist (from left to right and top to bottom). }
\label{fig:ETHOffDiag}
\end{center}
\end{figure}

\begin{figure}[t!]
\begin{center}
\includegraphics[width=0.45\textwidth]{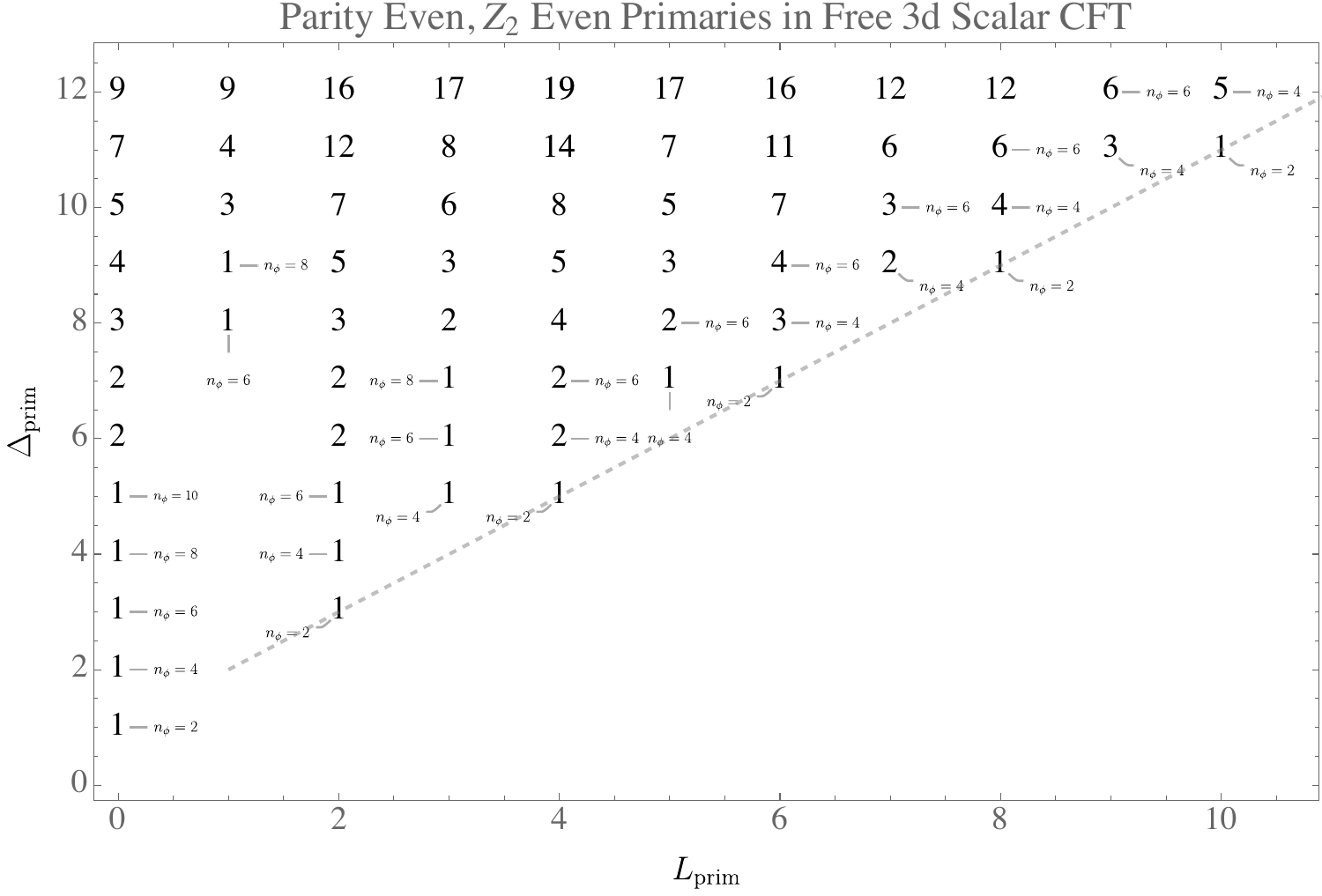}
\includegraphics[width=0.45\textwidth]{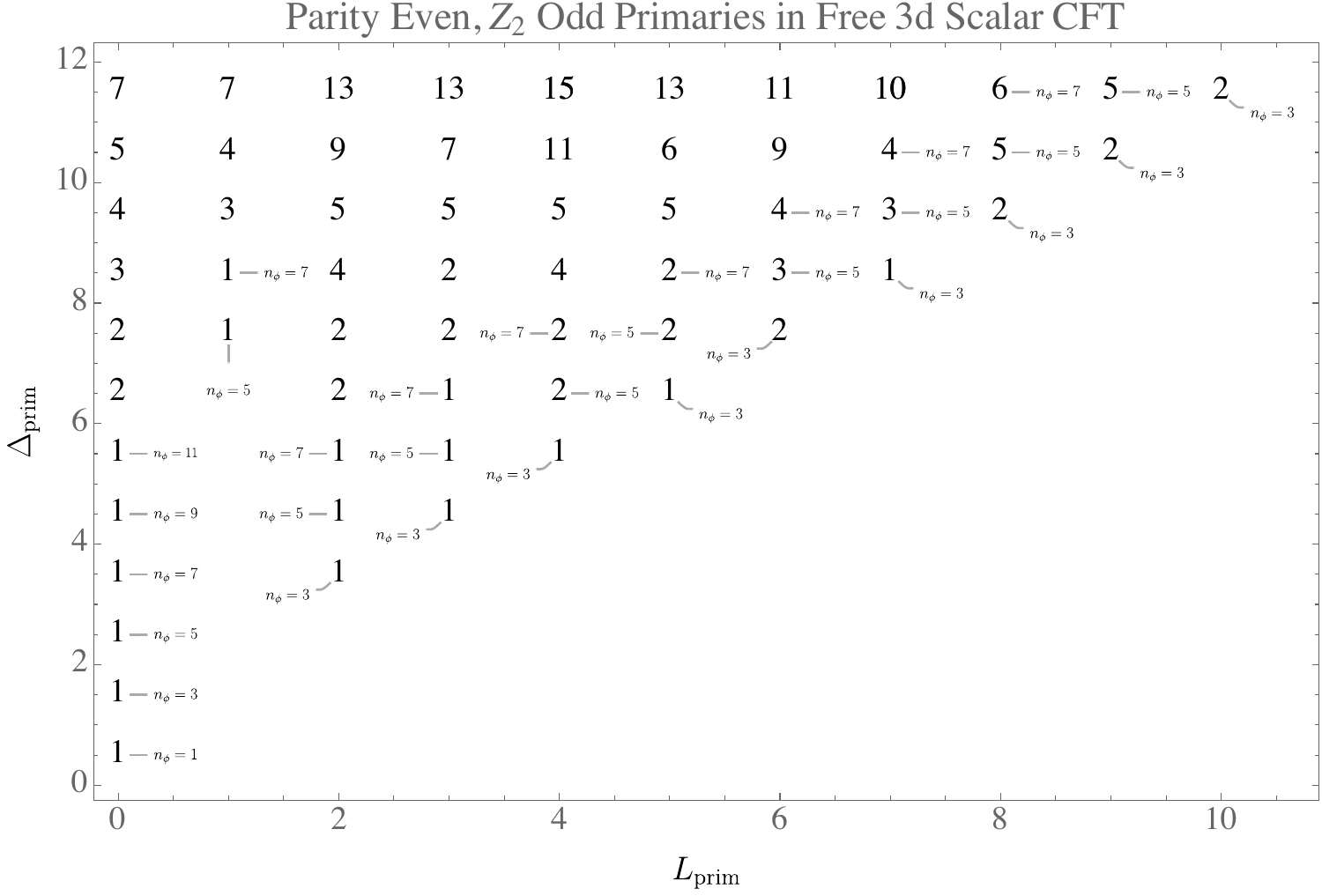}
\includegraphics[width=0.45\textwidth]{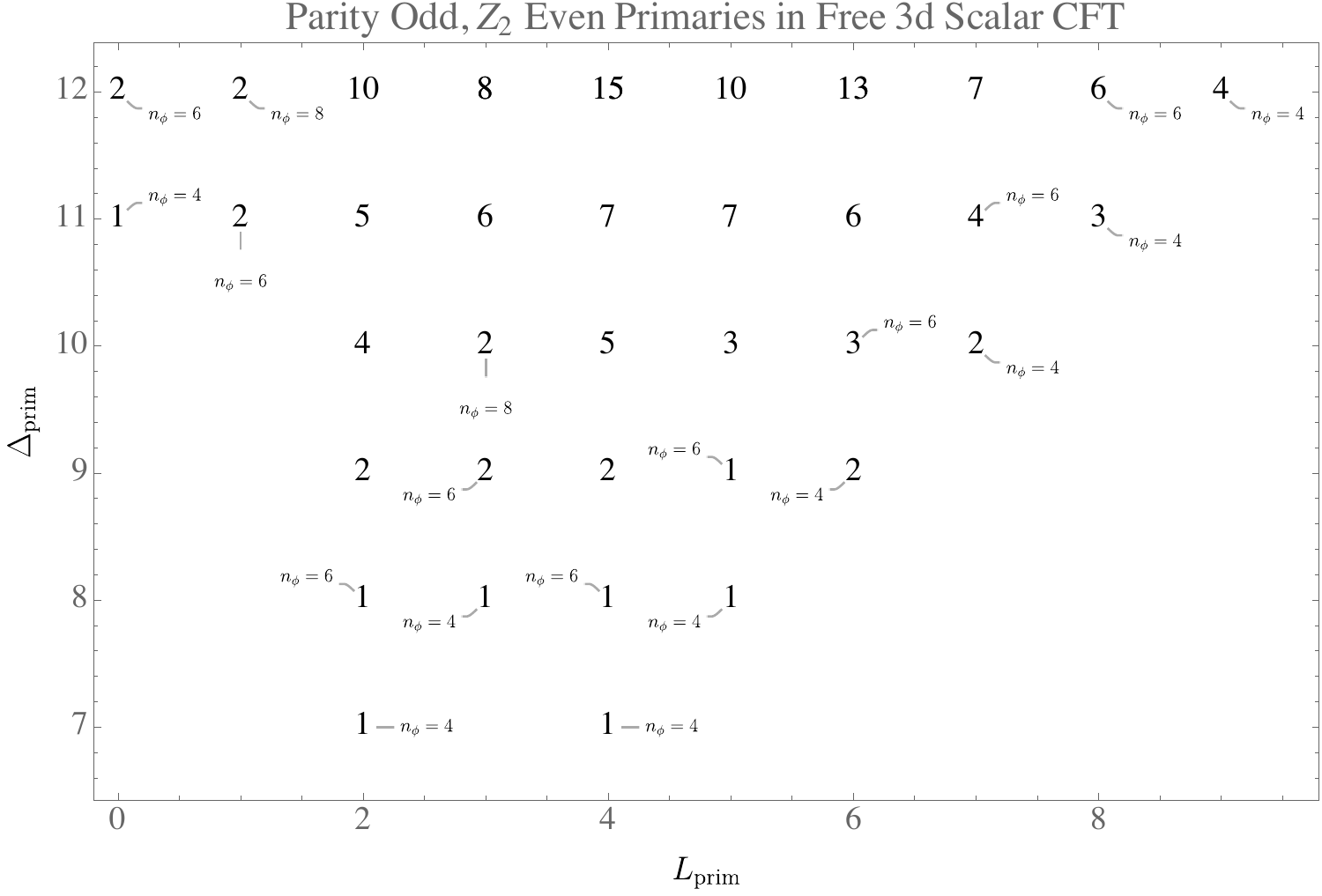}
\includegraphics[width=0.45\textwidth]{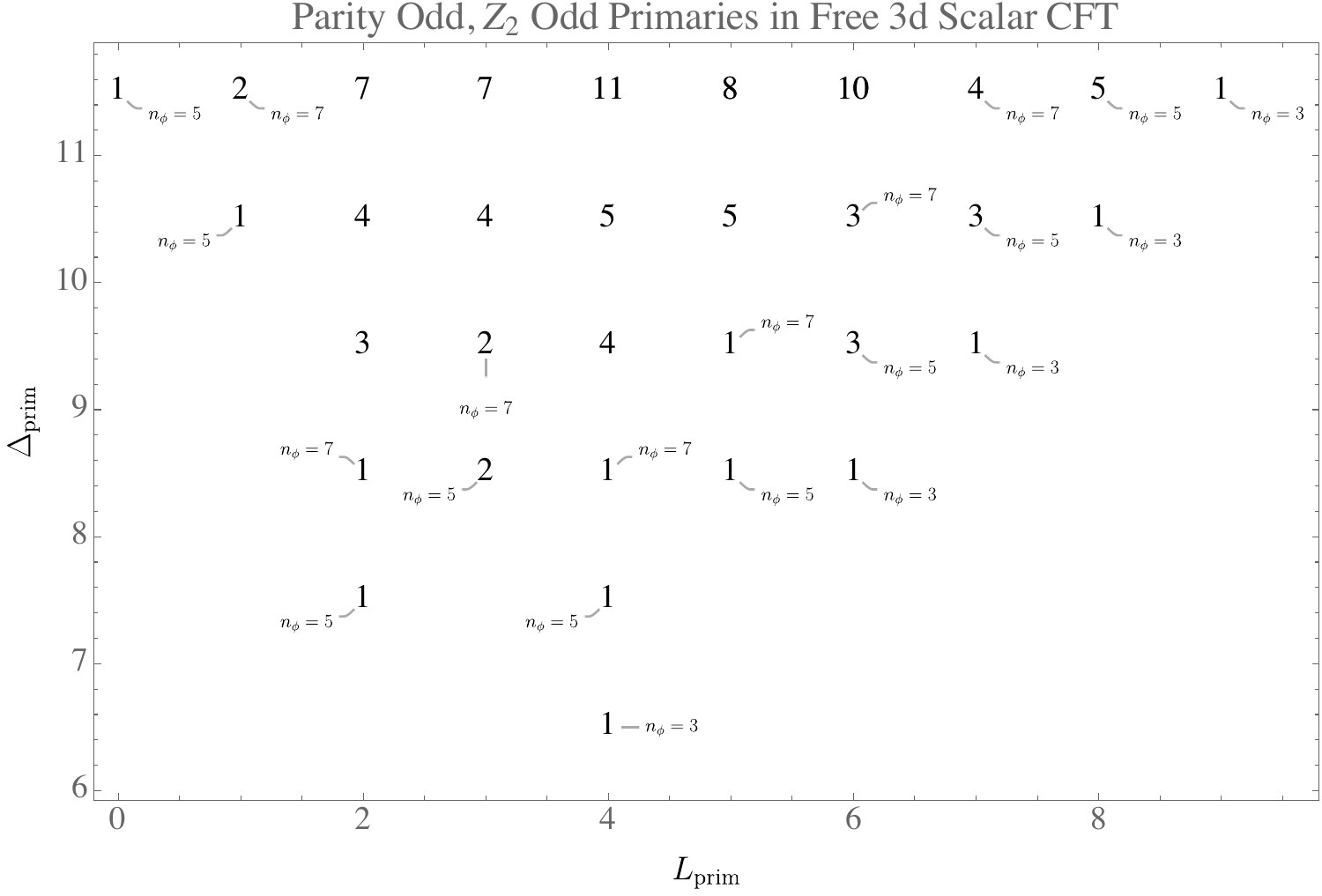}
\caption{Table of low-lying primary operators at a given spin $L$ and dimension $\Delta$ in the free 3d scalar CFT, sorted by parity and $\mathbb{Z}_2$.  The number at each value of $\Delta$ and $L$ is the number of primaries at that $\Delta$ and $L$.  Whenever every primary operator at a given $\Delta,L$, parity and $\mathbb{Z}_2$ is built from the same number $n_\phi$ of $\phi$s, we also indicate $n_\phi$.  The dashed line in the upper-right panel is the unitarity bound. %\noteGF{Giulia add figure!}
}
\label{fig:ParityOdd}
\end{center}
\end{figure}

Next, we consider the off-diagonal entries of $\epsilon$.  In Fig.~\ref{fig:ETHOffDiag}, we show the off-diagonal matrix elements $\< \CO_i | \epsilon | \CO_j\>$ divided by the geometric mean of the diagonal matrix elements $\< \CO_i | \epsilon | \CO_i\>$ and $\< \CO_j | \epsilon | \CO_j\>$.  The ETH prediction (\ref{eq:ETHPrediction}) is that these should be suppressed by the square root of the density of states, i.e. by $e^{-S(\bar{E})/2}$, and the function $f_\CO(\omega, \bar{E})$.    While the function $f_\CO(\omega, \bar{E})$ is unknown, in the hydrodynamic regime it is expected to decay like a power-law with increasing $\omega \equiv |E_i - E_j|$ \cite{Delacretaz:2020nit}. Moreover, by dimensional analysis at small $\omega$ we expect it to roughly cancel in the ratio with the diagonal matrix elements. 
 In contrast with the ETH prediction, the matrix elements close to the diagonal are not always suppressed, though some suppression becomes visible farther from the diagonal.  Overall, we conclude that the leading ETH prediction is not accurate in the energy regime accessible to us. Perhaps these states  still contain some nontrivial structure which could be exploited to calculate their energies and OPE coefficients within an EFT framework, even at the highest-twist primaries we can currently reach.\footnote{The rows and columns in Fig.~\ref{fig:ETHOffDiag} have been organized by twist rather than dimension, because on general grounds we expect many operators with low twist but high dimension to exhibit simple nearly-integrable structure. }

\section{Summary and Future Directions} \label{Sec:summaryandfuture}

The main message of this paper is that using the special conformal generators $K^A$ improves CFT data from the fuzzy sphere by reducing mixing between primaries and descendants.  We have focused on the case of the 3d Ising model, but the prescription should be completely general for any IR CFT fixed point.  The procedure involves the following steps:
\begin{enumerate}
\item Construct the components of $\Lambda \equiv K+P$ by integrating the Hamiltonian density $\CH$ against $\hat{x}^i$ on the sphere (equivalently, against the $\ell=1$ spherical harmonics $Y_{1 , m}$).  At this step, the parameters in $\CH$ should all be free parameters, and will be shifted from their values used in the Hamiltonian $H$ due to total derivative operators in $\CH$ that integrate to zero in $H$ but not in $\Lambda$.  
\item Improve $K+P$ by tuning the parameters in $\int d^2\Omega \hat{x}^i \CH(\hat{x})$ and its overall normalization in order to match a finite number of their matrix elements to the CFT prediction.  Separate out $K$ from $K+P$ by keeping only the matrix elements that lower the energy of the Hamiltonian eigenstates.
\item Construct the matrix $|K|^2 \equiv K_A^\dagger K_A$ and keep the space of states with small $|K|^2$ expectation values.
\item Rediagonalize the Hamiltonian $H$ within this reduced space.
\end{enumerate}
The underlying reason for the improvement with this procedure is that in the CFT limit, primaries all have $|K|^2=0$ whereas descendants have $|K|^2 \ge 1$  (and usually have $|K|^2 \gg 1$), outside of a small number of `almost-primary' descendants that arise when conformal multiplets are close to satisfying shortening conditions.  Consequently, irrelevant UV deformations on the fuzzy sphere do not strongly mix primaries and descendants until one reaches states with dimensions close to the cutoff.  There can still be strong mixing between primaries that are nearly descendant with each other, but the density of primaries grows more slowly than the total density of states and therefore leads to fewer degeneracies.

The procedure outlined here does not remove all UV effects from the fuzzy sphere, and can be improved further.  In particular, when computing OPE coefficients $\sim \< i | O |j\>$ with a fuzzy sphere operator $O$, using $|K|^2$ to improve the states does not help remove higher order irrelevant operators from $O$.  This kind of contamination gets worse the higher the dimension of the CFT operator that one is trying to construct with $O$, as one can already see from the fact that the OPE results are more accurate when using $n_z$ to obtain $\sigma$ than when using $n_x$ to obtain $\epsilon$.  Higher dimension operators, such as $\epsilon'$ and $\sigma'$, will be even more challenging.  It would be useful in such cases to have a cross-check for more of the OPE coefficients from the conformal bootstrap.  Trying to obtain such OPE coefficients from the conformal bootstrap using different sets of mixed correlators, say with $\epsilon'$ or $\sigma'$ as external operators, seems to be a promising direction, but one that we leave for future work.

One of the main motivations we have in mind for these OPE coefficients is an analysis of 3d Ising Field Theory (IFT) using the Truncated Conformal Space Approach (TCSA).  This theory is the vicinity of the 3d Ising CFT, i.e.~deformed by $\sigma$ and/or $\epsilon$.  To study this theory accurately requires as many OPE coefficients of $\sigma$ and $\epsilon$ as possible, as accurately as possible. In this context, knowing many of these OPE coefficients from both the fuzzy sphere and the conformal bootstrap would provide a welcome increase in confidence in this CFT data.

\acknowledgments{We are grateful to Yin-Chen He, Wei Li, Matthew Mitchell, David Poland, Noah Ring, Slava Rychkov, and Alessandro Vichi for helpful discussions,  to Yin-Chen He, Matthew Mitchell, and Alessandro Vichi  for comments on a draft, and in particular to Yuan Xin for helpful feedback throughout the project. Numeric results in this paper used the publicly available Julia code provided at \href{https://www.fuzzified.world/}{https://www.fuzzified.world}.  GF,  ALF, and EK  are supported by the US Department of Energy Office of Science under Award Number DE-SC0015845, and GF was partially supported by the Simons Collaboration on the Non-perturbative Bootstrap. }

\newpage

\appendices

\section{Curvature terms}\label{appendix:CurvatureTerms}

\subsection{General Form}

When we do matching calculations on $S^{2} \times \mathbb{R}$, we have to allow for curvature terms in the low-energy action.  This point has been discussed elsewhere (see e.g.~\cite{Lauchli:2025fii}), but here we will make some comments about the general form of such terms and show an explicit simple example where they are generated by UV modes. 

In the effective theory of the 3d Ising fixed point plus small deformations, in general the action contains all local terms $L_i$ allowed by symmetry:
\begin{equation}
S = S_{\rm CFT} + \int d^3 x \sqrt{g} \sum_i g_i  L_i .
\end{equation}
The local terms $L_i$ are scalars built from local operators in the 3d Ising CFT together with the spacetime metric $g$, for instance
\begin{equation}
\{ L_i \}  \supset \{ \epsilon,  \CR \epsilon,  \CR_{\mu \nu} T^{\mu\nu} , \dots \},
\end{equation}
where $\CR, \CR_{\mu\nu}, \CR_{\mu\nu\rho \sigma}$ are the Ricci scalar, Ricci curvature, and Riemann tensor, respectively.  

When we put the theory on  Euclidean $S^{2} \times \mathbb{R}$, 
\begin{equation}
ds^2 = dt_E^2 + R^2d \hat{x}^2,
\end{equation}
where $d\hat{x}^2$ is the metric of the unit two-sphere, we also map the CFT operators to the sphere.  The most convenient way to treat the two-sphere is using embedding space coordinates.  Since the embedding space is Euclidean $\mathbb{R}^3$, the components of tensors along the two-sphere in embedding space are described in the same way as the tensors in flat space.  All that changes is that there is now also a time component of space, and the time component of tensors are related to the components on $\mathbb{R}^3$ by
\begin{equation}
V_0 = \hat{x}^i V_i.
\end{equation}
The two-sphere is maximally symmetric and thus all curvature components can be reduced to the Ricci scalar and the sphere metric $\gamma_{ij}$. Contracting any $S^2$ indices with $\gamma_{ij}$ is proportional to taking a trace in the $\mathbb{R}^3$ indices and therefore vanishes for symmetric traceless tensors.  Consequently, without loss of generality all local terms $L_i$ can be built from the Ricci scalar $\CR$ and operators in the CFT that are scalars under rotations, which includes the time components of vectors (e.g.~$T^{00}$) on $S^2 \times \mathbb{R}$. 

As discussed in section \ref{sec:review}, we can set $R^2 = N$ without loss of generality, since any other choice for the subleading behavior of $R$ as a function of $N$ can be absorbed into the coefficients of curvature correction terms in the action and other observables.  At fixed $N$, $\mathcal{R}= 2/R^2$ is simply a constant, and therefore terms containing $\mathcal{R}$ are indistinguishable from shifts in the coefficients without factors of $\mathcal{R}$.  In other words, the combination of terms such as
\begin{equation}
g_\epsilon \epsilon + g_{\mathcal{R} \epsilon} \mathcal{R} \epsilon + \dots
\end{equation}
is equivalent to an $N$-dependent coefficient $g_\epsilon$:
\begin{equation}\label{eq:RFuncN}
g_\epsilon(N)  \epsilon , \qquad g_\epsilon(N) = g_\epsilon + g_{\mathcal{R} \epsilon} \mathcal{R} + \dots.
\end{equation}

\subsection{Curvature from Heavy Fields}

The following is a simple example where curvature terms are generated by integrating out a heavy field $\chi$ coupled to a light field $\phi$ through the interaction
\begin{equation}
g \phi \chi^2.
\end{equation}
We will just work to leading order in perturbation theory, where it does not matter that this potential is unstable for $\phi$.  The one-loop diagram with a $\chi$ loop generates a mass term for $\phi$.  Let $\phi$ be a constant in time and space, so the one-loop diagram generates the following term in the action:
\begin{equation}
\delta S = \frac{g^2}{2} \phi_0^2 \int \frac{d\omega}{2\pi} \sum_{\ell} \frac{1}{R^2}\frac{2\ell+1}{(\omega^2 + \ell(\ell+1)/R^2 + M^2)^2},
\end{equation}
where $M$ is the mass of $\chi$ and $R$ is the radius of the sphere.  The $\omega$ integral can be done in closed form:
\begin{equation}
\delta S = \frac{g^2}{2} \phi_0^2 \sum_{\ell} \frac{1}{R^2}\frac{2\ell+1}{4( \ell(\ell+1)/R^2 + M^2)^\frac{3}{2}},
\end{equation}
We can evaluate the $\ell$ sum in a large $M$ expansion using the Euler-Maclaurin formula:
\begin{equation}
\begin{aligned}
&\sum_{\ell=0}^\infty \frac{2\ell+1}{4( \ell(\ell+1)/R^2 + M^2)^\frac{3}{2}} = \int_0^\infty d\ell \frac{2\ell+1}{4( \ell(\ell+1)/R^2 + M^2)^\frac{3}{2}}\\
& \qquad + \frac{1}{2} \frac{1}{4M^3}  - \sum_{k=1}^\infty \frac{B_{2k}}{(2k)!} \left[ \left(\frac{d}{d\ell}\right)^{2k-1} \left( \frac{2\ell+1}{4( \ell(\ell+1)/R^2 + M^2)^\frac{3}{2}} \right)\right]_{\ell\rightarrow \infty} .
\end{aligned}
\end{equation}
The first line can be evaluated exactly
\begin{equation}
\begin{aligned}
 \int_0^\infty d\ell \frac{2\ell+1}{4( \ell(\ell+1)/R^2 + M^2)^\frac{3}{2}} & \stackrel{s \equiv \ell(\ell+1)}{=} \int_0^\infty ds \frac{1}{4 (s+M^2)^{3/2}} 
   = \frac{R^2}{2M},
  \end{aligned}
  \end{equation}
   and the second line can be Taylor expanded term-by-term.  The result is
\begin{equation}
\delta S = \frac{g^2}{2} \phi_0^2 \left( \frac{1}{2M} + \frac{1}{12 M^3 R^2} + \frac{1}{40 M^5 R^4} + \dots \right) .
\end{equation}
The Ricci curvature on $S^2$ is $\CR_{S^2} = \frac{2}{R^2}$, so we can write this in terms of curvature invariants as
\begin{equation}
\delta S = \frac{g^2}{2} \phi_0^2 \left( \frac{1}{2M} + \frac{1}{24 M^3} \CR + \frac{1}{160 M^5}\CR^2 + \dots \right) .
\end{equation}

\subsection{Casimir Energy}

The Casimir energy of a 3d CFT on $S^3$ -- or more precisely, its scheme-independent part --  is an important quantity for characterizing the theory, since it decreases monotonically under RG flows \cite{Casini:2012ei,Jafferis:2011zi,Jafferis:2010un}.  However, the corresponding quantity of a 3d CFT on $S^2 \times \mathbb{R}$ (and more generally, for any odd-dimension $d$ CFT on  $S^{d-1} \times \mathbb{R}$) vanishes by Weyl invariance, which has no anomaly in odd $d$.  Here we will review how to derive this explicitly for a free conformally coupled scalar in $d$ dimensions (see \cite{dowker1978finite} for an early derivation of this result and \cite{Giombi:2014yra} for a more recent one).  

The action for the conformally coupled scalar contains
\begin{equation}
S \supset \frac{1}{2} \int d^d x \sqrt{g} \xi \CR \phi^2,
\end{equation}
where $\xi = \frac{d-2}{4(d-1)}$ is the conformal coupling, and $\CR = \frac{(d-1)(d-2)}{R^2}$ is the curvature of the $(d-1)$-sphere with radius $R$.  The Casimir energy is
\eqna{
E_{\rm Cas} &= \frac{1}{2} \tr \log \left( -\partial_t^2 - \nabla^2 + \frac{(d-2)^2}{4R^2} \right) \\
&= \frac{1}{2} \int_{-\infty}^\infty \frac{d\omega}{2\pi} \sum_{\ell=0}^\infty N_\ell \log \left(\omega^2 + R^{-2} \left( \ell+ \frac{d-2}{2}\right)^2 \right),
}[]
where $N_\ell = (2\ell+d-2) \frac{(\ell+d-3)!}{\ell! (d-2)!}$ is the dimension of a spin $\ell$ representation of $SO(d-1)$.  The integral and sum  in this expression are divergent and must be regularized.  A simple way to regularize the $\omega$ integral is differentiate with respect to $R^{-2}$, perform the $d\omega$ integral, and then integrate with respect to $R^{-2}$.  This procedure gives the result
\begin{equation}
E_{\rm Cas} = \frac{1}{2} \sum_{\ell=0}^\infty N_\ell R^{-1} \left(\ell+\frac{d-2}{2}\right),
\end{equation}
up to an  integration constant which is independent of $R$ (and thus can be removed by a local counterterm).  Next we perform the sum on $\ell$ using a heat kernel regulator:
\begin{equation}
F_d(\alpha) \equiv \frac{1}{2} \sum_{\ell=0}^\infty N_\ell R^{-1} \left(\ell + \frac{d-2}{2}\right) e^{-\alpha R^{-1} (\ell+\frac{d-2}{2})}.
\end{equation}
It is important that the term in the exponent $\alpha \sqrt{-\nabla^2 + \xi \CR}$ is the combination that preserves Weyl transformations of the spatial metric \cite{Monin:2016bwf}.  The sum can be performed in closed form:
\begin{equation}
F_d(\alpha) =\frac{e^{-\frac{\alpha  (d+2)}{2 R}} \left(1-e^{-\frac{\alpha }{R}}\right)^{-d} \left(d
   \left(e^{\frac{\alpha }{R}}+1\right)^2-2 \left(e^{\frac{2 \alpha
   }{R}}+1\right)\right)}{4 R}.
   \end{equation}
   Expanding at small $\alpha$, we obtain:
   \begin{equation}
   \begin{aligned}
   F_2(\alpha) & = \frac{R}{\alpha^2} - \frac{1}{12 R} + O(\alpha) , \\
   F_3(\alpha) & =  \frac{2R^2 }{\alpha^3}  + O(\alpha), \\
   F_4(\alpha) & = \frac{3R^3}{\alpha^4} + \frac{1}{240 R} + O(\alpha) , \\
   F_5(\alpha) &= \frac{4R^4}{\alpha^5} - \frac{R^2}{12 \alpha^3} + O(\alpha).
   \end{aligned}
   \end{equation}
The regulated Casimir energy is the coefficient of $\alpha^0$, or equivalently the $R^{-1}$ term.  In $d=2$, it is $-\frac{1}{12 R}$, as expected for a single free scalar.  In odd $d$, it vanishes as claimed.  In even $d>2$, the $R^{-1}$ term is scheme-dependent since it may be removed by adding a local counterterm $\sim \CR^{d/2}$ to the action.

\section{Counting Free Theory States}
\label{app:Counting}

An efficient way to count states in each symmetry sector in the free 3d scalar theory is to construct the partition function of the theory and decompose it into characters.  The single-particle states correspond to the spherical harmonics $Y_{\ell , m}$ on $S^2$, which have dimension $\Delta(\ell) = \frac{1}{2} + \ell$.  We can therefore write the full partition function as 
\begin{equation}
Z(q,y,z,\omega) = \tr(q^D y^{J_z} z^{N_\phi} \omega^P) =  \prod_{\ell=0}^\infty \prod_{m=-\ell}^\ell \frac{1}{1- q^{\frac{1}{2} + \ell} y^m z \omega^\ell},
\end{equation}
where $q = e^{-\beta}$ keeps track of the dimension, $y$ keeps track of $J_z$, $z$ keeps track of the particle number, and $\omega$  keeps track of parity ($\omega^2 \equiv 1$).

We decompose $Z$ into characters:
\begin{equation}
Z(q,y,z,\omega) = \sum_{\Delta,\ell_0, n_\phi, p} n_{\Delta, \ell_0, n_\phi, p} \chi_{\Delta, \ell_0, n_\phi, p}(q, y, z, \omega),
\end{equation}
with $p=0,1$ for even and odd parity, respectively.
The character for a generic primary of spin $\ell_0$, dimension $\Delta> \ell_0+1$, parity $p$, and particle number $n_\phi$ is
\begin{equation}
\chi_{\Delta, \ell_0, n_\phi, p}(q, y, z, \omega) = q^\Delta z^{n_\phi} \omega^{\ell_0+p} \frac{\frac{y^{\ell_0} - 1/y^{\ell_0 + 1}}{1 - 1/y}}{(1-q \omega )(1-q \omega/y)(1-q \omega y)} .
\end{equation}
In the special case where $\Delta = \ell_0 +1 \ge 2$,
\begin{equation}
\chi_{ \ell_0+1, \ell_0, n_\phi, p}(q, y, z, \omega) = q^{\ell_0+1} z^{n_\phi} \omega^{\ell_0+p} \frac{\frac{y^{\ell_0} - 1/y^{\ell_0 + 1}}{1 - 1/y} - q \omega \frac{y^{\ell_0-1} - 1/y^{\ell_0 }}{1 - 1/y}}{(1-q \omega )(1-q \omega/y)(1-q \omega y)},
\end{equation}
and when $\Delta = 1/2, \ell_0=0$, 
\begin{equation}
\chi_{\frac{1}{2}, 0, n_\phi, p}(q, y, z, \omega) = q^\frac{1}{2} z^{n_\phi} \omega^{p} \frac{1-q^2}{(1-q \omega )(1-q \omega/y)(1-q \omega y)}.
\end{equation}

\section{Explicit Relations between CFT Basis and Fuzzy Sphere Harmonics}\label{app:CFTtoFS}

Here we review some of the basic formalism for efficiently connecting the matrix elements of operators evaluated  in radial quantization on the one hand, and a standard basis of structures of CFT three-point functions of spinning operators on the other hand, following the formalism of  \cite{Costa:2011mg}. We begin with the three-point functions in embedding space in the index-free formalism.  In this paper we are mainly interested in OPE coefficients where at least one of the operators $\CO_2$ is a scalar, in which case the correlator can be written as
\begin{equation}
\langle \CO_1(P_1, Z_1) \CO_2(P_2) \CO_3(P_3, Z_3)\rangle,
\end{equation}
where $P_i$ are embedding space positions and $Z_i$ keep track of the spin indices.  The standard basis of structures is built from the invariants $H_{13}, V_{1,32}$ and $V_{3,21}$, where
\eqna{
H_{ij} &\equiv -2[ (Z_i \cdot Z_j)(P_i \cdot P_j) - (Z_i \cdot P_j)(Z_j \cdot P_i)]\, , \\
 V_{i,jk} & \equiv \frac{(Z_i \cdot P_j)(P_i \cdot P_k) - (Z_i\cdot P_k)(P_i \cdot P_j)}{(P_j \cdot P_k)},
}[]
and the three-point function OPE coefficients $\lambda_{\CO_1 \CO_2 \CO_3}^{(n)}$ are defined as 
\begin{equation}
\langle \CO_1(P_1, Z_1) \CO_2(P_2) \CO_3(P_3, Z_3)\rangle = \frac{\sum_{a=0}^{{\rm min}(\ell_1, \ell_3)} \lambda_{\CO_1 \CO_2 \CO_3}^{(a)} H_{13}^a V_{1,32}^{\ell_1 - a} V_{3,21}^{\ell_3 - a}}{(2P_1 \cdot P_2)^{\frac{\bar{h}_1 + \bar{h}_2-\bar{h}_3}{2}} (2P_1 \cdot P_3)^{\frac{\bar{h}_1 + \bar{h}_3-\bar{h}_2}{2}} (2P_2 \cdot P_3)^{\frac{\bar{h}_2 + \bar{h}_3-\bar{h}_1}{2}}}, \label{eq:TensorBasis}
\end{equation}
where $\bar{h}_i \equiv \Delta_i + \ell_i$.  When the correlator is parity-odd, the following parity-odd invariant can appear:
\begin{equation}
\epsilon_{13,2} \equiv \epsilon( Z_1, Z_3, P_1, P_2, P_3) \left( \frac{2 P_1 \cdot P_3}{2 P_1 \cdot P_2 2 P_2 \cdot P_3} \right)^{1/2},
\end{equation}
and we define the parity-odd three-point function OPE coefficients $\lambda^{-(a)}_{\CO_1 \CO_2 \CO_3}$ as
\begin{equation}
\langle \CO_1(P_1, Z_1) \CO_2(P_2) \CO_3(P_3, Z_3)\rangle = \frac{\sum_{a=1}^{{\rm min}(\ell_1, \ell_3)}  \lambda_{\CO_1 \CO_2 \CO_3}^{-(a)} \epsilon_{13,2}  H_{13}^{a-1} V_{1,32}^{\ell_1 - a} V_{3,21}^{\ell_3 - a}}{(2P_1 \cdot P_2)^{\frac{\bar{h}_1 + \bar{h}_2-\bar{h}_3}{2}} (2P_1 \cdot P_3)^{\frac{\bar{h}_1 + \bar{h}_3-\bar{h}_2}{2}} (2P_2 \cdot P_3)^{\frac{\bar{h}_2 + \bar{h}_3-\bar{h}_1}{2}}}.
\end{equation}
 To connect the three-point function to the matrix elements of $\CO_2$ in radial quantization, we expand the point $P_3$ around the origin and the point $P_1$ around conformal infinity.   In some cases one might be interested in the matrix elements of $\CO_2$ between descendant states in radial quantization. For descendants of the ket state $|\CO_3\rangle$, taking descendants is equivalent to taking derivatives with respect to $P_3$ and therefore descendants at a fixed level naturally come out of a Taylor expansion.  By contrast, taking descendants of the bra state $\langle \CO_1|$ requires acting on $P_1$ with special conformal transformations. A significant advantage of embedding space is that it is trivial to perform a conformal inversion on the point $P_1$, so that descendants again emerge from a simple Taylor expansion.  With an eye toward such an application, we will take $P_1$ to be a conformal inversion of the real space point $x_1$ (and similarly $Z_1$ to be a conformal inversion of the embedding space polarization vector associated with the real space point $x_1$ and polarization vector $u_1^A$.  We also take the point $P_2$ where the scalar $\CO_2$ is inserted to lie on the unit sphere, $\hat{x}_2^2=1$.  Then, the embedding space coordinates in terms of the real space coordinates are 
\begin{equation}
\begin{aligned}
Z_1 &= (2 u_1 \cdot x_1, 0 ,u_1^A)  , \\
Z_3 &= (0 , 2 u_3 \cdot x_3, u_3^A) , \\
P_1 &= (x_1^2, 1 , x_1^A) , \\
P_3 & = (1, x_3^2, x_3^A) , \\
P_2 & = (1,1,\hat{x}_2^A) .
\end{aligned}
\end{equation}
  Now it is straightforward to evaluate all the inner products:
\begin{equation}
\begin{aligned}
Z_1 \cdot Z_3 &= 2 (u_1 \cdot x_1)(u_3 \cdot x_3) - u_1 \cdot u_3 , \\
P_1 \cdot P_3 & = \frac{1}{2} (1 - 2 x_1 \cdot x_3 + x_1^2 x_3^2) , \\
Z_1 \cdot P_3 & = u_1 \cdot x_1 x_3^2 - u_1 \cdot x_3 , \\
Z_3 \cdot P_1 & = u_3 \cdot x_3 x_1^2 - u_3 \cdot x_1 , \\
Z_1 \cdot P_2 & = u_1 \cdot x_1 - u_1 \cdot \hat{x}_2 , \\
Z_3 \cdot P_2 & = u_3 \cdot x_3 - u_3 \cdot \hat{x}_2 , \\
P_1 \cdot P_2 & = \frac{1}{2}(1+x_1^2 -2 x_1 \cdot \hat{x}_2 ) , \\
P_2 \cdot P_3 & = \frac{1}{2}(1+x_3^2 -2 x_3 \cdot \hat{x}_2 ) , \\
\epsilon(Z_1, Z_3, P_1, P_2, P_3) &= \det \left( \begin{array}{ccccc}  2 u_1 \cdot x_1 & 0 & x_1^2 &  1 & 1 \\
0 & 2 u_3 \cdot x_3 & 1 & x_3^2 & 1 \\
u_1 & u_3 & x_1 & x_3 & \hat{x}_2 \end{array} \right).
\end{aligned}
\end{equation}
For descendants of $\CO_1, \CO_3$  at level $n$ we series expand $x_1,x_3$, respectively, to the power $n$.  If we just want to relate the conformal structures to the spherical harmonic projections of the operator $\CO_2(\hat{x}_2)$ evaluated between primaries, then $n=0$ and we set $x_1 = x_3 = 0$.  In that case, the inner products are simply
\begin{equation}
\begin{aligned}
&Z_1 \cdot Z_3 \sim - u_1 \cdot u_3, \qquad  2 P_i \cdot P_j \sim 1, \qquad Z_1 \cdot P_3 \sim Z_3 \cdot P_1 \sim 0, \\
& Z_i \cdot P_2 \sim -u_i \cdot \hat{x}_2 , \quad \quad \epsilon(Z_1, Z_2, P_1, P_2, P_3)  \sim \det ( u_1 \  u_3 \  \hat{x}_2) = \hat{x}_2 \cdot (u_3 \times u_1).
\end{aligned}
\end{equation}
where `$\sim$' means at $x_1 = x_3 =0$.  

 Finally, the traceless symmetric projection of $\CO_i^{\mu_1 \dots \mu_\ell}$ is extracted by acting with the Todorov operator
\begin{equation}
D_i^A \equiv \left(\frac{d-2}{2} + u_i \cdot \frac{\partial}{\partial u_i}\right) \frac{\partial}{\partial u^A_i} - \frac{1}{2} u_i^A \frac{\partial^2}{\partial u_i \cdot \partial u_i},
\end{equation}
so that
\begin{equation}
\CO_i^{\mu_1 \dots \mu_\ell}(x_i) = \frac{1}{\ell!(\frac{d}{2}-1)_\ell} D_i^{\mu_1} \dots D_i^{\mu_\ell} \CO_i(x_i, u_i).
\end{equation}
The $j_z=0$ components are easily picked out by contracting free indices with $\hat{z} = (0,0,1)$. 
Consider for instance the following case. Say  we want the contribution to the matrix element of $\CO_2$ between the $j_z=0$ components of two spin-two primary states, from the structure $a=0$:
\begin{equation}
\begin{aligned}
\langle \CO_1(P_1, Z_1) \CO_2(P_2) \CO_3(P_3, Z_3)\rangle &= \lambda_{\CO_1 \CO_2 \CO_3}^{(0)} V_{1,32}^2 V_{3,21}^2 \\
 & = (u_1 \cdot \hat{x}_2)^2 (u_3 \cdot \hat{x}_2)^2 + (\textrm{higher orders in } x_1, x_3).
\end{aligned}
\end{equation}
Applying the Todorov operator and contracting free indices with $\hat{z}$, we obtain (now taking $d=3$)
\begin{equation}
\begin{aligned}
\langle \CO_1, j_z=0 | \CO_2(\hat{x}_2) | \CO_3, j_z =0\rangle & = \frac{1}{N_1 N_3} \left( \cos^2\theta - 1/3\right)^2,
\end{aligned}
\end{equation}
where we have chosen coordinates on the sphere so $\hat{x}_2\cdot \hat{z} = \cos \theta$, and the normalizations $N_1, N_3$ are chosen so that the states
\begin{equation}
| \CO_i, j_z=0\rangle = \frac{1}{N_i} \hat{z}_{\mu_1} \dots \hat{z}_{\mu_\ell} \CO_i^{\mu_1 \dots \mu_\ell}(0) | {\rm vac} \rangle
\end{equation}
is unit normalized; concretely, if the two-point function of $\CO_i$ is normalized to be
\begin{equation}
\langle \CO_i(P_1, Z_1) \CO_i(P_2, Z_2)\rangle = c_{\CO_i} \frac{H_{12}^{\ell_i}}{(2 P_1 \cdot P_2)^{\bar{h}_i}}
\end{equation}
then
\begin{equation}
N_i^2 = c_{\CO_i} 2^{\ell_i} \frac{(\ell_i!)^2}{(2\ell_i)!}.
\end{equation}
In this convention, the  scalar-scalar-(spin $\ell$) OPE coefficient is related to the  matrix element as follows:\footnote{Notice that in the convention of~\eqref{eq:TensorBasis}, if the operator $\CO_1$ is a scalar and $\CO_3$ has spin $\ell$, then $\lambda^{(0)}_{\CO_1\CO_2\CO_3}=(-1)^{\ell} \lambda^{(0)}_{\CO_3\CO_2\CO_1}$. } 
\eqna{
\sqrt{4\pi } \langle \CO_1 |  \int d^2\Omega  \lsp Y_{\ell,0 }(\Omega)\CO_2(\Omega)|\CO_3^{\mu_1 \cdots \mu_\ell}, j_z=0\rangle= (-1)^{\ell} \frac{2^{\frac{\ell}{2}}\Gamma(\ell+1)}{\sqrt{\Gamma(2\ell+2)}}\lambda^{(0)}_{\CO_1\CO_2\CO_3}\, ,
}[]
Note that this convention (used also in eg.~\cite{Poland:2023vpn, Poland:2023bny,Poland:2025ide}) differs from the one used in \cite{Simmons-Duffin:2016wlq}, related by $\lambda^{(0)} ({\rm there})=2^{-\frac{\ell}{2}}\lambda^{(0)}({\rm here})$.
Given the OPE coefficient for the spinning primary, it is possible to obtain the OPE coefficient for any of its scalar descendants of the schematic form $\square^n \partial_{\mu_1}\cdots \partial_{\mu_\ell}\CO_3^{\mu_1\cdots \mu_\ell}$ 
\eqna{
&\frac{\lambda^{(0)}_{\CO_1\CO_2\square^n \partial^\ell \CO_3}}{\lambda^{(0)}_{\CO_1\CO_2\CO_3} }= \frac{2^{2n+\frac{3}{2}\ell}}{\sqrt{n!}}{\left(\frac{ \Delta_1+\Delta_2-\Delta_3-\ell -1}{2}\right)_{n+\ell }
   \left(\frac{\Delta_1+\Delta_2-\Delta_3+\ell }{2} \right)_n}\times\\
   &\quad\, \sqrt{\frac{2 \Gamma (\ell +1)^2 \Gamma (n+\ell +2)}{\Gamma (2 \ell +1) \Gamma (2 n+2 \ell +3)}}\sqrt{\left| \frac{\Gamma (2 \Delta_1-1) \Gamma (n+\Delta_1) \Gamma (-\ell +\Delta_1-1)
   \Gamma (\ell +\Delta_1)}{\Gamma (\Delta_1-1) \Gamma (\Delta_1) \Gamma (2
   n+2 \Delta_1-1) \Gamma (n+\ell +\Delta_1)}\right|}\,.
}[]
For a three-point function of two spinning operators and a scalar, let us define
\eqna{
\CY_{\ell}^{(\ell_1, \ell_3)}=\sqrt{4\pi } \langle \CO_1^{\nu_1\cdots \nu_{\ell_1}} |  \int d^2\Omega  \lsp Y_{\ell,0 }(\Omega)\CO_2(\Omega)|\CO_3^{\mu_1 \cdots \mu_{\ell_3}}\rangle\, ,
}[]
where the only non vanishing cases are for $\ell=|\ell_3-\ell_1|, |\ell_3-\ell_1|+2, \cdots, \ell_3+\ell_1$.  Then the change of matrix from these basis to the usual CFT invariants defined in~\eqref{eq:TensorBasis} is
\eqna{
\CY_{\ell}^{(\ell_1, \ell_3)}=(-1)^{\ell_3}2^{\frac{\ell_1+\ell_3}{2}}M^{(\ell_1, \ell_3)}_{\ell, i}\lambda_{\CO_1\CO_2\CO_3}^{(i)}\, .
}[]
Some examples of these matrices are
\begingroup
\allowdisplaybreaks
\begin{align*}
M^{(2,2)}&=
\left(
\begin{array}{ccc}
 \frac{1}{30} & -\frac{1}{12} & \frac{1}{4} \\
 \frac{1}{21 \sqrt{5}} & -\frac{1}{12 \sqrt{5}} & 0 \\
 \frac{1}{35} & 0 & 0 \\
\end{array}
\right)\, ,  &&M^{(2,3)}=\left(
\begin{array}{ccc}
 \frac{3}{70 \sqrt{10}} & -\frac{1}{10 \sqrt{10}} & \frac{1}{4 \sqrt{10}} \\
 \frac{\sqrt{\frac{2}{105}}}{15} & -\frac{1}{5 \sqrt{210}} & 0 \\
 \frac{\sqrt{\frac{5}{66}}}{21} & 0 & 0 \\
\end{array}
\right)\, , \\
M^{(2,4)}&=\left(
\begin{array}{ccc}
 \frac{1}{35 \sqrt{21}} & -\frac{3 \sqrt{\frac{3}{7}}}{140} & \frac{\sqrt{\frac{3}{7}}}{20} \\
 \frac{2 \sqrt{\frac{5}{21}}}{231} & -\frac{\sqrt{\frac{5}{21}}}{84} & 0 \\
 \frac{\sqrt{\frac{5}{273}}}{22} & 0 & 0 \\
\end{array}
\right)\, ,  &&M^{(2,5)}=\left(
\begin{array}{ccc}
 \frac{5}{693 \sqrt{6}} & -\frac{1}{63 \sqrt{6}} & \frac{1}{28 \sqrt{6}} \\
 \frac{5}{117 \sqrt{462}} & -\frac{1}{18 \sqrt{462}} & 0 \\
 \frac{\sqrt{\frac{7}{10}}}{286} & 0 & 0 \\
\end{array}
\right)\, ,\\
M^{(2,6)}&= \left(
\begin{array}{ccc}
 \frac{5}{286 \sqrt{154}} & -\frac{5}{132 \sqrt{154}} & \frac{1}{12 \sqrt{154}} \\
 \frac{\sqrt{\frac{7}{286}}}{165} & -\frac{\sqrt{\frac{7}{286}}}{132} & 0 \\
 \frac{\sqrt{\frac{14}{187}}}{195} & 0 & 0 \\
\end{array}
\right)\, ,  &&M^{(3,3)}=\left(
\begin{array}{cccc}
 \frac{1}{140} & -\frac{1}{60} & \frac{1}{24} & -\frac{1}{8} \\
 \frac{1}{105 \sqrt{5}} & -\frac{2}{105 \sqrt{5}} & \frac{1}{30 \sqrt{5}} & 0 \\
 \frac{3}{770} & -\frac{1}{210} & 0 & 0 \\
 \frac{5}{231 \sqrt{13}} & 0 & 0 & 0 \\
\end{array}
\right)\, , \\
M^{(3,4)}&=\left(
\begin{array}{cccc}
 \frac{\sqrt{\frac{2}{21}}}{105} & -\frac{\sqrt{\frac{3}{14}}}{70} & \frac{1}{10 \sqrt{42}} & -\frac{1}{4 \sqrt{42}} \\
 \frac{1}{385 \sqrt{2}} & -\frac{1}{210 \sqrt{2}} & \frac{1}{140 \sqrt{2}} & 0 \\
 \frac{\sqrt{\frac{2}{77}}}{91} & -\frac{1}{42 \sqrt{154}} & 0 & 0 \\
 \frac{\sqrt{\frac{35}{6}}}{858} & 0 & 0 & 0 \\
\end{array}
\right) \, ,  &&M^{(3,5)}=\left(
\begin{array}{cccc}
 \frac{5}{1386 \sqrt{7}} & -\frac{1}{126 \sqrt{7}} & \frac{1}{56 \sqrt{7}} & -\frac{1}{24 \sqrt{7}} \\
 \frac{\sqrt{\frac{5}{7}}}{1001} & -\frac{2}{231 \sqrt{35}} & \frac{1}{84 \sqrt{35}} & 0 \\
 \frac{\sqrt{\frac{7}{65}}}{396} & -\frac{\sqrt{\frac{7}{65}}}{396} & 0 & 0 \\
 \frac{2 \sqrt{\frac{7}{85}}}{429} & 0 & 0 & 0 \\
\end{array}
\right)\, , \\
M^{(3,6)}&=\left(
\begin{array}{cccc}
 \frac{5 \sqrt{\frac{5}{33}}}{3003} & -\frac{5 \sqrt{\frac{5}{33}}}{1386} & \frac{\sqrt{\frac{5}{33}}}{126} & -\frac{\sqrt{\frac{5}{33}}}{56} \\
 \frac{\sqrt{\frac{7}{15}}}{1716} & -\frac{\sqrt{\frac{35}{3}}}{5148} & \frac{\sqrt{\frac{7}{15}}}{792} & 0 \\
 \frac{6 \sqrt{\frac{7}{11}}}{12155} & -\frac{\sqrt{\frac{7}{11}}}{2145} & 0 & 0 \\
 \frac{\sqrt{\frac{21}{1045}}}{221} & 0 & 0 & 0 \\
\end{array}
\right) \, ,  &&M^{(4,4)}=\left(
\begin{array}{ccccc}
 \frac{1}{630} & -\frac{1}{280} & \frac{1}{120} & -\frac{1}{48} & \frac{1}{16} \\
 \frac{2 \sqrt{5}}{4851} & -\frac{\sqrt{5}}{1176} & \frac{\sqrt{5}}{588} & -\frac{\sqrt{5}}{336} & 0 \\
 \frac{27}{35035} & -\frac{27}{21560} & \frac{3}{1960} & 0 & 0 \\
 \frac{2}{693 \sqrt{13}} & -\frac{5}{1848 \sqrt{13}} & 0 & 0 & 0 \\
 \frac{7}{1287 \sqrt{17}} & 0 & 0 & 0 & 0 \\
\end{array}
\right)\, , 
\end{align*}
\begin{align*}
M^{(4,5)}&=\left(
\begin{array}{ccccc}
 \frac{\sqrt{\frac{5}{6}}}{1386} & -\frac{1}{126 \sqrt{30}} & \frac{1}{56 \sqrt{30}} & -\frac{1}{24 \sqrt{30}} & \frac{\sqrt{\frac{5}{6}}}{48} \\
 \frac{\sqrt{\frac{10}{7}}}{3003} & -\frac{1}{154 \sqrt{70}} & \frac{1}{84 \sqrt{70}} & -\frac{1}{56 \sqrt{70}} & 0 \\
 \frac{1}{273 \sqrt{110}} & -\frac{1}{182 \sqrt{110}} & \frac{1}{168 \sqrt{110}} & 0 & 0 \\
 \frac{10 \sqrt{\frac{2}{3}}}{21879} & -\frac{1}{1287 \sqrt{6}} & 0 & 0 & 0 \\
 \frac{21}{2431 \sqrt{190}} & 0 & 0 & 0 & 0 \\
\end{array}
\right) \, ,  \\
M^{(4,6)}&=\left(
\begin{array}{ccccc}
 \frac{5}{2002 \sqrt{66}} & -\frac{5}{924 \sqrt{66}} & \frac{1}{84 \sqrt{66}} & -\frac{\sqrt{\frac{3}{22}}}{112} & \frac{1}{16 \sqrt{66}} \\
 \frac{\sqrt{\frac{10}{33}}}{3003} & -\frac{5 \sqrt{\frac{5}{66}}}{4004} & \frac{\sqrt{\frac{5}{66}}}{462} & -\frac{\sqrt{\frac{5}{66}}}{336} & 0 \\
 \frac{\sqrt{\frac{2}{2145}}}{187} & -\frac{1}{66 \sqrt{4290}} & \frac{1}{66 \sqrt{4290}} & 0 & 0 \\
 \frac{6 \sqrt{\frac{6}{935}}}{2717} & -\frac{\sqrt{\frac{3}{1870}}}{286} & 0 & 0 & 0 \\
 \frac{\sqrt{\frac{35}{22}}}{4199} & 0 & 0 & 0 & 0 \\
\end{array}
\right)\, , \\
M^{(5,5)}&=\left(
\begin{array}{cccccc}
 \frac{1}{2772} & -\frac{1}{1260} & \frac{1}{560} & -\frac{1}{240} & \frac{1}{96} & -\frac{1}{32} \\
 \frac{5 \sqrt{5}}{54054} & -\frac{2}{2079 \sqrt{5}} & \frac{1}{504 \sqrt{5}} & -\frac{1}{252 \sqrt{5}} & \frac{1}{144 \sqrt{5}} & 0 \\
 \frac{1}{6006} & -\frac{3}{10010} & \frac{3}{6160} & -\frac{1}{1680} & 0 & 0 \\
 \frac{20}{35343 \sqrt{13}} & -\frac{8}{10395 \sqrt{13}} & \frac{1}{1386 \sqrt{13}} & 0 & 0 & 0 \\
 \frac{35}{48906 \sqrt{17}} & -\frac{7}{12870 \sqrt{17}} & 0 & 0 & 0 & 0 \\
 \frac{3 \sqrt{21}}{46189} & 0 & 0 & 0 & 0 & 0 \\
\end{array}
\right) \, .
\end{align*}
\endgroup
In the case where one of the operator is a conserved current, the $\lambda^{(i)}$ coefficients need to satisfy certain constraint. Suppose that $\CO_1=T_{\mu\nu}$ and $\CO_3$ a spin-$\ell>0$ primary with dimension $\Delta_3$,  then \cite{Costa:2011mg}
\eqna{
\lambda^{(1)}_{T\CO_2\CO_3}&= \frac{2(2\Delta_2-2\Delta_3-\ell)}{\Delta_2-\Delta_3+\ell+3}\lambda^{(0)}_{T\CO_2\CO_3}\, , \\
\lambda^{(2)}_{T\CO_2\CO_3}&= \frac{2(\Delta_2-\Delta_3)^2-\ell(\ell+1)}{(\Delta_2-\Delta_3+\ell+1)(\Delta_2-\Delta_3+\ell+3)}\lambda^{(0)}_{T\CO_2\CO_3}\, .\\
}[WIconstraint]
In the case of $\ell=0$, the three-point function $\langle T \CO_2 \CO_3\rangle$ is non-vanishing if and only if $\CO_2=\CO_3$.

\newpage
\section{First Discarded Primaries}
\label{app:discarded}
In this appendix we report the scaling dimensions and the corresponding values of $|K|^2$, together with their fits, for the first operators that we have discarded as being primaries despite having relatively small values of $|K|^2$. 
These operators were excluded either because $|K|^2$ was not yet sufficiently small, or because it was unclear whether it was approaching zero as a function of $N$. 
In most cases, diagonalizing a larger system leads to a further decrease of $|K|^2$, suggesting that these operators are likely to correspond to genuine primaries.
\begin{figure}[h!]
  \centering
  \subfloat[$\mathbb{Z}_2$ odd, spin 4 operator]{
    \includegraphics[width=0.41\textwidth]{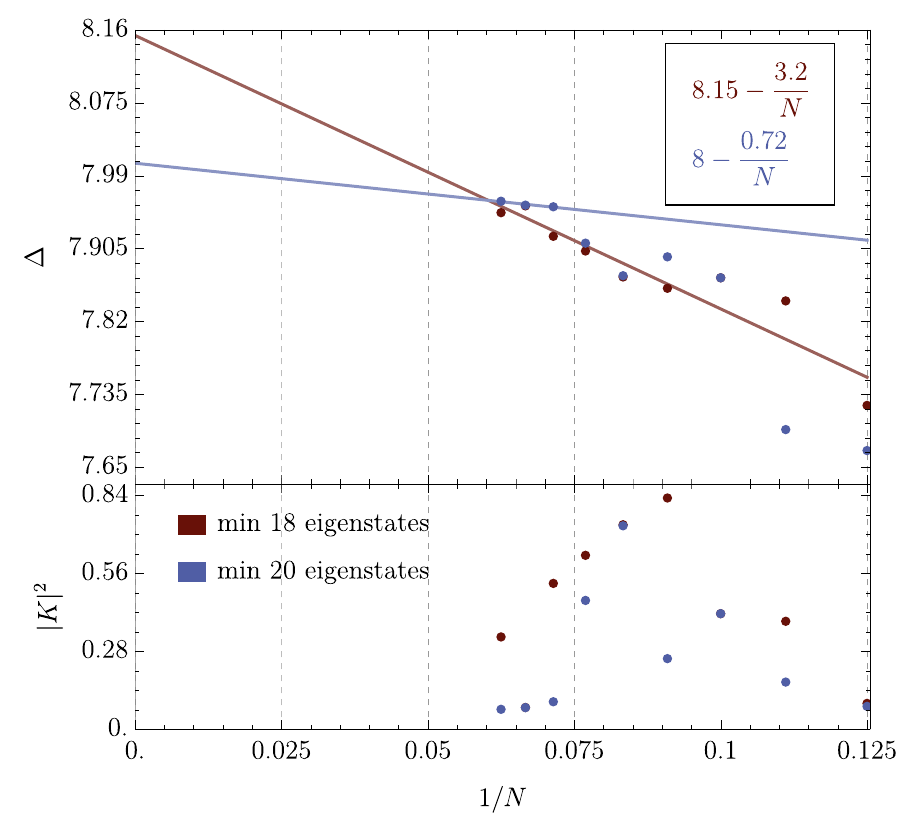}
  }
  \hfill
  \subfloat[$\mathbb{Z}_2$ even, spin 6 operator]{
    \includegraphics[width=0.41\textwidth]{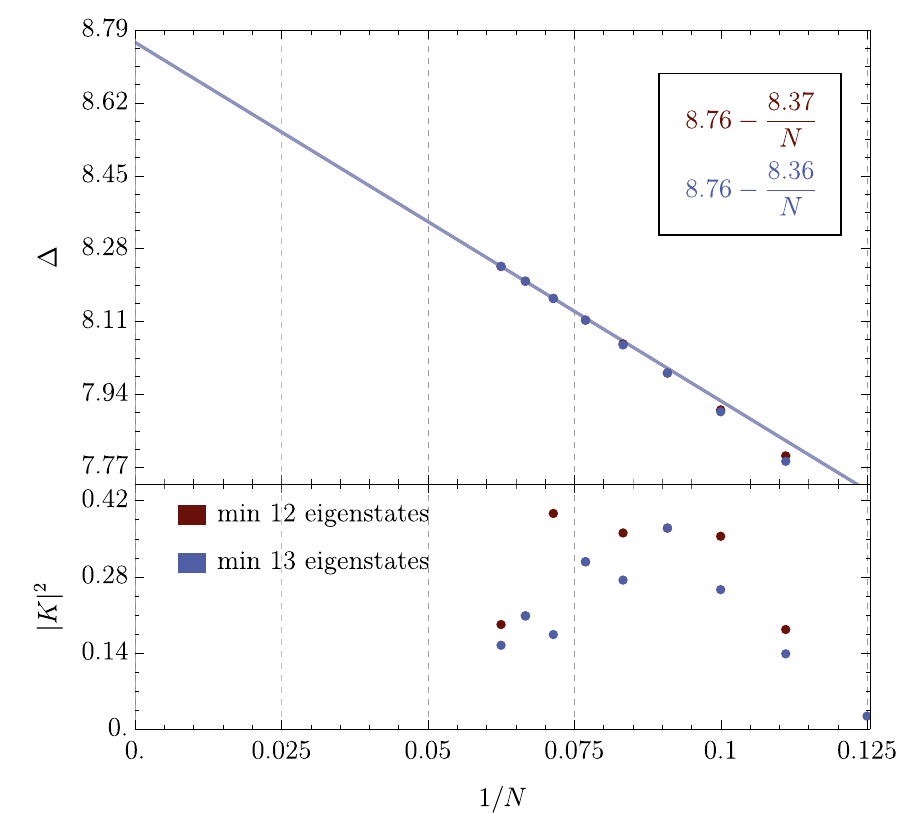}
  }
  \vspace{0.1cm}
  \subfloat[$\mathbb{Z}_2$ even, spin 5 operator]{
    \includegraphics[width=0.41\textwidth]{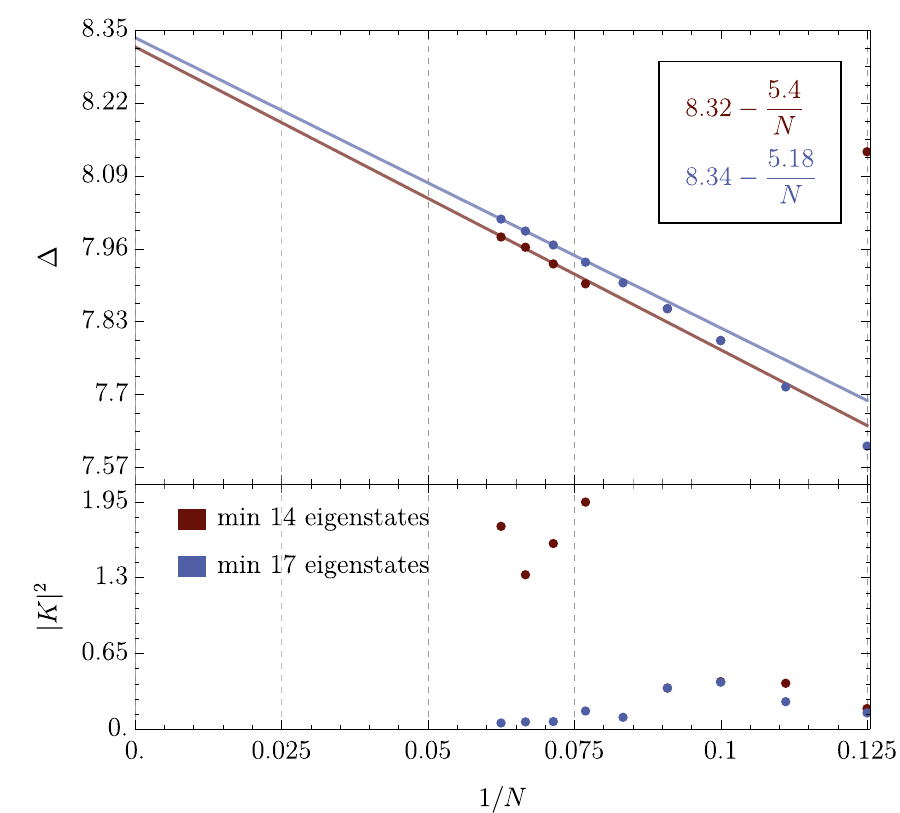}
  }
  \hfill
    \subfloat[$\mathbb{Z}_2$ even,  parity odd, spin 5 operator]{
    \includegraphics[width=0.41\textwidth]{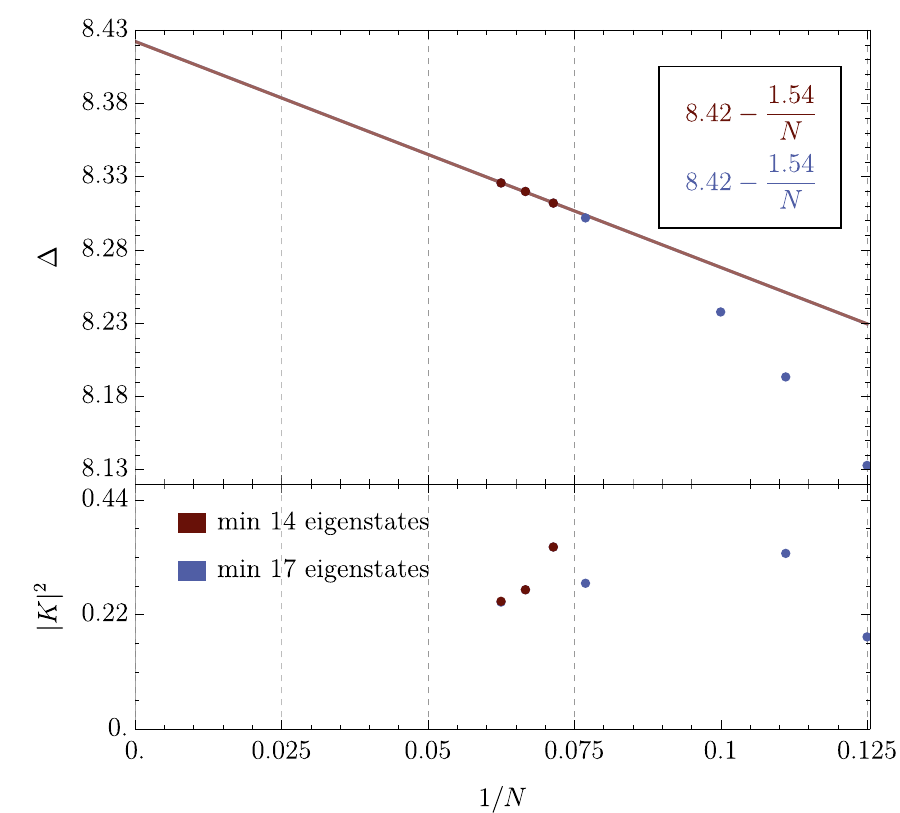}
  }
  \caption{Scaling dimensions and $|K|^2$ values for operators with $|K|^2 < 1$ that we have excluded as primaries. 
  The two colors correspond to systems with different numbers of eigenvalues included in the diagonalization.  For a fixed spin the number of eigenstates depends on $N$ and we report the minimum number of eigenstates among all the $N$ from 8 to 16. 
  In both cases we retain 150 eigenstates in exact diagonalization, working either in the $j_z=0$ or in the $j_z=1$ sector. 
  Although the total number of eigenstates is fixed, working in the $j_z=1$ sector (which eliminates spin-0 states) allows access to a larger number of eigenstates in the spinning sectors. }
  \label{fig:excludedPrimaries}
\end{figure}

\newpage
\section{OPE Coefficient Tables}
\begin{figure}
\begin{center}
\begin{minipage}{0.45\textwidth}
        \centering
        \includegraphics[width=0.95\textwidth]{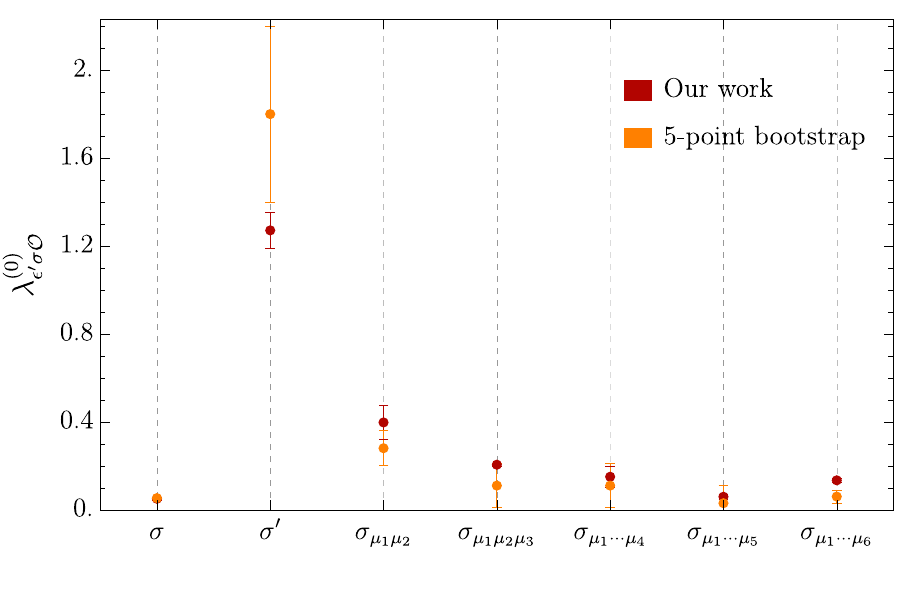}
    \end{minipage}
    \hfill
\begin{minipage}{0.45\textwidth}
        \centering
        \includegraphics[width=1\textwidth]{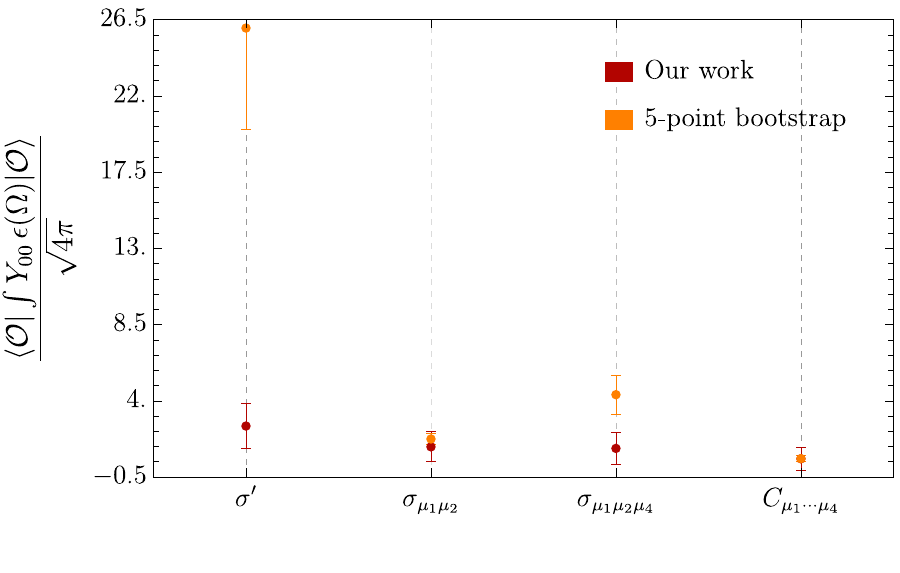}
    \end{minipage}
\caption{\textit{(Left)}: OPE coefficients $\lambda^{(0)}_{\epsilon^\prime\sigma \CO}$ obtained from matrix elements of $n_z$ (shown in red) and from~\cite{Poland:2025ide} (shown in orange).  The quoted value in our work corresponds to the intercept of a linear extrapolation in $1/N$. \textit{(Right)}: integration against the $Y_{0,0}$ spherical harmonic for various diagonal matrix elements obtained from matrix elements of $n_x$ (shown in red) and from~\cite{Poland:2025ide} (shown in orange).  The quoted value in our work corresponds to the large $N$ value obtained from a fit $a+b/N^{\frac{3-\Delta_\epsilon}{2}}$ using the data at $N=14,15,16$. In both cases, our uncertainty is estimated as half of the difference between the extracted value at $N= \infty$ and the value at $N=16$. }
\label{fig:EpsPSigmaO}
\end{center}
\end{figure}
This appendix collects all  results for the OPE coefficients involving the primary operators identified in this work and the operators $\sigma$ and $\epsilon$,  extracted respectively from matrix elements of  $n_z$ and $n_x$.  When, in addition to $\sigma$ and $\epsilon$,  at least one of the   external operators is a scalar, we report the coefficient $\lambda^{(0)}$.  For OPE coefficients involving two spinning external operators, we instead present the large-$N$ extrapolation of the spherical-harmonic integrals defined in~\eqref{intnz} and~\eqref{intnx}.\footnote{We  compute and report results involving $Y_{\ell,0}$ for $\ell \leq 10$. Contributions from higher values of $\ell$ are in general small and comparable to our  precision. Moreover they can only be accessed for $N>\ell$, making it difficult to observe a clear dependence on $N$ for larger   values of $\ell$.} 
All results shown in this appendix are obtained from a linear extrapolation in $1/N$ using  data at $N=14,15,16$.  We also employ a linear fit for matrix elements involving $\epsilon$, where the leading correction arises from contamination by $T^{00}$ in $n_x$. This choice ensures a homogeneous treatment across different observables and is motivated by the fact that multiple effects generate corrections with powers of $1/N$ that are numerically close to one another, making a linear extrapolation the most robust option.
Readers interested in a more detailed, case-by-case analysis can use the data provided in the accompanying Mathematica ancillary file. Finally, we highlight the matrix elements corresponding to OPE coefficients that were previously known from the four-point bootstrap analyses of~\cite{Simmons-Duffin:2016wlq,Chang:2024whx} and the five-point bootstrap results in~\cite{Poland:2025ide}. As an illustrative example,   Fig.~\ref{fig:EpsPSigmaO} compares our result with the five-point bootstrap for OPE coefficients involving $\epsilon^\prime$, $\sigma$ and  various $\mathbb{Z}_2$-odd operators,as well as  the integration against the $Y_{0,0}$ spherical harmonic for several diagonal matrix elements involving $\epsilon$.
We find overall  good agreement with the available bootstrap data.  Matrix elements extracted from $n_z$ show significantly better agreement than those obtained from $n_x$, which we attribute primarily to contamination in $n_x$ from operators other than $\epsilon$.  In particular the matrix elements $\langle \sigma| \int Y_{\ell,0}\lsp  n_x |\CO_{\mu_1 \cdots \mu_\ell }\rangle  $ for $\ell > 2$ deviate significantly from the known bootstrap results, while the corresponding matrix elements extracted from $n_z$, i.e. $\langle \epsilon| \int Y_{\ell,0} n_z \lsp | \CO_{\mu_1 \cdots \mu_\ell }\rangle$, exhibit  very good agreement. We interpret this behavior as arising mainly from the presence of descendant operators of the form $\nabla^{2n} \epsilon$ inside $n_x$.  When integrated against $Y_{\ell,0}$ these contributions are enhanced by factors of order $\sim \ell^{2n}$, leading to  larger deviations at higher spin.

\begin{landscape}
\begin{table}[!h]
\scriptsize
\setlength{\tabcolsep}{5pt}
\renewcommand{\arraystretch}{1.3}
\hspace{-1.5cm}
\begin{tabular}{c||cccc}
$ \boldsymbol{ \sigma}$    & $\sigma$ & $\sigma^\prime$ & $\sigma_{\mu_1\mu_2}$ & $\sigma^\prime_{\mu_1\mu_2}$ \\
\hline\hline
$\mathds{1} $   & \toTeal{ 1}   &   $\left(-0.19+\frac{4.13}{N}\right)10^{-3}$ &  $\left(-0.06+\frac{1.2}{N}\right)10^{-2}$  &  $\left(0.1-\frac{3.6}{N} \right)10^{-4}$       \\
$\epsilon$    &   \toTeal{$1.05-\frac{0.31}{N}$} &  \toTeal{ $-0.05+\frac{0.42}{N}$} & \toTeal{$0.79 -\frac{1.68}{N}$}    &  \toTeal{ $0.07 -\frac{0.76}{N}$  }     \\
$\epsilon^\prime$ & \toTeal{$-0.05+\frac{0.19}{N}$}   & \toOrange{$1.27-\frac{2.57}{N}$}   &    \toOrange{$0.4 +\frac{2.43}{N}$} &  $-0.82+\frac{3.17}{N}$      \\
$\epsilon^{\prime\prime}$   & $\left(-0.54+\frac{3.36}{N}\right)10^{-3}$   &   $1.35-\frac{3.46}{N} $ & $-0.12+\frac{1.24}{N}$    &  $0.41 +\frac{5.3}{N}$       \\
$\epsilon^{\prime\prime\prime}$   & $\left(-0.26-\frac{3.03}{N}\right)10^{-3}$     & $0.18 -\frac{1.27}{N}$   &  $0.43 -\frac{2.03}{N}$  &   $0.03 -\frac{1.26}{N}$      \\
$ T $   &  \toTeal{$0.66-\frac{0.76}{N}$ }  &  $-0.005+\frac{0.02}{N}$  & \toOrange{$\left\{0.73 -\frac{0.94}{N},-0.08-\frac{0.32}{N},-0.005-\frac{0.07}{N}\right\}$}   & $\left\{0.06 -\frac{0.64}{N},-0.01+\frac{0.1}{N},\lesssim 10^{-5}\right\}$       \\
$ T^\prime $   &  \toTeal{$0.02-\frac{0.12}{N}$}  &  $0.39 +\frac{4.28}{N}$  &  $\left\{0.9 -\frac{1.03}{N},-0.08-\frac{0.47}{N},-0.01 -\frac{0.25}{N}\right\}$  &   $\left\{1.16 -\frac{3.48}{N},-0.07-\frac{0.3}{N},-\frac{0.29}{N}-0.03\
\right\}$      \\
$T^{\prime\prime} $   &  \toTeal{$\sim 6.5 \cdot 10^{-4}$   } &  $-0.01-\frac{0.25}{N}$  &   $\left\{-0.01-\frac{0.18}{N},-0.23+\frac{0.88}{N},0.02 +\frac{0.07}{N\
}\right\}$ &  $\left\{-0.09-\frac{0.28}{N},0.06 +\frac{0.99}{N},0. 002-\frac{0.05}{N}\right\}$       \\
$ T^- $   &   0 &  0  & 0   & 0    \\
$C $   &  \toTeal{$0.26-\frac{0.82}{N}$}  &  \toOrange{$0.03 +\frac{0.18}{N}$}  &  $\left\{-0.08-\frac{0.66}{N},-0.01-\frac{0.07}{N},-0.001+\frac{0.001}{N}\right\}$  &  $-\left\{0. 001-\frac{0.03}{N},0.003 -\frac{0.02}{N},0. 001-\frac{0.01}{N}\right\}$     \\
$C^\prime $   & \toTeal{ $\left(0.89-\frac{5.83}{N}\right)10^{-2}$}   &  $0.08 +\frac{0.57}{N}$  & $\left\{-0.19+\frac{0.04}{N},0.01 +\frac{0.09}{N},0.003-\frac{0.001}{N}\right\}$   &$\left\{0.01 +\frac{0.38}{N},-0.01-\frac{0.12}{N},-0.002 -\frac{0.01}{N}\right\}$         \\
$ C^{\prime\prime} $   & $\left(0.77-\frac{4.44}{N}\right)10^{-2}$    &  $0.2 +\frac{2.28}{N}$  &  $\left\{0.2 -\frac{1.12}{N},-0.01-\frac{0.12}{N},-0.004 -\frac{0.02}{N}\right\}$  &  $-\left\{0.1+\frac{1.82}{N},0.02-\frac{0.03}{N},0.01-\frac{0.05}{N}\right\}$       \\
$ C^- $   & 0   & 0   &  0  &  0       \\
$ \epsilon_{ \mu_1\cdots\mu_6} $   & \toTeal{$0.1-\frac{0.49}{N}$}   &  $0.04 -\frac{0.22}{N}$  &   $-\left\{0.02+\frac{0.19}{N},0.004 -\frac{0.01}{N},\left(1.2+\frac{4.3}{N} \right)10^{-4}\right\}$ &  $\left\{-0.01+\frac{0.02}{N},-0.002 +\frac{0.02}{N},-0.0003+\frac{0.003}{N}\right\}$       \\
\end{tabular}
\end{table}

\begin{table}[!htb]
\scriptsize
\setlength{\tabcolsep}{5pt}
\renewcommand{\arraystretch}{1.3}
\hspace{-1.5cm}
\begin{tabular}{c||ccc}
$ \boldsymbol{ \sigma}$    & $\sigma_{\mu_1\mu_2\mu_3}$ &  $\sigma_{\mu_1\mu_2\mu_3}^\prime$ &  $\sigma_{\mu_1\mu_2\mu_3\mu_4}$ \\
\hline\hline
$\mathds{1} $    & $0.003-\frac{0.17}{N}$   & $\sim 0.0004$  & $-0.001+\frac{0.03}{N}$       \\
$\epsilon$      & \toTeal{$0.41 -\frac{0.93}{N}$}   &  $\sim -0.015$ &  \toTeal{ $0.41 -\frac{1.91}{N}$ }     \\
$\epsilon^\prime$  & \toOrange{ $\sim 0.22 $} &  $-0.68+\frac{2.71}{N}$  &\toOrange{$0.15 +\frac{1.56}{N}$  }    \\
$\epsilon^{\prime\prime}$     &   $\sim 0.004$   &  $-0.25 -\frac{4.2}{N}$  & $0.08 -\frac{0.1}{N}$        \\
$\epsilon^{\prime\prime\prime}$    &  $\sim -0.18 $  & $-0.19 -\frac{3.77}{N}$   & $-0.11-\frac{1.36}{N}$        \\
$ T $     & \toOrange{$\left\{0.67 -\frac{0.34}{N},-0.01-\frac{0.16}{N},-0. 001-\frac{0.01}{N}\right\}$}   &   $\sim \left\{-0.04, 0.005, -0.0003 \right\}$ & \toOrange{ $-\left\{-0.14+\frac{0.55}{N},0.01+\frac{0.11}{N},0.001+\frac{0.004}{N}\right\}$ }      \\
$ T^\prime $   &  $\left\{-0.12+\frac{0.71}{N},0.02 +\frac{0.28}{N},0. 004+\frac{0.04}{N}\right\}$  &  $-\left\{0.46 -\frac{0.98}{N},0.04 -\frac{0.01}{N},0.004 +\frac{0.01}{N}\right\}$  & $-\left\{0.1+\frac{1.2}{N},0.02+\frac{0.03}{N},0.01-\frac{0.01}{N}\right\}$      \\
$T^{\prime\prime} $   & $\left\{0.17 -\frac{0.62}{N},0. 004+\frac{0.08}{N},-0.004-\frac{0.004}{N}\right\}$   & $-\left\{0.06 -\frac{0.02}{N},0.03 +\frac{0.74}{N},0. 001-\frac{0.09}{N}\right\}$   & $\left\{0.05 +\frac{0.38}{N},0.01 +\frac{0.27}{N},0.01 -\frac{0.04}{N\
}\right\}$      \\
$ T^- $      &  0  &   0 &  0      \\
$C $    & $-\left\{0.15+\frac{0.99}{N},0. 004+\frac{0.24}{N},0.002 +\frac{0.02}{N},0.0005+\frac{0.004}{N}\right\}$    &    $-\left\{-0.07+\frac{0.4}{N},0.004 +\frac{0.02}{N},-0. 0003+\frac{0.01}{N}, \lesssim 10^{-5}\right\}$& $-\left\{-0.7 +\frac{1.35}{N},0.07+\frac{0.4}{N},0.01+\frac{0.07}{N},0.003 -\frac{0.01}{N},0.0002+\frac{0.002}{N}\right\}$         \\
$C^\prime $      &  $-\left\{-0.75 +\frac{1.13}{N},\frac{0.23}{N}+0.03,0. 001 +\frac{0.02}{N},0.0007-\frac{0.005}{N}\right\}$  &  $-\left\{0.05 -\frac{1.06}{N},0.02 +\frac{0.3}{N},0.00002-\frac{0.04}{N},0.0007 -\frac{0.01}{N}\right\}$  &  $\left\{0.09 -\frac{0.21}{N},0. 002+\frac{0.1}{N},-0. 002+\frac{0.11}{N},0.002+\frac{0.001}{N},0. 001-\frac{0.01}{N}\right\}$      \\
$ C^{\prime\prime} $    &   $\left\{0.04 -\frac{0.03}{N},0.02 -\frac{0.09}{N},0.01 +\frac{0.05}{N},0.001-\frac{0.003}{N}\right\}$ &  $-\left\{-0.09-\frac{0.61}{N},0. 003-\frac{0.27}{N},0.01 -\frac{0.06}{N},0.0002+\frac{0.01}{N}\right\}$  & $-\left\{-0.92 +\frac{2.23}{N},0.08+\frac{0.49}{N},0.02+\frac{0.07}{N},0.01-\frac{0.04}{N},0. 001-\frac{0.01}{N}\right\}$        \\
$ C^- $     & 0   & 0    &   0      \\
$ \epsilon_{ \mu_1\cdots \mu_6} $     &  $-\left\{0.03+\frac{0.34}{N},0. 001+\frac{0.06}{N},0.0007-\frac{0.002}{N}, \lesssim 10^{-5}\right\}$  &  $\sim \left\{-0.003,0. 0003,\lesssim 10^{-5},\lesssim 10^{-6}\right\}$  &$-\left\{0.09+\frac{0.74}{N},0.02+\frac{0.03}{N},0. 005-\frac{0.03}{N},0. 001-\frac{0.01}{N},\lesssim 10^{-5}\right\}$        \\
\end{tabular}
\end{table}

\begin{table}[!htb]
\scriptsize
\setlength{\tabcolsep}{5pt}
\renewcommand{\arraystretch}{1.3}
%\hspace{-1.5cm}
\begin{tabular}{c||cc}
$ \boldsymbol{ \sigma}$    & $\sigma_{\mu_1\mu_2\mu_3\mu_4}^-$ & $\sigma_{\mu_1\mu_2\mu_3\mu_4\mu_5}$  \\
\hline\hline
$\mathds{1} $      & 0    &    $0. 003-\frac{0.07}{N}$      \\
$\epsilon$      &   0  &  \toTeal{$0.22 -\frac{0.97}{N}$}       \\
$\epsilon^\prime$     &  0  & \toOrange{ $0.06 +\frac{0.41}{N}$   }    \\
$\epsilon^{\prime\prime}$        & 0   &   $0.05 -\frac{0.31}{N}$       \\
$\epsilon^{\prime\prime\prime}$       & 0    &    $-0.1+\frac{0.35}{N}$  \\
$ T $      & 0    & \toOrange{$-\left\{-0.15 +\frac{0.44}{N},0. 003+\frac{0.04}{N},0.001-\frac{0.004}{N}\right\}$      }   \\
$ T^\prime $    &   0 &   $\left\{0.02 +\frac{0.1}{N},0.01 +\frac{0.02}{N},0. 002-\frac{0.01}{N}\right\}$       \\
$T^{\prime\prime} $      &   0  &    $-\left\{0.03+\frac{0.28}{N},0.003 +\frac{0.02}{N},0.002 -\frac{0.02}{N}\right\}$      \\
$ T^- $      &  $\left\{0.09 +\frac{0.41}{N},-0.02-\frac{0.16}{N},0.001-\frac{0.004}{N}\right\}$   &   0       \\
$C $      &  0   & $-\left\{-0.5 +\frac{0.37}{N},0.01+\frac{0.29}{N},0.004 +\frac{0.04}{N},0.001 -\frac{0.01}{N},\lesssim 10^{-5}\right\}$       \\
$C^\prime $       &  0  & $-\left\{0.15+\frac{1.27}{N},0.003 +\frac{0.38}{N},0.004 +\frac{0.03}{N},0.001 -\frac{0.01}{N},0.0001-\frac{0.001}{N}\right\}$         \\
$ C^{\prime\prime} $    & 0   & $\left\{-0.07 -\frac{0.03}{N},0.02 +\frac{0.15}{N},0.01 -\frac{0.01}{N},0. 002-\frac{0.02}{N},0.0003-\frac{0.003}{N}\right\}$         \\
$ C^- $     & $-\left\{0.64-\frac{1.85}{N},-0.02 -\frac{0.15}{N},0. 002
-\frac{0.07}{N},0.01 - \frac{0.02}{N},\lesssim 10^{-5}\right\}$  & 0        \\
$ \epsilon_{ \mu_1\cdots \mu_6} $       & 0    &  $-\left\{0.13 +\frac{1.68}{N},0.01+\frac{0.27}{N},0.01+\frac{0.02}{N},0.002 -\frac{0.01}{N},0.0002-\frac{0.002}{N}\right\}$        \\
\end{tabular}
\end{table}
\begin{table}[!htb]
\scriptsize
\setlength{\tabcolsep}{5pt}
\renewcommand{\arraystretch}{1.3}
%\hspace{-1.5cm}
\begin{tabular}{c||cc}
$ \boldsymbol{ \sigma}$     & $\sigma_{\mu_1\mu_2\mu_3\mu_4\mu_5\mu_6}^\prime$ & $\sigma_{\mu_1\mu_2\mu_3\mu_4\mu_5\mu_6}$ \\
\hline\hline
$\mathds{1} $      &  $ 0.001-\frac{0.02}{N}$   &   $\lesssim 10^{-5}$  \\
$\epsilon$        &   $-0.01-\frac{0.31}{N}$  & \toTeal{$0.18 -\frac{1.2}{N}$}    \\
$\epsilon^\prime$        &  $0.01 -\frac{0.72}{N}$   & \toOrange{$0.13 -\frac{0.27}{N}$}    \\
$\epsilon^{\prime\prime}$         &   $-0.002-\frac{0.05}{N}$  &  $0.07 -\frac{0.63}{N}$   \\
$\epsilon^{\prime\prime\prime}$       & $-0. 003+\frac{0.24}{N}$ &   $-0.12+\frac{0.36}{N}$  \\
$ T $         &   $\left\{0.1 -\frac{0.54}{N},-0. 002+\frac{0.04}{N},0.0004-\frac{0.004}{N}\right\}$   &    \toOrange{$\left\{0.04 -\frac{0.03}{N},0.0003-\frac{0.07}{N},-0.002+\frac{0.02}{N}\right\}$ }\\
$ T^\prime $      &  $\left\{-0.02+\frac{0.24}{N},0.002-\frac{0.003}{N}, \lesssim 10^{-4}\right\}$   &   $-\left\{0.02+\frac{0.4}{N},0.01-\frac{0.08}{N},0.002-\frac{0.02}{N}\right\}$  \\
$T^{\prime\prime} $      & $\left\{-0. 003-\frac{0.15}{N},0. 004-\frac{0.05}{N},-0.0002+\frac{0.002}{N}\right\}$    &  $\left\{0.02 -\frac{0.14}{N},0.01 +\frac{0.02}{N},0. 002-\frac{0.03}{N}\right\}$   \\
$ T^- $      &    0 &   0  \\
$C $         &  $\left\{-0.42+\frac{0.68}{N},-0.01+\frac{0.33}{N},0. 0003+\frac{0.02}{N},\lesssim 10^{-5},\lesssim 10^{-6} \right\}$   &   $-\left\{-0.19 +\frac{2.67}{N},0.02+\frac{0.01}{N},0.01-\frac{0.05}{N},0.002
-\frac{0.02}{N},\lesssim 10^{-6} \right\}$  \\
$C^\prime $         &   -$\left\{0.08+\frac{0.79}{N},0.004 +\frac{0.13}{N},0.001+\frac{0.01}{N},\lesssim 10^{-5},\lesssim 10^{-6}\right\}$      & $\left\{0.05 -\frac{0.83}{N},0. 004-\frac{0.04}{N},-0.001
+\frac{0.03}{N},\lesssim 10^{-5},0.0001-\frac{0.002}{N}\right\}$     \\
$ C^{\prime\prime} $    &  $\left\{-0.05+\frac{1.37}{N},-0.01+\frac{0.04}{N},0.0005-\frac{0.01}{N},0.0004-\frac{0.005}{N},\lesssim 10^{-6}\right\}$  &    $-\left\{0.11+\frac{1.33}{N},0.02-\frac{0.04}{N},0.
01-\frac{0.07}{N},0.001-\frac{0.01}{N},\lesssim 10^{-5}\right\}$     \\
$ C^- $       & 0     &     \\
$ \epsilon_{ \mu_1\cdots \mu_6} $      &  $\left\{0.45 \
-\frac{3.8}{N},-0.04+\frac{1.06}{N},-0.01+\frac{0.35}{N},0. 0001
+\frac{0.04}{N},0.0005-\frac{0.003}{N},\lesssim 10^{-6} \right\}$    &  $-\left\{-0.69 +\frac{0.97}{N},0.07+\frac{0.32}{N},0.02-\frac{0.11}{N},0.01-\frac{0.08}{N},0.001-\frac{0.01}{N},\lesssim 10^{-5}\right\}$   \\
\end{tabular}
\caption{OPE coefficients involving $\sigma$.  Known matrix elements are highlighted in teal (from~\cite{Simmons-Duffin:2016wlq,Chang:2024whx}) and in orange (from the five-point bootstrap~\cite{Poland:2025ide}).  All extrapolations are performed using data at $N=14,15,16$. When no clear trend as a function $N$ can be identified,  we report the value at $N=16$ indicated by the symbol $\sim$.}\label{Tab: sigma}
\end{table}
\begin{table}[!htb]
\scriptsize
\setlength{\tabcolsep}{5pt}
\renewcommand{\arraystretch}{1.3}
\hspace{-1.5cm}
\begin{tabular}{c||ccccc}
$ \boldsymbol{ \epsilon}$    & $\sigma$ & $\sigma^\prime$ & $\sigma_{\mu_1\mu_2}$ & $\sigma^\prime_{\mu_1\mu_2}$ &  $\sigma_{\mu_1\mu_2\mu_3}$ \\
\hline\hline
$\sigma$ & \toTeal{ $1.06 +\frac{4.64}{N}$} &  \toTeal{$-0.06+\frac{0.4}{N}$} & \toTeal{$0.76 -\frac{1.85}{N}$} & \toTeal{ $0.07 -\frac{0.68}{N}$} & \toTeal{$-0.08+\frac{3.23}{N}$}\\
$\sigma^\prime$ &   &\toOrange{ $3.07 +\frac{32.79}{N}$} & \toOrange{$0.48 -\frac{3.79}{N}$} & $-3.5-\frac{10.9}{N}$ & \toOrange{$\sim 0.188$ } \\
$\sigma_{\mu_1\mu_2}$ &  &   & \toOrange{$\left\{1.66 +\frac{22.17}{N},0.36 +\frac{1.31}{N},0.35 -\frac{1.33}{N}\right\}$ } & $\left\{0.89 -\frac{2.52}{N},-0.13+\frac{0.55}{N},-0.01-\frac{0.07}{N}\right\}$ & \toOrange{ $\left\{0.41 +\frac{3.97}{N},-0.37-\frac{0.05}{N},-0.1+\frac{0.59}{N}\right\}$} \\
$\sigma^\prime_{\mu_1\mu_2}$  &  &   &   & $\left\{3.07 +\frac{34.65}{N},0.26 -\frac{0.85}{N},0.49 -\frac{3.93}{N}\right\}$ & $\sim \left\{-0.078, 0.01, 0.008\right\}$ \\
$\sigma_{\mu_1\cdots\mu_3}$ &   &   &   &   & \toOrange{$\left\{1.59 +\frac{23.4}{N},0.8 +\frac{2.3}{N},0.04 -\frac{0.53}{N},0.02 -\frac{0.22}{N}\right\}$ }\\
$\sigma_{\mu_1\cdots\mu_3}^\prime$ &   &   &   &   &     \\
$\sigma_{\mu_1\cdots\mu_4}$ &   &   &   &   &    \\
$\sigma_{\mu_1\cdots\mu_4}^-$ &   &   &   &   &   \\
$\sigma_{\mu_1\cdots\mu_5}$ &   &   &   &   &    \\
$\sigma_{\mu_1\cdots\mu_6}^\prime$ &   &   &   &   &   \\
$\sigma_{\mu_1\cdots\mu_6}$ &   &   &   &   &   
\end{tabular}
\end{table}

\begin{table}[!htb]
\scriptsize
\setlength{\tabcolsep}{5pt}
\renewcommand{\arraystretch}{1.3}
\hspace{-1.5cm}
\begin{tabular}{c||ccc}
$ \boldsymbol{ \epsilon}$     &  $\sigma_{\mu_1\mu_2\mu_3}^\prime$ &  $\sigma_{\mu_1\mu_2\mu_3\mu_4}$& $\sigma_{\mu_1\mu_2\mu_3\mu_4}^-$ \\
\hline\hline
$\sigma$ & $\sim -0.014$ & \toTeal{$0.29 -\frac{1.5}{N}$ } & 0 \\
$\sigma^\prime$ & $4.07 -\frac{3.86}{N}$& $-0.63+\frac{5.55}{N}$& 0\\
$\sigma_{\mu_1\mu_2}$ & $-\left\{0.99-\frac{5.54}{N},-0.06 +\frac{0.61}{N},0.02-\frac{0.09}{N}\right\}$ & $\left\{0.93 +\frac{3.89}{N},0.2 -\frac{1.28}{N},0.08 -\frac{0.67}{N}\right\}$ & 0  \\
$\sigma^\prime_{\mu_1\mu_2}$   & $\left\{-0.54 -\frac{4.02}{N},0.3 -\frac{0.96}{N},0.26 -\frac{1.71}{N}\right\}$ &  $\left\{-0.05-\frac{0.82}{N},0.02 -\frac{0.37}{N},0.01 -\frac{0.19}{N}\right\}$ & 0 \\
$\sigma_{\mu_1\cdots\mu_3}$ & $\sim \left\{-0.34, 0.05, 0.02, 0.004\right\}$ & $-\left\{-0.09 -\frac{0.23}{N},0.06+\frac{0.07}{N},0.1-\frac{0.5}{N},0.02-\frac{0.15}{N}\right\}$ & 0  \\
$\sigma_{\mu_1\cdots\mu_3}^\prime$ & $\left\{2.98 +\frac{34.75}{N},0.31 +\frac{0.57}{N},0.24 -\frac{0.83}{N},0.11 -\frac{0.91}{N}\right\}$ & $\left\{0.32 +\frac{0.88}{N},0.02 -\frac{0.46}{N},0.01 -\frac{0.12}{N},-0.003+\frac{0.02}{N}\right\}$& 0  \\
$\sigma_{\mu_1\cdots\mu_4}$ &   & $\left\{1.69 +\frac{22.61}{N},0.55 -\frac{0.37}{N},0.3 -\frac{2.34}{N},0.06 -\frac{0.65}{N},0.01 -\frac{0.13}{N}\right\}$ &0 \\
$\sigma_{\mu_1\cdots\mu_4}^-$ &   &   & $\left\{1.54 +\frac{23.77}{N},0.27 +\frac{0.14}{N},-0.01+\frac{0.41}{N},-0.1+\frac{0.87}{N},0.001-\frac{0.01}{N}\right\}$ \\
$\sigma_{\mu_1\cdots\mu_5}$ &   &   &     \\
$\sigma_{\mu_1\cdots\mu_6}^\prime$ &   &   &   \\
$\sigma_{\mu_1\cdots\mu_6}$ &   &   &  
\end{tabular}
\end{table}

\begin{table}[!htb]
\scriptsize
\setlength{\tabcolsep}{5pt}
\renewcommand{\arraystretch}{1.3}
%\hspace{-1.5cm}
\begin{tabular}{c||cc}
$ \boldsymbol{ \epsilon}$    & $\sigma_{\mu_1\mu_2\mu_3\mu_4\mu_5}$ & $\sigma_{\mu_1\mu_2\mu_3\mu_4\mu_5\mu_6}^\prime$ \\
\hline\hline
$\sigma$  & \toTeal{ $-0.02+\frac{1.2}{N}$} & $0.12+\frac{1.47}{N}$  \\
$\sigma^\prime$ & $-0.38+\frac{3.59}{N}$ & $-0.01 -\frac{0.84}{N}$ \\
$\sigma_{\mu_1\mu_2}$  & $-\left\{0.04-\frac{0.001}{N},0.13-\frac{0.73}{N},0.02-\frac{0.2}{N}\right\}$ & $\left\{ 0.32-\frac{2.65}{N},-0.02+\frac{0.23}{N}, -0.001+\frac{0.0001}{N} \right\}$   \\
$\sigma^\prime_{\mu_1\mu_2}$   &  $-\left\{-0.03 +\frac{0.26}{N},0.03-\frac{0.34}{N},0.005 -\frac{0.06}{N}\right\}$ & $\left\{-0.02+\frac{0.01}{N},-0.002+\frac{0.07}{N},0.0003+\frac{0.0002}{N}\right\}$   \\
$\sigma_{\mu_1\cdots\mu_3}$ & $\left\{0.78 +\frac{3.65}{N},0.23 -\frac{1.26}{N},0.01 -\frac{0.19}{N},0.002-\frac{0.02}{N}\right\}$ & $\left\{0.65 -\frac{0.82}{N},0.09 -\frac{0.45}{N},-0.002+\frac{0.04}{N},0.001-\frac{0.01}{N}\right\}$  \\
$\sigma_{\mu_1\cdots\mu_3}^\prime$  & $\sim \left\{0.04, -0.01, 0.001, \lesssim 10^{-5} \right\}$  & $\left\{-0.08, 0.01, 0.0003, \lesssim 10^{-5} \right\}$  \\
$\sigma_{\mu_1\cdots\mu_4}$ & $-\left\{-0.52 -\frac{3.05}{N},0.2+\frac{0.06}{N},0.09-\frac{0.7}{N},0.03-\frac{0.25}{N},0.002-\frac{0.03}{N}\right\}$ &$\left\{-0.07-\frac{8.82}{N},0.12 -\frac{0.51}{N},0.005+\frac{0.04}{N},-0.002+\frac{0.02}{N},\lesssim 10^{-6}\right\}$  \\
$\sigma_{\mu_1\cdots\mu_4}^-$ &  0 & 0 \\
$\sigma_{\mu_1\cdots\mu_5}$  & $\left\{1.58 +\frac{25.04}{N},0.92 +\frac{0.39}{N},0.24 -\frac{2.45}{N},0.04 -\frac{0.43}{N},0.001-\frac{0.01}{N},\lesssim 10^{-5}\right\}$& $\left\{-0.41-\frac{3.11}{N},0.03 -\frac{0.54}{N},0.09 -\frac{0.54}{N},0.01 -\frac{0.12}{N},0.001-\frac{0.01}{N}\right\}$   \\
$\sigma_{\mu_1\cdots\mu_6}^\prime$ &   & $\left\{1.49 +\frac{24.28}{N},0.98 +\frac{0.98}{N},0.32 -\frac{2.22}{N},0.03 -\frac{0.46}{N},0.002 -\frac{0.03}{N},\lesssim 10^{-5}\right\}$\\
$\sigma_{\mu_1\cdots\mu_6}$ &    &   
\end{tabular}
\end{table}

\begin{table}[!htb]
\scriptsize
\setlength{\tabcolsep}{5pt}
\renewcommand{\arraystretch}{1.3}
%\hspace{-1.5cm}
\begin{tabular}{c||c}
$ \boldsymbol{ \epsilon}$    & $\sigma_{\mu_1\mu_2\mu_3\mu_4\mu_5\mu_6}$\\
\hline\hline
$\sigma$   & \toTeal{$\sim 0.07$}\\
$\sigma^\prime$ &  $-0.43+\frac{5.04}{N}$\\
$\sigma_{\mu_1\mu_2}$   & $\left\{0.44 -\frac{0.65}{N},0.07 -\frac{0.79}{N},0.01 -\frac{0.13}{N}\right\}$  \\
$\sigma^\prime_{\mu_1\mu_2}$   &  $\left\{0.08 -\frac{1.21}{N},0.004-\frac{0.07}{N},0.002-\frac{0.03}{N}\right\}$ \\
$\sigma_{\mu_1\cdots\mu_3}$ & $\left\{-0.08+\frac{3.08}{N},-0.02+\frac{0.22}{N},-0.002+\frac{0.11}{N},-0.01+\frac{0.08}{N}\right\}$\\
$\sigma_{\mu_1\cdots\mu_3}^\prime $  & $\left\{0.05 -\frac{0.51}{N},-0.02+\frac{0.22}{N},0.003-\frac{0.03}{N},-0.0004+\frac{0.005}{N}\right\}$ \\
$\sigma_{\mu_1\cdots\mu_4}$  & $\left\{1.26 -\frac{0.7}{N},0.32 -\frac{2.8}{N},0.07 -\frac{0.85}{N},0.01 -\frac{0.08}{N},0.001-\frac{0.01}{N}\right\}$ \\
$\sigma_{\mu_1\cdots\mu_4}^-$  & 0\\
$\sigma_{\mu_1\cdots\mu_5}$   & $-\left\{-0.24 +\frac{2.95}{N},0.02+\frac{0.5}{N},0.09-\frac{0.79}{N},0.02-\frac{0.29}{N},0.003-\frac{0.04}{N}\right\}$  \\
$\sigma_{\mu_1\cdots\mu_6}^\prime$ & $\left\{0.16 -\frac{1.82}{N},-0.24+\frac{3.82}{N},0.003
+\frac{0.57}{N},0.02 -\frac{0.05}{N},0. 002-\frac{0.01}{N},\lesssim 10^{-5}\right\}$\\
$\sigma_{\mu_1\cdots\mu_6}$ &   $\left\{1.83 +\frac{21.98}{N},0.78 -\frac{3.85}{N},0.38 -\frac{4.16}{N},0.07 -\frac{0.9}{N},0.005 -\frac{0.07}{N},\lesssim 10^{-5}\right\}$
\end{tabular}
\caption{OPE coefficients involving $\epsilon$ and $\mathbb{Z}_2$-odd primaries.  Known matrix elements are highlighted in teal (from~\cite{Simmons-Duffin:2016wlq,Chang:2024whx}) and in orange (from the five-point bootstrap~\cite{Poland:2025ide}).  All extrapolations are performed using data at $N=14,15,16$. When no clear trend as a function $N$ can be identified,  we report the value at $N=16$ indicated by the symbol $\sim$. Blank entries are fixed by symmetry. } \label{Tab: epsilonOdd}
\end{table}

%%%%%%%%%%%%%%%%%%%%%%%%

\begin{table}[!htb]
\scriptsize
\setlength{\tabcolsep}{5pt}
\renewcommand{\arraystretch}{1.3}
%\hspace{-1.5cm}
\begin{tabular}{c||ccccccc}
$ \boldsymbol{ \epsilon}$ &   $\mathds{1}$ & $\epsilon$ &  $\epsilon^{\prime}$ &  $\epsilon^{\prime\prime}$ &   $\epsilon^{\prime\prime\prime}$ & $T_{\mu_1\mu_2}$ & $T^\prime_{\mu_1\mu_2}$ \\
\hline\hline
$\mathds{1}$ &\toTeal{ 0} & \toTeal{1} &  $-0.01-\frac{0.13}{N}$ & $0.002-\frac{0.01}{N}$& $-0.0001-\frac{0.03}{N}$ & $0.12 +\frac{2.97}{N}$& $-0.002+\frac{0.1}{N}$\\
$\epsilon$ &   & \toTeal{$1.57 +\frac{10.51}{N}$ } & \toTeal{ $-1.54+\frac{3.44}{N}$} & $-0.1+\frac{0.71}{N}$ & $0.11 -\frac{0.95}{N}$& \toTeal{$2.1 +\frac{4.84}{N}$} & $1.24 -\frac{6.25}{N}$ \\
$\epsilon^{\prime}$ &  &   &  \toTeal{$2.54 +\frac{24.92}{N}$} & $1.9 -\frac{8.29}{N}$ & $0.5 -\frac{3.44}{N}$ & $-0.12-\frac{0.84}{N}$ & $-3.07-\frac{10.89}{N}$\\
$\epsilon^{\prime\prime}$  &  &   &   &  $3.57 +\frac{41.97}{N}$ & $0.22 -\frac{0.22}{N}$& $-0.01-\frac{0.01}{N}$ & $0.21 +\frac{0.88}{N}$  \\
$\epsilon^{\prime\prime\prime}$ &   &   &   &   & $2.15 +\frac{31.27}{N}$ & $0.08 +\frac{0.83}{N}$& $\sim 1.4$ \\
$T$ &   &   &   &   &   &  \toTeal{ $\left\{1.1 +\frac{14.42}{N},0.65 +\frac{2.21}{N},0.17 -\frac{0.63}{N}\right\}$}  & $\left\{0.67 -\frac{1.98}{N},-0.09+\frac{0.48}{N},0.001-\frac{0.09}{N}\right\}$\\
$T^\prime$ &   &   &   &   &   &   & $\left\{2.38 +\frac{28.06}{N},0.24 -\frac{0.02}{N},0.43 -\frac{2.36}{N}\right\}$ \\
$T^{\prime\prime}$  &   &   &   &   &   &    &   \\
$T^{-}$  &   &   &   &   &   &    &  \\
$C$ &   &   &   &   &   &    &  \\
$C^{\prime}$ &   &   &   &   &   &    &  \\
$C^{\prime\prime}$ &   &   &   &   &   &    &  \\
$C^{-}$ &   &   &   &   &   &    &  \\
$\epsilon_{\mu_1\cdots \mu_6} $ &   &   &   &   &   &    &  
\end{tabular}
\end{table}

\begin{table}[!htb]
\scriptsize
\setlength{\tabcolsep}{5pt}
\renewcommand{\arraystretch}{1.3}
%\hspace{-1.5cm}
\begin{tabular}{c||ccc}
$ \boldsymbol{ \epsilon}$  & $T^{\prime\prime}_{\mu_1\mu_2}$ & $T^-_{\mu_1\mu_2}$ & $C_{\mu_1\mu_2\mu_3\mu_4}$ \\
\hline\hline
$\mathds{1}$ & $-0.001+\frac{0.04}{N}$ & 0 & $-0.06+\frac{0.92}{N}$ \\
$\epsilon$ & $-0.25+\frac{2.44}{N}$ & 0 & \toTeal{$1.28 -\frac{3.83}{N}$}\\
$\epsilon^{\prime}$ & $\sim 0.15$ & 0 & \toOrange{$-0.71+\frac{4.2}{N}$ } \\
$\epsilon^{\prime\prime}$  & $0.42 +\frac{1.14}{N}$ & 0 &  $0.02 +\frac{0.07}{N}$ \\
$\epsilon^{\prime\prime\prime}$ & $-1.96-\frac{0.68}{N}$ & 0 & $-0.17-\frac{1.06}{N}$ \\
$T$ & $\left\{0.42 -\frac{2.79}{N},-0.13+\frac{0.69}{N},0. 005+\frac{0.05}{N}\right\}$ & 0 & \toOrange{$\left\{0.78 +\frac{3.5}{N},0.24 -\frac{0.94}{N},0.03 -\frac{0.24}{N}\right\}$} \\
$T^\prime$ & $-\left\{0.05+\frac{0.03}{N},0.79+\frac{1.5}{N},-0.02 -\frac{0.79}{N}\right\}$ & 0 & $\left\{-0.06+\frac{0.13}{N},0.03 -\frac{0.33}{N},0.01 -\frac{0.09}{N}\right\}$ \\
$T^{\prime\prime}$  & $\left\{2.12 +\frac{31.3}{N},0.2 +\frac{0.66}{N},0.06 -\frac{1.57}{N}\right\}$ & 0 & $-\left\{0.01+\frac{0.48}{N},-0.02 +\frac{0.06}{N},0.002 -\frac{0.03}{N}\right\}$\\
$T^{-}$  &   & $\left\{2.14 +\frac{31.47}{N},-0.39-\frac{1.28}{N},-0.34+\frac{0.48}{N}\right\}$ & 0 \\
$C$ &   &   & \toOrange{$\left\{0.87 +\frac{17.19}{N},0.72 +\frac{1.15}{N},0.3 -\frac{1.84}{N},0.06 -\frac{0.51}{N},0. 003-\frac{0.04}{N}\right\}$} \\
$C^{\prime}$ &   &   &   \\
$C^{\prime\prime}$  &   &   &   \\
$C^{-}$  &   &   &  \\
$\epsilon_{\mu_1\cdots \mu_6} $  &   &   &   
\end{tabular}
\end{table}

\begin{table}[!htb]
\scriptsize
\setlength{\tabcolsep}{5pt}
\renewcommand{\arraystretch}{1.3}
%\hspace{-1.5cm}
\begin{tabular}{c||cc}
$ \boldsymbol{ \epsilon}$  &  $C^{\prime}_{\mu_1\mu_2\mu_3\mu_4}$ & $C^{\prime\prime}_{\mu_1\mu_2\mu_3\mu_4}$ \\
\hline\hline
$\mathds{1}$ & $0. 003-\frac{0.18}{N}$ & $-0.002 +\frac{0.06}{N}$ \\
$\epsilon$ & $0.01 +\frac{2.28}{N}$ & $0.59 -\frac{4.16}{N}$ \\
$\epsilon^{\prime}$ & $-1.36+\frac{3.8}{N}$ & $-2.84+\frac{10.14}{N}$\\
$\epsilon^{\prime\prime}$  & $0.04 -\frac{1.38}{N}$ & $-0.81+\frac{9.68}{N}$\\
$\epsilon^{\prime\prime\prime}$  & $-0.17+\frac{6.67}{N}$ & $2.12 -\frac{17.49}{N}$\\
$T$ & $\left\{0.39 -\frac{2.33}{N},0. 003+\frac{0.19}{N},-0.001+\frac{0.03}{N}\right\}$  & $\left\{0.22 -\frac{1.37}{N},-0.02+\frac{0.16}{N},0.0001-\frac{0.01}{N}\right\}$  \\
$T^\prime$ & $\left\{-0.36-\frac{1.28}{N},0.17 -\frac{0.07}{N},0.04 -\frac{0.22}{N}\right\}$ & $\left\{1.08 +\frac{3.55}{N},0.2 -\frac{1.75}{N},0.11 -\frac{1.12}{N}\right\}$\\
$T^{\prime\prime}$&  $\left\{0.63 +\frac{1.14}{N},0.19 -\frac{0.76}{N},-0.03+\frac{0.12}{N}\right\}$ & $\sim \left\{-0.08, -0.17, 0.02\right\}$\\
$T^{-}$ & 0 & 0 \\
$C$ &  $\left\{-0.18-\frac{2.68}{N},-0.07+\frac{0.12}{N},0.02 +\frac{0.2}{N},0.003-\frac{0.003}{N},\lesssim 10^{-5}\right\}$ & $\left\{0.5 -\frac{2.01}{N},-0.09+\frac{0.28}{N},-0.003 -\frac{0.06}{N},0.002-\frac{0.04}{N},0.001-\frac{0.01}{N}\right\}$ \\ 
$C^{\prime}$& $-\left\{-2.29 -\frac{29.49}{N},-0.89 -\frac{2.53}{N},0.03+\frac{0.48}{N},0.003 +\frac{0.05}{N},0.002-\frac{0.01}{N}\right\}$ & $\left\{0.07 -\frac{0.35}{N},0.01 +\frac{0.35}{N},0.05 -\frac{0.2}{N},0.04 -\frac{0.18}{N},0.01 -\frac{0.07}{N}\right\}$ \\
$C^{\prime\prime}$ &   & $\left\{2.4 +\frac{29.21}{N},0.48 -\frac{1.97}{N},0.31 -\frac{3.01}{N},0.06 -\frac{0.79}{N},0.02 -\frac{0.21}{N}\right\}$ \\
$C^{-}$ &   &    \\
$\epsilon_{\mu_1\cdots \mu_6} $ &   &   
\end{tabular}
\end{table}

\begin{table}[!htb]
\scriptsize
\setlength{\tabcolsep}{5pt}
\renewcommand{\arraystretch}{1.3}
%\hspace{-1.5cm}
\begin{tabular}{c||cc}
$ \boldsymbol{ \epsilon}$  & $C^-_{\mu_1\mu_2\mu_3\mu_4}$ & $\epsilon_{\mu_1\mu_2\mu_3\mu_4\mu_5\mu_6}$\\
\hline\hline
$\mathds{1}$ & 0 & $0.01 -\frac{0.06}{N}$\\
$\epsilon$ & 0 & \toTeal{$0.55 -\frac{2.98}{N}$}\\
$\epsilon^{\prime}$ & 0 &\toOrange{ $-0.52+\frac{5}{N}$}\\
$\epsilon^{\prime\prime}$  & 0 & $\sim -0.01$\\
$\epsilon^{\prime\prime\prime}$  & 0 & $\sim -0.08$ \\
$T$ & 0 & $\left\{0.36 -\frac{0.74}{N},0.06 -\frac{0.42}{N},0. 003-\frac{0.02}{N}\right\}$  \\
$T^\prime$ & 0 & $\left\{0.05 -\frac{0.55}{N},0.01 -\frac{0.15}{N},0. 002-\frac{0.03}{N}\right\}$ \\
$T^{\prime\prime}$&  0 & $\sim \left\{-0.01, 0.001, -0.0002\right\}$ \\
$T^{-}$ & $\left\{0.18 -\frac{0.76}{N},0.12 +\frac{0.3}{N},-0.06+\frac{0.46}{N}\right\}$ & 0 \\
$C$ & 0 & $\left\{0.88 +\frac{4.03}{N},0.38 -\frac{2.69}{N},0.08 -\frac{0.85}{N},0.01 -\frac{0.1}{N},0.0002-\frac{0.002}{N}\right\}$ \\
$C^{\prime}$& 0 & $\left\{0.01 -\frac{0.81}{N},0.01 -\frac{0.07}{N},0.01 -\frac{0.04}{N},0.0004-\frac{0.003}{N},\lesssim 10^{-6} \right\}$  \\
$C^{\prime\prime}$ & 0 & $\left\{-0.13+\frac{0.69}{N},0.02 -\frac{0.38}{N},0.004-\frac{0.07}{N},0.0003-\frac{0.01}{N},\lesssim 10^{-6} \right\}$ \\
$C^{-}$ & $\left\{2.18 +\frac{30.71}{N},0.26 -\frac{0.65}{N},-0.11+\frac{1.17}{N},-0.11+\frac{1.04}{N},-0.001+\frac{0.01}{N}\right\}$ & 0   \\
$\epsilon_{\mu_1\cdots \mu_6} $ &   &$\left\{0.96 +\frac{17.38}{N},0.88 -\frac{1.61}{N},0.44 -\frac{4.28}{N},0.1 -\frac{1.2}{N},0.01 -\frac{0.15}{N},0.001-\frac{0.01}{N}\right\}$
\end{tabular}
\caption{OPE coefficients involving $\epsilon$ and $\mathbb{Z}_2$-even primaries.  Known matrix elements are highlighted in teal (from~\cite{Simmons-Duffin:2016wlq,Chang:2024whx}) and in orange (from the five-point bootstrap~\cite{Poland:2025ide}).  All extrapolations are performed using data at $N=14,15,16$. When no clear trend as a function $N$ can be identified,  we report the value at $N=16$ indicated by the symbol $\sim$. Blank entries are fixed by symmetry. }\label{Tab: epsilonEven}
\end{table}

\end{landscape}
\Bibliography[refs.bib]

\end{document}